\documentclass[letterpaper, 11pt]{article}
\usepackage{jheppub} 
            
\usepackage{braket,slashed,bm,bbm,mathrsfs,amssymb}
\usepackage{array,multirow}
\usepackage{color}
\usepackage{graphics} 
\usepackage{longtable} 
\usepackage{amsmath}
\usepackage{wrapfig}
\usepackage{enumerate}
\usepackage[toc,page]{appendix}
\usepackage{supertabular}
\usepackage{subcaption}
\usepackage{pdflscape}
\usepackage{mathtools}
\usepackage{comment}
\usepackage{mathdots}
\usepackage{xparse}
\usepackage{etoolbox}
\usepackage{booktabs}
\usepackage{multirow}
\usepackage{blkarray}
\usepackage[svgnames]{xcolor}
\usepackage{changepage}
\usepackage{enumitem}
\usepackage{hyperref}

\usepackage[framemethod=TikZ]{mdframed}
\definecolor{darkcerulean}{rgb}{0.03, 0.27, 0.49}

\newcounter{ex}[section]

\usepackage{tikz}
\usetikzlibrary{arrows,decorations.pathmorphing,backgrounds,positioning,fit,petri,automata,shadows,calendar,mindmap,decorations.markings,calc}
\usetikzlibrary{arrows.meta,bending,shapes.geometric}

\def\nn{\nonumber\\ }

\newcommand{\ba}[1]{\begin{align} #1 \end{align} }

\def\tr{{\text{Tr}}}

\def\CA{{\cal A}}
\def\CB{{\cal B}}

\def\CD{{\cal D}}

\def\CI{{\cal I}}

\def\CL{{\cal L}}
\def\CM{{\cal M}}
\def\CN{{\cal N}}
\def\CO{{\cal O}}
\def\CP{{\cal P}}

\def\CR{{\cal R}}
\def\CS{{\cal S}}
\def\CT{{\cal T}}

\def\sfb{{\mathsf{b}}}
\def\sff{{\mathsf{f}}}
\def\sfc{{\mathsf{c}}}

\def\ra{{\mathrm{a}}}

\newcommand{\specialcell}[2][c]{%
\begin{tabular}[#1]{@{}c@{}}#2\end{tabular}}

\newcommand{\udl}[1]{\mathrm{d} #1 \,}

\newcommand{\qfac}[1]{\left( #1; q \right)_\infty}

\newcommand{\thetafunc}[1]{\theta\left( #1;q\right)}

\newcommand{\bZ}{\mathbb{Z}}

\newcommand{\Rep}{\mathrm{Rep}}

\usepackage[normalem]{ulem}
\usepackage{cancel}

\newcommand{\be}{\begin{equation}}
\newcommand{\ee}{\end{equation}}
\newcommand{\bea}{\begin{eqnarray}}
\newcommand{\eea}{\end{eqnarray}}

\title{Dimensionally Reducing Generalized Symmetries from (3+1)-Dimensions}

\author[a]{Emily Nardoni,}
\author[b]{Matteo Sacchi,}
\author[c]{Orr Sela,}
\author[d]{Gabi Zafrir,}
\author[a,e,f]{Yunqin Zheng}

\affiliation[a]{Kavli Institute for the Physics and Mathematics of the Universe, University of Tokyo,  Kashiwa, Chiba 277-8583, Japan}
\affiliation[b]{Mathematical Institute, University of Oxford, Andrew-Wiles Building, Woodstock Road, Oxford, OX2 6GG, United Kingdom}
\affiliation[c]{Mani L. Bhaumik Institute for Theoretical Physics, Department of Physics and Astronomy, University of California, Los Angeles, CA 90095, USA}
\affiliation[d]{Department of Mathematics and Physics, University of Haifa at Oranim, Kiryat Tivon 3600600, Israel}
\affiliation[e]{Institute for Solid State Physics, University of Tokyo,  Kashiwa, Chiba 277-8583, Japan}
\affiliation[f]{C.~N.~Yang Institute for Theoretical Physics,  Stony Brook University, Stony Brook, NY 11794-3840, USA}

\abstract{Recently there has been an increasing interest in the study of generalized symmetries in dimensions higher than two. This has lead to the discovery of various manifestations of generalized symmetries, notably higher-group and non-invertible symmetries, in four dimensions. In this paper we shall examine what happens to this structure when the 4d theory is compactified to lower dimensions, specifically to 3d and 2d, where we shall be mainly interested in generalized symmetry structures whose origin can be linked to mixed flavor-gauge anomalies. We discuss several aspects of the compactification, and in particular argue that under certain conditions the discussed generalized symmetry structure may trivialize in the infrared. Nevertheless, we show that even when this happens the presence of the 4d generalized symmetry structure may still leave an imprint on the low-energy theory in terms of additional 't Hooft anomalies or by breaking part of the symmetry. We apply and illustrate this using known examples of compactifications from four dimensions, particularly, the reduction of  4d $\mathcal{N}=1$ $U(N_c)$ SQCD on a circle to 3d and on a sphere to 2d.}

\begin{document} 
\maketitle

\section{Introduction}
\label{sec:intro}

Compactification is a fruitful tool in the study of quantum field theory. Specifically, starting with a given higher-dimensional theory we can put it on a manifold with compact directions, resulting at low energies in a quantum system in lower dimensions. The emerging relation between the two theories can then be used to extract properties of both the higher and lower dimensional theories. One of the prime examples of such a relation is the class S construction \cite{Gaiotto:2009we,Gaiotto:2009hg}, in which a 6d  superconformal field theory (SCFT) is compactified on a Riemann surface, resulting in an SCFT in four spacetime dimensions. The ensuing connection allows us to learn much about the four-dimensional SCFTs---indeed, this method can be used to realize many 4d SCFTs whose construction by inherently 4d methods would otherwise be strenuous. Furthermore, this construction allows us to compute and constrain many of their properties, be it their global symmetries, operator spectrum, conformal manifolds, or dual descriptions.\footnote{~These aspects have been studied extensively for the compactification of 6d $(1,0)$ theories on Riemann surfaces to 4d $\mathcal{N}=1$ generalizing the class S construction (see \emph{e.g.}~\cite{Razamat:2022gpm} for a review of some of these directions), and also more recently for 5d $\mathcal{N}=1$ theories to 3d $\mathcal{N}=2$ \cite{Sacchi:2021afk,Sacchi:2021wvg,Sacchi:2023rtp,Sacchi:2023omn}.
}

Of particular importance is the relationship between the symmetries of the higher and lower dimensional theories.  The realization through compactification leads to a relation between the symmetries of the two theories, where knowledge of the symmetries of one theory can teach us about the symmetries of the other. Unfortunately, such relations are complicated primarily due to two phenomena. One is the appearance of accidental symmetries that might emerge at the end of a renormalization group flow. Such accidental symmetries can lead to the lower-dimensional theory having \emph{more} symmetries than its higher-dimensional parent. 
A common example of this is when the lower-dimensional theory acquires a symmetry due to an isometry of the compactification surface; such a symmetry is accidental from the perspective of the lower-dimensional theory, although not from the higher-dimensional viewpoint, and so can be anticipated when analyzing the higher-dimensional system on the compact surface. In general, however, it is not possible to predict nor completely rule out the appearance of accidental symmetries.\footnote{~In certain cases, symmetries that are accidental from the perspective of the lower-dimensional theory can be identified from the full compactification geometry---for instance, this has been studied in the context of the class S construction in \cite{Bah:2019jts,Bah:2019rgq,Bah:2019vmq}.}

 Another complicating feature is that some symmetries present in the higher-dimensional theory may act trivially in the infrared (IR), in which case the lower-dimensional theory would have \emph{less} symmetry than its higher-dimensional parent. While it is not possible to completely rule out the former possibility, it is possible to rule out the latter in certain cases: when 't Hooft anomalies are present.
Specifically, symmetries possessing 't Hooft anomalies cannot act completely trivially in the IR, as there must be some sector in the low-energy theory that reproduces the anomaly. Fortuitously, the 't Hooft anomalies of the lower and higher dimensional theories are also related, by integration over the compactification surface of the anomaly polynomial for continuous symmetries \cite{Benini:2009mz,Alday:2009qq} and the anomaly theory for discrete symmetries (see {\it e.g.}~\cite{Sacchi:2023omn}). This allows us to infer whether certain symmetries can end up acting trivially in the lower-dimensional theory, leading to a clearer relation between the symmetries of the two theories.

Recently, there has been a renewed interest in the subject of symmetries, motivated by the discovery of what are now called generalized symmetries \cite{Gaiotto:2014kfa}. These include both higher-form symmetries \cite{Gaiotto:2014kfa,Gaiotto:2017yup}, and the more recent non-invertible symmetries (see \emph{e.g.}~\cite{Bhardwaj:2017xup,Tachikawa:2017gyf, Kaidi:2021xfk, Choi:2021kmx, Choi:2022zal, Chang:2018iay, Bhardwaj:2022yxj, Thorngren:2019iar, Bhardwaj:2023fca, Bhardwaj:2023bbf, Bhardwaj:2024qrf, Cordova:2022ieu, Choi:2022jqy, Cordova:2022qtz}). Higher $p$-form symmetries refer to symmetries acting only on non-local ($p$-dimensional) operators, unlike ordinary 0-form symmetries which can also act on local (0-dimensional) operators. Even though these do not act on local operators, their presence can lead to interesting constraints on the dynamics \cite{Gaiotto:2017yup, Lee:2021obi, Cordova:2016emh, Cordova:2019bsd, Cordova:2020tij, Genolini:2022mpi, Choi:2023pdp}. This is especially the case when they possess 't Hooft anomalies, or mix with 0-form symmetries through what is now known as a higher-group structure \cite{Cordova:2018cvg,Benini:2018reh}. Non-invertible symmetries, broadly speaking,  refer to symmetries that do not form a group, and in particular the combination of two symmetries yields a direct sum of operations rather than a single one. These symmetries have a long history of study in the context of topological line operators in 2d theories, while in more recent years they have been understood as a generalization of ordinary symmetries. We refer the reader to the reviews \cite{Cordova:2022ruw,Schafer-Nameki:2023jdn,Bhardwaj:2023kri,Shao:2023gho} for a more complete list of citations.
Below we shall be more explicit regarding the specific generalized symmetry structure that we will consider here.   

Given this recent appreciation for more general symmetry structures, it is natural to ask how the aforementioned relation between standard symmetries and their anomalies in compactification can be extended to also involve generalized symmetries. One approach in an attempt to tackle this question is to try to understand how non-invertible symmetries and higher-group structures arise from the compactification of a higher-dimensional theory. This has been studied, notably, in the context of the class S construction by various authors \cite{Bah:2020uev,Bhardwaj:2021pfz,Bhardwaj:2021zrt,Bhardwaj:2021mzl,Bashmakov:2022jtl,Bashmakov:2022uek,Antinucci:2022cdi}. Here, we shall explore what happens to some of the generalized symmetry structures discovered in four spacetime dimensions when we compactify to lower dimensions, either to three dimensions on $S^1$ or to two dimensions on $S^2$. Our motivation is to better understand the implication of the presence of various generalized symmetry structures in the higher (in this case, four) dimensional theory for the properties of the lower-dimensional theory. Another motivation meriting the exploration of compactification to 2d is that non-invertible symmetries are quite ubiquitous there, and it is interesting if these can be related to 4d symmetry structures.

As previously pointed out, compactification naturally leads to a relation between the 't Hooft anomalies of the two theories, and this can be extended also to anomalies involving higher-form symmetries \cite{Sacchi:2023omn}. It is known that both non-invertible and higher-group structures can sometimes be related to 't Hooft anomalies, via certain topological manipulations \cite{Tachikawa:2017gyf}. It is therefore natural to expect that the relation between the anomalies can facilitate the understanding of the fate of a generalized symmetry structure upon compactification. Indeed, we shall rely here on the relation between anomalies to understand various features of the compactification of a generalized symmetry structure. 
We also point out that a natural generalization of the relation between 't Hooft anomalies through compactification is that the symmetry topological field theory (Symmetry TFT) of the higher and lower dimensional theories can be related by a similar reduction. The Symmetry TFT is a topological field theory defined in one higher dimension which, along with the specification of its topological boundary conditions, captures the generalized symmetry structure of the field theory of interest---see \cite{Gaiotto:2020iye,Apruzzi:2021nmk,Freed:2022qnc,Kaidi:2022cpf}.  Such a correspondence would then automatically provide a relation encompassing all such structures related through topological manipulations, since the Symmetry TFT is invariant under topological manipulations of the field theory. Indeed, we shall use this approach to motivate some of the relations we observe (see also \cite{Bashmakov:2022uek,Chen:2023qnv} for some recent applications of this approach).

We proceed now to specify in more detail the problems we address in this work.
We shall concentrate on specific classes of four-dimensional gauge theories, compactified on $S^1$ to three dimensions, or on $S^2$ to two dimensions. 
Notably, the 4d theories we consider possess a magnetic $U(1)^{(1)}_M$ 1-form symmetry that acts on the 't Hooft lines associated with a $U(1)_G$ part of the gauge group, which we will generally take to be continuous and non-simply connected; the case of gauge group $U(N_c)$ serves as our quintessential example. 

We in particular consider cases in which $U(1)_G$  participates in an Adler--Bell--Jackiw (ABJ) anomaly with a classical $U(1)_F$ flavor symmetry, as captured by the following term in the six-form anomaly polynomial,
	\begin{align}
	\label{fg2}
	\CI_6 \,\supset \,\frac{k}{2} c_1(F) c_1(G)^2\,.
	\end{align}
Here, $c_1(F)$ is the first Chern class for the $U(1)_F$ flavor symmetry and  $c_1(G)$ for the $U(1)_G$ gauge symmetry, and we are suppressing wedge products---see Appendix \ref{sec:appconventions} for more details on our conventions. This term~\eqref{fg2} is equivalent to the presence of an ABJ anomaly of the form $\tr\, U(1)_F U(1)_G^2 =k$.
The invertible part of $U(1)_F$ that survives the anomaly \eqref{fg2} in the quantum theory is $\mathbb{Z}_{k}$,\footnote{~Throughout this work, we assume the spacetime manifold to be a spin manifold.} but as has been recently understood \cite{Choi:2022jqy,Cordova:2022ieu}, there is actually a larger non-invertible symmetry that is preserved labeled by rational numbers $\mathbb{Q}/\mathbb{Z}$. As we will review in more detail below, this non-invertible symmetry can be understood either by stacking the $\mathbb{Z}_{k}$ symmetry defect with a suitable TQFT, or by a half-space gauging technique that involves gauging a subgroup of the magnetic 1-form symmetry. Many other types of non-invertible symmetries have been discussed recently in the literature, but for the sake of keeping our discussion tractable we will focus herein on non-invertible symmetries constructed in this way,  arising from an ABJ anomaly.   

We will furthermore consider 4d field theories in which various 0-form symmetries are extended by the magnetic $U(1)_M^{(1)}$ 1-form symmetry, leading to a 2-group. As discussed in \cite{Cordova:2018cvg}, this structure is signaled by the presence of gauge-global-global anomalies. In particular, a 2-group involving a $U(1)_F$ abelian global symmetry arises from the following anomaly term,\footnote{~Throughout we will use the letter $k$ to denote a gauge-gauge-global anomaly coefficient, and $\kappa$ to denote a gauge-global-global anomaly. Later, we also introduce $k_{F^3}$ for a global-global-global anomaly, \emph{i.e.}~the 't Hooft anomaly for the $U(1)_F$ global symmetry. } 
 	\begin{align} \label{f2g}
	\CI_6 \, \supset\,  \frac{\kappa}{2} c_1(F)^2 c_1(G) \,,
	\end{align}
which is equivalent to the presence of a mixed anomaly of the form $\tr \, U(1)_F^2 U(1)_G = \kappa$. 
Under a background gauge transformation $A_F\to A_F+d\lambda$ of the 1-form background gauge field associated to the $U(1)_F$ flavor symmetry, the action changes by $\delta S = {\frac{\kappa}{2}} \lambda c_1(F) c_1(G)$, which can be canceled by a source term for the 2-form $U(1)_M^{(1)}$  background connection $B$. This leads to the following transformation laws for the background fields (see Appendix \ref{sec:abjconventions} for details on our conventions),
	\begin{align} \label{2grouptr}
	A_F\to A_F+ d\lambda \,,\qquad
	B\to B + d\Lambda - \frac{\kappa}{{4}\pi} \lambda\, dA_F\,,
	\end{align}
with $A_F, B$ satisfying the gauge bundle constraint,
\begin{eqnarray}\label{eq:2groupbundle}
H=dB+ {\frac{\kappa}{2}}  A_F c_1(F)\,, \qquad dH = \pi \kappa c_1(F)^2 \,,
\end{eqnarray}
where $H$ is a $U(1)$ 3-form, and is invariant under \eqref{2grouptr}. 

Evidently, the non-invertible symmetry and 2-group structure are closely tied with  anomalies involving gauge symmetries: anomalies involving a gauge symmetry and two global symmetries lead to the formation of a 2-group, while anomalies involving two abelian gauge symmetries and a global symmetry lead to the global symmetry becoming non-invertible. It is well known that in a compactification, the 't Hooft anomalies of the lower-dimensional theory are related to those of the higher-dimensional one by integrating the anomaly polynomial on the compact surface; as we emphasized earlier, this prescription generally yields the contribution to symmetries of the lower-dimensional theory that are manifest in the higher-dimensional one. Below we shall see that something similar also holds for anomalies involving gauge symmetries. However, gauge symmetries are ultimately a redundancy in the description, and so are not expected to hold physical meaning. As such, how should we understand the observed mapping of anomalies involving gauge symmetries under compactification? The fact that many of these anomalies can be interpreted as leading to a generalized symmetry structure suggests that the proper interpretation of these observations should be in terms of said generalized symmetry structures. This would then also suggest the generalization of these to other cases where the anomaly interpretation is not readily available. We shall try to motivate these observations from the point of view of generalized symmetries.

\subsection{Summary of the Results}

We will now provide a summary of our main results and a roadmap for the case studies considered throughout this work. 
We restrict to classes of four-dimensional gauge theories with a $U(1)_G$ part of the (continuous) gauge group, which possess a $U(1)_M^{(1)}$ 1-form magnetic symmetry acting on the 't Hooft lines associated with $U(1)_G$, as well as various classical  0-form flavor symmetries, both abelian ($U(1)_i$) and non-abelian ($H_i$).  The anomalies involving the $U(1)_G$ gauge symmetry and the flavor symmetries are captured by the following terms in the anomaly polynomial,
\begin{align}\label{allcoeff}
	\CI_6\, \supset\,\frac{k_\ell}{2} c_1(\ell) c_1(G)^2 +  \sum_{i}  \kappa_{H_i} c_2(H_i) c_1(G)+ \sum_{i,j}\frac{{\kappa_{ij}}}{2}    c_1(i) c_1(j)c_1(G) \,,
\end{align}
where for the purpose of this discussion we have slightly generalized \eqref{fg2} and \eqref{f2g} to incorporate multiple flavor symmetries.
Theories with non-zero $k_\ell$ possess a non-invertible symmetry associated with $U(1)_\ell$ rotations by rational angles $\frac{2\pi p}{k_\ell q}$ for coprime $p$ and $q$. Meanwhile, theories with non-zero mixing coefficients $\kappa_{ij}=\kappa_{ji}$ or $\kappa_{H_i}$ enjoy a 2-group which extends the participating 0-form flavor symmetries by the magnetic 1-form symmetry. 
When both $k_\ell$ and $\kappa_{\ell i}$ are non-zero, the non-invertible defect acquires higher structure associated with junctions among defects \cite{Copetti:2023mcq}.
 Our hallmark example, whose generalized symmetries we review in Section~\ref{sec:4d}, is $\CN=1$ supersymmetric quantum chromodynamics (SQCD) with gauge group $U(N_c) = [SU(N_c) \times U(1)_V]/\mathbb{Z}_{N_c}$, with $N_f$ fundamental and anti-fundamental chiral superfields, and two additional chiral fields in the determinant and anti-determinant representations of the gauge group whose charges are designed to preserve the continuous $U(1)_R$ symmetry. We will also consider the infrared dual with gauge group $U(N_f-N_c)$. For this particular case, $U(1)_G$ is the baryonic symmetry $U(1)_B=U(1)_V/\mathbb{Z}_{N_c}$  of SQCD, under which the baryons $Q^{N_c}$ have unit charge.

First, we will consider 4d theories on $M_3\times S^1$,\footnote{~One can in general introduce a twist by a global symmetry when performing a circle compactification, even by a non-invertible symmetry (see \emph{e.g.}~\cite{Kaidi:2022uux,Giacomelli:2024sex}), however here we shall  limit ourselves to ordinary (untwisted) circle compactifications.} where the 4d gauge field $a$ has the following holonomy on the circle,
\begin{align}
	\sigma = \int_{S^1} a\,.
\end{align} 
Probing the system at high enough energy scales, $\sigma$ is a dynamical, compact scalar with $2\pi$ periodicity. The $U(1)_M^{(1)}$ 1-form symmetry reduces to the 1-form winding symmetry whose corresponding charged line operators are the winding defects generating the symmetry $\sigma\to\sigma+2\pi n$ for $n\in\mathbb{Z}$, and a magnetic 0-form symmetry that acts on monopole operators in the effective 3d theory. The non-invertible symmetry that was constructed by gauging a subgroup of the winding 1-form  symmetry reduces at these scales to the same non-invertible symmetry, now associated with gauging the winding 1-form and magnetic 0-form symmetry, so that in particular the $\mathbb{Q}/\mathbb{Z}$ symmetry associated with the ABJ anomaly in \eqref{allcoeff} reduces to the same symmetry in  this effective theory on the circle. Furthermore, the 2-group labeled by coefficients $\kappa_{ij}$ reduces to a 2-group in 3d labeled by the same $\kappa_{ij}$. 

There will be some scale below which the scalar $\sigma$ decompactifies, which generally will coincide with the KK scale $E\sim1/R$ for $R$ the circle radius. Below this decompactification scale, the magnetic 1-form symmetry in the effective theory acts trivially, and both the non-invertible symmetry and 2-group structure disappear along with it. In particular, in the deep IR limit where $E\ll1/R$, 
we can consider expanding around a sector of fixed $\sigma$; the winding defects lead to a change in these vacua and are not present in the IR theory, and the 1-form winding symmetry trivializes. 

These general observations are presented in more detail in Section~\ref{sec:discussion}, and then in Section~\ref{sec:3d} we examine a series of examples in further detail, including SQCD and its dual, and the electric-magnetic duality defects of pure Maxwell theory. Our case studies are summarized in Table~\ref{tab:cases3d}.  

	\def\arraystretch{1.1}
	\begin{table}[t!]
	\centering
\begin{tabular}{|c|c|c|c|c|}
\hline
Sec. & 4d Theory    & \specialcell[c]{Non-invertible \\ symmetry} & 2-group & \specialcell[c]{Scale of $\sigma$-\\ decompactification}  \\ \hline \hline
 \ref{sec:3dmaxwell} & $U(1)$ gauge theory  &\specialcell[t]{ S-duality defect \\ for $\frac{2\pi}{e_{4d}^2} \in \mathbb{Q}$;  \\ Condensation defect} & ----- & $R\to 0$   \\ \hline
\ref{sec:3dsqcd}  & \specialcell[t]{$\CN=1$ $U(N_c)$ SQCD \\ with det. matter }& $k_I = 2$ & \specialcell[t]{$\kappa_{Rt}, \kappa_{It}$, \\ $\kappa_{SU(N_f)_{L,R}} $\vspace{0.1cm}}     & $E \ll 1/R$ \\ \hline
\end{tabular}
\caption{A summary of the 4d gauge theories reduced on $S^1$ that are analyzed in the main text. The third column highlights the non-invertible symmetries that we discuss in the  text, and the fourth column lists the non-zero 2-group mixing coefficients as defined in \eqref{allcoeff}, where the index $R$ denotes the $U(1)_R$ symmetry, $I$ denotes the $U(1)_I$ symmetry that becomes non-invertible, and other indices refer to other flavor symmetries. For $\CN=1$ SQCD there is also an  axial symmetry which is broken to $\mathbb{Z}_{2N_f}$~by gauge anomalies.
\label{tab:cases3d}
}
\end{table}
	\def\arraystretch{1}

We then consider the (possibly twisted) compactification  of 4d gauge theories on $S^2$. In general this compactification will result in a direct sum of theories with different integer values of the gauge flux $\int_{S^2}c_1(G)$, so that sectors of fixed  flux are labeled by their charge under a magnetic 1-form symmetry that is inherited from the four-dimensional one. As each fixed-charge vacuum defines a ``universe" in 2d, even at finite volume $\text{vol}(M_2)$ of the 2d manifold we can restrict to a single sector in which the 1-form symmetry acts trivially; then,  neither the 2-group structure nor the non-invertible symmetry from \eqref{allcoeff} survive. We examine the imprint of these structures in the 2d fixed-charge sectors after twisted compactification, with gauge flux $m_G$ and flavor flux $m_{i}$ for a global symmetry $U(1)_i$,
\begin{align} \label{fluxes}
\int_{S^2} c_1(G) = m_G \in \mathbb{Z}\,,\qquad \int_{S^2}c_1(U(1)_i) = m_{i} \in \mathbb{Z}\,,
\end{align}
with the following results. 

In the presence of a non-zero gauge flux $m_G$ on the sphere, the 2-group in 4d leads to purely global 't Hooft anomalies in the 2d theory, whose form is given by substituting $c_1(G) \to m_G$ in \eqref{allcoeff}.
Furthermore, the non-invertible symmetry labeled by $k_\ell$ leads to a 2d ABJ anomaly, so that what remains of the 4d non-invertible symmetry is a  discrete invertible $\mathbb{Z}_{|m_Gk_\ell|}$ symmetry. When both $k_\ell$ and a mixing coefficient $\kappa_{\ell i}$ are nonzero, this discrete $\mathbb{Z}_{|m_Gk_\ell|}$ also possesses a global 't Hooft anomaly; for instance when $\kappa_{\ell \ell}$ is non-trivial, this is a self-anomaly of the form
 $\CI_4\supset \frac{1}{2}\kappa_{\ell \ell} m_G \beta(\ell)^2$, where $\beta(\ell)$ denotes the Bockstein map which acts on $\mathbb{Z}_{|m_Gk_\ell|}$-valued cocycles as $\beta = \delta / (|m_Gk_\ell|)$, for $\delta$  the coboundary. 
 
\def\arraystretch{1.05}
\setlength{\tabcolsep}{4.2pt}
\begin{table}[t!]
\centering
\begin{tabular}{|c|c|c|c|c|c|}
\hline
Sec. & \specialcell[c]{4d  Theory}   & \specialcell[c]{Non-invertible \\ symmetry} & 2-group & Flux  & Comments on IR \\ \hline \hline
\ref{sec:undets2}  &  \specialcell[c]{{\it twist 1}$_{[n>0]}$ \\ Table~\ref{tab:s2red}  \vspace{0.01cm}} & $k_I=2$ & $\kappa_{R,\{R,A,t,x,y\}}$ & \specialcell[c]{$m_G=0$  \\ $m_{R}=-1$}  & SUSY-breaking   \\ 
\hline
\ref{sec:undets2}  &  \specialcell[c]{{\it twist 1}$_{[n=0]}$ \\ Table~\ref{tab:s2red2} }  & $k_I=2$ & $\kappa_{R,\{A,t,z\}}$ &  \specialcell[c]{$m_G=0$  \\ $m_{R}=-1$} & \specialcell[c]{flow to 2d SCFT 
} 
\\ \hline
\ref{sec:undetsneg}   &  
\specialcell[c]{{\it twist 2} \\ Table~\ref{tab:charges4dnew} }   &  
$k_I=2$ & 
\specialcell[c]{$\kappa_{R,\{R,A,t,x,y\}}$ \\ $\kappa_{It},\kappa_{x,\{x,y\}},\kappa_{yy}$ \\ $\kappa_{SU(N_{1,2,3,f})}$ }  
&    \specialcell[c]{$m_G=0$  \\ $m_{R}=-1$ \vspace{-0.15cm}   \\ \dotfill \\ $m_G=N_c$  \\ $m_{R}=-1$   \vspace{-0.15cm}  \\ \dotfill \\ $m_{G_1}=1$ \\ $m_{G_2}=-\frac{1}{N_c}$  \\ $m_{R}=-1$}  & \specialcell[c]{ accidental $SU(2)_{\text{ISO}}$; \\   examples highlight \\  matching of gauge \\ anomalies between \\ 4d and 2d  }  \\    
\hline
 \ref{sec:triality}   &  \specialcell[c]{ {\it twist 3} \\ Table~\ref{tab:4dchargetri} }  & $k_I=2$ & \specialcell[c]{$\kappa_{R,\{R,A,t,I,2,3\}}$ } & \specialcell[c]{$m_G=0$  \\ $m_R=-1$}
 &
  \specialcell[c]{ accidental $SU(2)_{\text{ISO}};$  \\  $(\mathbb{Q}/\mathbb{Z})_I\to U(1)$;\qquad  \\  4d duality $\to$ 2d triality } \\
  \hline
\end{tabular}
\caption{A summary of case studies presented in Section~\ref{sec:2d}, from reducing  4d $\CN=1$ $U(N_c)$ SQCD with determinant matter on the sphere with different R-symmetry twists.
The third column highlights the $(\mathbb{Q}/\mathbb{Z})_I$ non-invertible symmetry that we discuss in the  text, which arises from a classical $U(1)_I$ symmetry. The fourth column lists the non-zero 2-group mixing coefficients, where the index $R$ denotes the $U(1)_R$ symmetry, $A$ denotes the  axial symmetry which is broken to $\mathbb{Z}_{2N_f}$ by gauge anomalies, and  other indices refer to other flavor symmetries present in the 4d theory.  The values of the gauge and R-symmetry fluxes defined in \eqref{fluxes}  are denoted in the fifth column. The last column highlights some features of the IR 2d theories.  \label{tab:cases2d}
}
\end{table}
	\def\arraystretch{1}
	\setlength{\tabcolsep}{6pt}
 
Meanwhile in the zero flux sector $m_G=0$, the 4d non-invertible symmetry labeled by $k_\ell$ reduces in general to an invertible $\mathbb{Q}/\mathbb{Z}$ symmetry in 2d, which enacts (now invertible) rotations by the same angles $\frac{2\pi p}{k_{\ell} q}$ that were present in 4d.\footnote{~In 2d this symmetry might accidentally enhance back to a full $U(1)$ or become part of another $U(1)$ symmetry---we will see examples of this in Section~\ref{sec:2d}.} Then for any $m_G$, the non-invertible symmetry always reduces to an invertible symmetry in the sectors with fixed gauge flux, as anticipated. 

In the presence of a flavor flux $m_i$, a non-zero mixed anomaly coefficient $\kappa_{ij}$ leads to a 2d ABJ anomaly that suggests the descendent of the $U(1)_j$ symmetry is broken to  $\mathbb{Z}_{|m_i\kappa_{ij}|}$. (Note that we only consider compactifications with flavor flux $m_{i\neq \ell}$, since twisting with respect to a subgroup of the 4d non-invertible symmetry leads to a pure gauge anomaly in 2d which renders the theory inconsistent). More generally, if multiple coefficients $\kappa_{ij}$ are nonzero, then one linear combination of these symmetries will be anomalous in two dimensions. We will specifically consider supersymmetric examples in which there is a flux for the 4d R-symmetry on the sphere, and the infrared R-symmetry at the putative 2d fixed point can be determined as mixing with the remaining non-anomalous symmetries inherited from 4d.

In Section~\ref{sec:discussion} we present arguments for these conclusions from a variety of perspectives, including from integrating the perturbative anomaly polynomials, from the transformations of the background fields, as well as from reducing the topological defects themselves. In Section~\ref{sec:symtft} we show that the same conclusions may be obtained by reducing the 5d Symmetry TFT. Then, in Section~\ref{sec:2d} we study a series of examples that arise from  4d $\CN=1$ $U(N_c)$ SQCD with determinant matter. We reduce this theory and its infrared dual on $S^2$, mixing the R-symmetry with various subgroups of the abelian flavor symmetries (which we label as {\it twists 1,2,3}), and threading R-symmetry flux through the sphere so as to obtain  $\CN=(0,2)$ theories in two dimensions.
Doing so provides interesting examples for which accidental symmetries of the 2d theories can be identified from the compactification geometry, for example from the $SU(2)_{\text{ISO}}$ isometry of the sphere,  and furthermore allows us to reproduce the 2d triality of \cite{Gadde:2013lxa}.  A summary of our case studies appears in Table~\ref{tab:cases2d}.

\section{Reducing Generalized Symmetries}
\label{sec:discussion}

In this section, we discuss how a non-invertible symmetry arising from an ABJ anomaly of the form \eqref{fg2}, and a 2-group arising from an anomaly of the form \eqref{f2g}, reduce from four dimensions to three or two dimensions, on $S^1$ or $S^2$ respectively. This section serves as  a general overview of the methods and subtleties involved, as well as a summary of the main conclusions which arise from the examples studied in later sections.

\subsection{Compactifying with Gauge Holonomy}
\label{sec:holonomy}

We consider the reduction of 4d theories on $S^1$ to 3d, and on $S^2$ to 2d, in general with some flux or holonomy {of either gauge symmetry or global symmetry} threading the compact space. {The compactification with a flux or holonomy of the global symmetry is the standard twisted compactification, enacted by demanding the twisted boundary condition of various fields along the compact direction. However,  compactification with a fixed flux or holonomy of a \emph{gauge symmetry} is more subtle. }
The issue is as follows. Consider a gauge theory with gauge group $G$ on $M_4$. In the path integral, we should sum over all $G$ gauge field configurations. Different $G$-configurations can be organized by (for example) the gauge holonomy on a cycle $S^1$, the gauge flux on $S^2$, and similar quantities defined on higher-dimensional manifolds. Therefore in the path integral, one should sum over all possible gauge holonomies and fluxes.  However, it is often the case in the literature of compactifications on $S^1$ and $S^2$ that one only sums over gauge field configurations with \emph{fixed} gauge holonomy or gauge flux, respectively.  Let us now comment on the validity of such an operation.

\subsubsection{Gauge theories on $S^1$ for $G$ continuous}

We first discuss the case when the gauge group $G$ is a compact, continuous group. When we compactify the theory on $M_3\times S^1$ and take $S^1$ to be small compared to $M_3$, one can define the Polyakov loop operator,
	\begin{eqnarray}\label{eq:poly}
	\Omega(\vec x) = \tr_{\CR}\CP\exp\left(i \int_{S^1} a\right)
	\end{eqnarray}
where $\CP$ is the path ordering operator, and $a$ is the dynamical $G$ gauge field.\footnote{~We generally reserve lowercase letters for dynamical gauge fields, and uppercase for background fields.} The Polyakov operator is a local and generically non-topological operator for the 3d compactified theory that depends on the 3d coordinate $\vec x$, although it is an extended operator in the 4d theory. Since we integrate over all possible configurations of $a$ in the 4d path integral, in the 3d path integral we should also integrate over all possible $\Omega$ configurations. 

Does it make sense to consider a theory for which we only sum over a subset of $\Omega$ configurations? We argue that it is possible in a special situation, as follows. Generically, a potential $V[\Omega]$ for the scalar $\Omega(\vec x)$ can be dynamically generated, and one can solve for its minimum. Consider a situation in which there are multiple minima of the potential labeled by $\omega_i$, so that in the infinite volume limit $\text{Vol}(M_3)\to \infty$, all the states in the Hilbert space split into the direct sum of several superselection sectors according to the $\omega_i$'s. In this case it makes sense to consider a sub-Hilbert space, spanned by the states $\ket{\psi,i}$ with fixed $\Omega(\vec x)$ expectation value, $\bra{\psi, i} \Omega(\vec x) \ket{\psi, i} = \omega_i$. 
However, when we deviate away from the deep infinite volume limit the energy barrier between different superselection sectors becomes finite, and hence states belonging to different sectors are no longer orthogonal. Thus, for finite volume one should sum over all possible $\Omega$ configurations.

\subsubsection{Gauge theories on $S^1$ for $G$ finite and abelian}

When $G$ is finite and abelian, we can still consider the loop operator \eqref{eq:poly}. The new feature here is that the Polyakov operator is independent of the coordinate $\vec{x}$, and hence is a topological local operator in 3d, which generates a $2$-form symmetry $G$.\footnote{~Strictly speaking the symmetry should be $\Rep(G)$, but since we are considering abelian $G$, $\Rep(G)\simeq G$ and we do not distinguish them. } This 2-form symmetry descends from the 2-form symmetry generated by the Wilson line of the finite $G$ gauge theory. 

It is known that the existence of a topological $(d-1)$-form $G$ symmetry implies $|G|$ number of universes (see {\it e.g.}~\cite{Sharpe:2022ene} and references therein). The universe is a strengthened concept of the superselection sector, where the states belonging to different universes have zero overlap even at finite volume. Therefore, for any $M_3$ (which can have finite volume), the Hilbert space splits into a direct sum of orthogonal sub-Hilbert spaces, distinguished by the expectation value of the topological local operator $\Omega$. In summary, it is valid to fix the value of $\Omega$ in the path integral.

\subsubsection{Gauge theories on $S^2$ for $G$ continuous}

When $G$ is continuous, the flux operator, 
	\begin{eqnarray}\label{gamma}
	\Gamma= \exp\left(i \alpha \int_{S^2} c_1(G)\right) 
	\end{eqnarray}
is a topological operator, and generates a 1-form symmetry. For example, the magnetic 1-form symmetry current  for gauge group $G=U(1)$ is  $j_M\sim \star c_1(G)$, so that \eqref{gamma} is the exponentiation of the conserved Noether's charge.
If we put the theory on $M_2\times S^2$ and take the size of $S^2$ to be much smaller than $M_2$, then $\Gamma$ generates a 1-form symmetry in 2d. For example, when $G=U(1)$ the 1-form symmetry is $U(1)^{(1)}$,  and when $G=SO(3)$ the 1-form symmetry is $\mathbb{Z}_2^{(1)}$. 
Since $\Gamma$ is topological, it also defines universes in the 2d theory. The discussion in the previous case therefore applies, and it is valid to path integrate over fixed values of $\Gamma$.

\subsection{Reducing from 4d to 3d}
\label{sec:intro3d}

We next discuss general features regarding the reduction of non-invertible symmetries and 2-group structure from 4d to 3d. Our discussion here is focused on theories which possess an ultraviolet (UV) Lagrangian description, as is the case in all the examples we study in later sections. As we comment at the end of this subsection, we stress that more general theories lacking a Lagrangian description might exhibit a different behavior.   

It is crucial that we carefully specify the different possible limits in this setup. Throughout, we will distinguish between the following regimes.  

\begin{itemize}
\item First we examine the effective 4d theory on $S^1$  with finite radius $R$, corresponding to energy scales of order $1/R$. This coincides with the standard compactification to lower dimensions, in which one formulates the quantum theory on $M_{3}\times S^1$, so that by taking $R\to \infty$ one recovers the full four-dimensional theory. As can perhaps be anticipated, we will see that this effective theory retains much of the generalized symmetry structure of the 4d parent. 
\item We next consider the system at energy scales in the deep infrared, much smaller than $1/R$. In this limit we probe the system below the scale at which we have integrated out the KK tower, keeping $R$ small but finite. As we will discuss in detail below, we in general expect that some of the generalized symmetry structure trivializes in this limit.     
\item Finally, the limit $R\to 0$ coincides with the strict dimensional-reduction limit, in which the least hint of the 4d generalized symmetry structure is retained by the lower-dimensional theory.  
\end{itemize}

\subsubsection{Reducing the 1-form symmetry}

In compactifying a 4d field theory on $M_3\times S^1$, the holonomy of the 4d vector field $a$ on the circle yields a dynamical compact scalar $\sigma$,
	\begin{align} \label{sigma}
	\sigma = \int_{S^1} a\,,\qquad \sigma \sim \sigma + 2\pi\,.
	\end{align}
The Wilson line wrapping the circle becomes the exponentiated scalar $e^{i\sigma}$, which is the analogue of the Polyakov operator \eqref{eq:poly}. {As discussed in the previous subsection, $\sigma$ is a non-topological local operator and should not be assigned a fixed value.  }

The magnetic 1-form symmetry plays a principal part in determining the reduction of the generalized symmetry structure, and so we first consider the reduction of the $U(1)^{(1)}_M$ symmetry, as discussed in \cite{Gaiotto:2014kfa}.\footnote{~When the theory lacks matter, there is also a $U(1)_E^{(1)}$ electric 1-form symmetry under which the Wilson line changes by a phase. This symmetry will play a role in the example in Section~\ref{sec:3dmaxwell}.} 
Let us first consider the effective theory compactified on $M_3\times S^1$, at finite circle radius $R$. 
The 't Hooft lines wrapping the circle reduce to monopole operators in 3d, and the symmetry acting on them becomes a 0-form symmetry acting on said monopole operators. Meanwhile, an 't Hooft line not wrapping the circle is expected to reduce to an 't Hooft line in 3d, which has the property that if we take the Wilson line operator that wraps the circle around them, then it transforms by a phase $e^{i\sigma}\to e^{i \sigma + 2\pi i n}$. This means that the 3d 't Hooft lines are defined by the property that the scalar undergoes a $2\pi n$ rotation when going around them.  
Therefore the magnetic 1-form symmetry reduces in the 3d effective theory to a magnetic 1-form winding symmetry that acts on the winding defect operators (whose conserved current is $j_w\sim \star d\sigma$), and a ``topological'' 0-form symmetry acting on the 3d monopole operators. In particular, the compactness of the scalar is crucial to the survival of the 1-form winding symmetry in 3d, since it leads to the winding lines. 

In the deep IR, far below the KK scale $1/R$, there are two possible scenarios. If there is no charged matter in the theory, then the scalar $\sigma$ remains compact and the conclusion of the previous paragraph continues to hold (although this too is lost in the strict dimensional-reduction limit where the radius is taken to zero). 
However, in theories with matter the scalar will generally decompactify below the scale in which the KK tower is integrated out, even when $R$ is kept finite. For a concrete example (discussed in Section~\ref{sec:3dmatter}) consider $U(1)$ gauge theory with a charged Weyl fermion, and periodic boundary conditions for the fermion around the compactification circle.
After integrating out the KK massive fermions the action is no longer invariant under shifts of $\sigma \to \sigma + 2\pi$, {\it i.e.}~the compactness of the scalar is lost in the restriction to zero modes on the circle, so that the winding lines accounting for the 1-form symmetry are not present.

The natural interpretation of this is as follows. In the compactification, we can introduce a holonomy on the circle for the 4d gauge field, which sets the background around which we expand in the IR. The low-energy theory has a moduli space of vacua associated to this background value, so that the $\bZ$ shift symmetry $\sigma \to \sigma+ 2\pi$ of the scalar is spontaneously broken at low energies. Note that this is compatible with the discussion in the previous subsection regarding superselection sectors: generically a shift-symmetry-preserving potential  $V[\sigma]$ is generated, and its minima spontaneously break the shift symmetry. Now, the 3d winding defects lead to a change in vacua as we go around them. We expect that they become massive and are thus not present at low energies, so that the 1-form winding symmetry trivializes.

\subsubsection{Consequences for generalized symmetry structure}

Let us examine the consequence of the possible trivialization of the 1-form symmetry for the reduction of the 2-group  \eqref{2grouptr}, involving the 0-form $U(1)_F$ flavor symmetry and $U(1)^{(1)}_M$ magnetic 1-form symmetry.
At the level of the background fields, the 4d vector field reduces to a 3d vector field, $a_{4d}\to a_{3d}$; the magnetic 1-form symmetry background field $B_{4d}$ reduces to the field which enacts the would-be 3d 1-form symmetry, $B_{3d}$, and its holonomy on the circle reduces to a 1-form gauge field that enacts the 3d 0-form symmetry, $\int_{S^1} B_{3d}\to \tilde{A}_{3d}$. The 3d 1-form symmetry associated with the winding lines forms a 2-group with the 3d 0-form symmetry coming from the 4d 0-form symmetry, where the transformation of the background fields is inherited from the 4d transformation laws \eqref{2grouptr}. As such, the survival of the 2-group structure in 3d requires the scalar $\sigma$ to be compact, and trivializes in the limit that it decompactifies. This is consistent with what is expected for a 2-group, in the following sense. The 2-group is an extension of a 0-form symmetry by a 1-form symmetry, so that the 1-form symmetry is the subgroup while the 0-form symmetry is the quotient. When a symmetry acts trivially, the remaining faithfully acting symmetry is the quotient of the full symmetry by the one that acts trivially, consistent with the fact that we lose the 1-form symmetry without losing the 0-form symmetry. 

We next consider the consequence for the 4d non-invertible symmetry arising from \eqref{fg2}.  Again, whether or not it survives the dimensional reduction depends on whether the scalar decompactifies and the 1-form symmetry trivializes, since the 4d non-invertible symmetry arises from gauging a discrete subgroup of the magnetic 1-form symmetry. We expect it to reduce in 3d to a non-invertible symmetry which involves gauging a discrete subgroup of the 3d winding symmetry, and thus its fate depends on whether or not this symmetry trivializes.

An example that we will study in some detail is $U(N_c)$ $\CN=1$ supersymmetric QCD.  This theory possesses a non-invertible symmetry whose discrete invertible part is $\mathbb{Z}_2$, as well as a 2-group that combines the magnetic 1-form symmetry with various 0-form symmetries.\footnote{~We will actually focus mainly on the theory with additional matter  designed to cancel the ABJ anomaly involving the R-symmetry, but both this and the original theory have these structures.} In the effective 3d theory at energy scales of order $1/R$, the four-dimensional 2-group structure reduces to a 2-group involving the 3d 1-form symmetry and the various 0-form symmetries, and the non-invertible symmetry similarly reduces to a non-invertible symmetry in the effective theory on $M_3\times S^1$. Both of these structures trivialize in the low energy limit where $E\ll 1/R$, in which the compactness of the scalar is lost and the magnetic 1-form symmetry trivializes. Then, the 4d defects become invertible in 3d.

While we largely focus on non-invertible symmetry defects that arise from an ABJ anomaly of the form \eqref{fg2}, it should be noted that other types of non-invertible defects might exhibit different behavior under circle compactification.
One example in which the non-invertible symmetry survives to the deep IR of the compactification is the non-invertible electric-magnetic duality symmetry of pure Maxwell theory. In reducing this theory on a circle, the non-invertible symmetry at special values of the coupling reduces to a non-invertible symmetry in the  effective theory of 3d Maxwell plus a periodic scalar, which can be constructed purely in three dimensions by gauging discrete subgroups of the shift and winding symmetries. We will discuss this example in more detail in Section~\ref{sec:3dmaxwell}.

\subsubsection{Reducing the anomaly polynomial}

These features can be studied at the level of reducing the anomaly polynomial. This is of course somewhat subtle since there are no fermion anomalies in three spacetime dimensions, but as we will see it is still a useful exercise that reproduces the same conclusions. 

Let us first consider the compactification of the ABJ anomaly \eqref{fg2} on the circle, which is achieved by introducing the compact scalar $\sigma = \int_{S^1} a$.  As commented in Section~\ref{sec:holonomy}, $\sigma$ is a dynamical field and should not be assigned a fixed value. This leads to the following inflow action for the effective 3d theory\footnote{~Note that further introducing a twist by the $U(1)_F$ flavor symmetry (\emph{i.e.}~$U(1)_F$ twisted compactification) does not lead to an additional anomaly. To see this, suppose the holonomy of $U(1)_F$ along $S^1$ is $\Sigma_F = \int_{S^1} A_F$. Then \eqref{fg2} induces the 5d term $\frac{k \Sigma_F}{2}\int c_1^2(G)$, which is not an anomaly in the absence of time reversal symmetry.} (our conventions for this action are discussed around \eqref{inflow}),
	\begin{align} \label{inf3da}
	\CA^{\text{inf}}_4 \, \supset \,\frac{k}{2\pi} \int_{M_4} {A_F}\wedge c_1(G)\wedge d\sigma \,.
	\end{align}
This can be interpreted as an ``ABJ-like'' anomaly involving the dynamical $G$-gauge symmetry, the $(-1)$-form $U(1)$ {gauge} symmetry associated to the dynamical compact scalar $\sigma$, and the $U(1)_F$ flavor symmetry. 
This  anomaly implies that again in the effective theory on $M_3\times S^1$ (that is at energies of order $1/R$), the $U(1)_F$ flavor symmetry is broken to a non-invertible symmetry whose invertible subgroup is $\mathbb{Z}_{k}$. In the regime that the scalar is compact and the 1-form symmetry survives, this non-invertible symmetry can be understood by several complementary perspectives: (1) starting from this ABJ-like anomaly and stacking with the appropriate TQFT; (2) half-space gauging a finite subgroup of the 1-form winding symmetry generated by $d\sigma$ and magnetic 0-form symmetry generated by $c_1(G)$ directly in the effective 3d theory; and (3) by directly compactifying the topological defect on the circle. 
Notably,  in the $R\to 0$ limit this term explicitly trivializes, and as we will explain in more detail in Section~\ref{sec:3dsqcd},  these features are no longer present in the IR theory. 

We similarly consider the reduction of the global-global-gauge anomaly \eqref{f2g}. Integrating over the circle yields the following effective inflow action, 
	\begin{align} \label{inf3db}
	{\CA_4^{\text{inf}}\, \supset\, \frac{\kappa}{4\pi} \int_{M_4} A_F\wedge c_1(F) \wedge d\sigma\,.}
	\end{align}
This term is canceled by precisely the same 2-group structure \eqref{eq:2groupbundle}. However, we emphasize again that this term explicitly goes to zero in the limit that the radius of the circle is taken to zero.

\subsubsection{Regarding the fate of symmetries in dimensional reduction}

The above discussion suggests that most of the generalized symmetry structure we observe in 4d is lost once we compactify to 3d. Central to this claim is that the shift 0-form and winding 1-form symmetries are usually lost once we flow to the IR (specifically, integrating out the KK tower). However, we emphasize that this conclusion is based on a Lagrangian gauge-theory description, and still leaves open the question of what happens when there is no such description. 

For instance, consider 4d $\CN=4$ super Yang-Mills. This theory possesses a conformal manifold parameterized by the complex coupling, and electric-magnetic duality generically maps the theory with one value of the coupling to another with a different value, changing the global form of the gauge group.   
The theory with gauge group $SU(N)$ has an electric $\mathbb{Z}_{N}^{(1)}$  1-form symmetry, while the $SU(N)/\mathbb{Z}_N$ version instead has a magnetic $\mathbb{Z}_{N}^{(1)}$ symmetry. If we compactify the weakly-coupled $SU(N)$ version to 3d, then we expect the low-energy theory to have a $\mathbb{Z}_N^{(1)}$ 1-form symmetry, with the 0-form shift symmetry broken. However, if we compactify the weakly-coupled $SU(N)/\mathbb{Z}_N$ version, then we expect to get a $\mathbb{Z}_N^{(0)}$ 0-form symmetry, now losing the 1-form winding symmetry.\footnote{~Notice that these two weakly-coupled theories (which have the familiar Lagrangian descriptions) are not dual to each other, and neither are the theories obtained by compactifying them to 3d. Instead, the dual of each version of the weakly-coupled theory is given by the other version at strong coupling.} 
So, what symmetry do we expect to get if we reduce this 4d SCFT on some random point on the 4d conformal manifold where no Lagrangian description exists?  The naive expectation from reducing the gauge theory is not sufficient to answer this question.

\subsection{Reducing  from 4d to 2d}
\label{sec:intro2d}

We now proceed with  a discussion of the compactification of 4d theories to 2d on $S^2$ in the presence of fluxes for flavor symmetries $U(1)_F$ and abelian gauge symmetries $U(1)_G$, whose mixed anomalies imply non-invertible and 2-group symmetries in 4d. In Section~\ref{sec:2d}, we will then consider the example of the reduction of the 4d $\mathcal{N}=1$ $U(N_c)$ SQCD to 2d on a sphere, showing how all the features discussed in this section manifest.

\subsubsection{Reducing the 1-form symmetry}

First let us consider the fate of the 1-form symmetry when compactifying the theory on $M_2\times S^2$.  The 4d gauge connection $f_{4d}$ reduces to its 2d counterpart $f_{2d}$, and its flux $\int_{S^2} c_1(G)$ on the compact space yields a (would-be) dynamical integer-valued scalar. The 2-form background gauge field $B_{4d}$ reduces to a 2-form $B_{2d}$ which enacts the would-be 2d 1-form symmetry, and its holonomy $\int_{S^2}B_{4d}$ reduces to a scalar which generates a $(-1)$-form symmetry, whose corresponding current is $j\sim \star f_{2d}$. 

Generically, as we argued in Section~\ref{sec:holonomy}, the full 2d theory consists of the direct sum of theories with all different integer-valued gauge fluxes \cite{Gadde:2015wta}. In the deep IR these different vacuum sectors should have zero overlap, and one can consider a vacuum labeled by a fixed gauge flux. In fact, for abelian gauge groups, it is possible to physically realize this choice as follows. We can consider gauging a $\mathbb{Z}_p$ subgroup of the magnetic 1-form symmetry. This eliminates monopole configurations from the sum whose charge is not divisible by $p$. Additionally, we gain a new electric $\mathbb{Z}_p$ 1-form symmetry, and can consider introducing a holonomy for it on the sphere. This holonomy fixes the value of the magnetic flux modulo $p$. This then allows us to fix the magnetic charge if $p$ is taken to be sufficiently large.\footnote{~In practice, the sum over monopole charges appears to eventually truncate so only a finite value of $p$ is needed, see \cite{Gadde:2015wta}.} Note that this would not work for monopoles in non-abelian symmetries such as $SU(N)$, where the 't Hooft lines do not carry a 1-form symmetry charge. While one still expects the fixed-charge vacuum to give a well defined theory in such cases, it is not clear how one can isolate this vacuum in a physical construction.

It is interesting to consider what happens to the 2d 1-form symmetry in the fixed-flux sector. Recall that the topological operator for it is given by the 4d one \eqref{gamma}, $\Gamma \sim \exp (i \alpha \int_{S^2} c_1(G))$, wrapping the 2-sphere. However, $\int_{S^2} c_1(G)$ is fixed in this sector so that the topological operator appears to become non-dynamical. This suggests that the 2d 1-form symmetry acts trivially once we focus on a vacuum labeled by a fixed gauge flux. We therefore expect that neither the 2-group structure nor the non-invertible symmetry will be present in the IR limit of the compactification, though we shall next show that they still leave an imprint on the resulting 2d theory.

\subsubsection{Reducing the anomaly polynomial}

Let us next consider what we can learn from the reduction of the anomaly polynomial of the 4d theories on $S^2$. 
We note the following possible behaviors:\footnote{~For the purpose of streamlining the discussion we will take \eqref{fg2} and \eqref{f2g} as our starting point; the generalization to multiple flavor symmetries of the form \eqref{allcoeff} is straightforward.}

\begin{enumerate}

\item \label{item1}
First consider a 4d theory with a 2-group originating from the anomaly term \eqref{f2g}, and compactify it on $S^2$ with a fixed gauge flux,
\begin{align}
\label{gfluxx}
\int_{S^2} c_1(G)=m_G\,.
\end{align}
Integrating  \eqref{f2g} in the presence of \eqref{gfluxx} yields the following  term in the 2d anomaly polynomial, 
\begin{align}
\CI_4\, \supset  \, \frac{1}{2} {\kappa m_G}c_1(F)^2\,.
\end{align}
In other words, the 2-group structure in 4d  has reduced in the presence of a gauge flux to an 't Hooft anomaly for the 2d remnant of the $U(1)_F$ flavor symmetry.  

\item  \label{item2}
Next consider a 4d theory with the same anomaly \eqref{f2g}, and this time compactify with a flavor flux,
\begin{align}
\int_{S^2} c_1(F)=m_F\,. 
\end{align}
Integrating the anomaly polynomial this time yields the 2d gauge anomaly term,
\begin{align}
\CI_4\, \supset\,  {\kappa m_F}c_1(F) c_1(G)\,. 
\end{align}
In other words, the 2-group structure in 4d has reduced after compactification with a flavor symmetry flux to the anomalous breaking,
\begin{align}
U(1)_F\to \mathbb{Z}_{|m_F \kappa|}\,.
\end{align}

\item \label{item3}
Now consider the non-invertible symmetry from \eqref{fg2}, whose invertible part is $\mathbb{Z}_{k}$. In the presence of the ABJ anomaly \eqref{fg2} it does not make sense to compactify with a non-trivial flavor flux $\int_{S^2} c_1(F)= m_F$, as this gives a pure gauge anomaly of the 2d theory $\CI_4\supset \frac{1}{2}k m_F c_1(G)^2$, and so we consider only the case of non-trivial gauge flux. When we compactify with a gauge flux  $\int_{S^2} c_1(G)=m_G$, integrating the anomaly polynomial yields,
\begin{align}
\CI_4\, \supset \, {km_G}c_1(F) c_1(G)\,,
\end{align}
and therefore an ABJ anomaly in 2d. In other words, if we compactify with gauge flux a 4d theory with a non-invertible symmetry resulting from an anomaly \eqref{fg2}, we obtain in 2d the following anomalous breaking to a discrete invertible symmetry, 
\begin{align}
U(1)_F\to \mathbb{Z}_{|m_Gk|}\,.
\end{align}

\item \label{item4}
When compactifying the 4d theory with non-invertible symmetry from \eqref{fg2} and no flux for either $U(1)_F$ or the gauge symmetry, there is no general statement and the result will depend on the details. If the 1-form symmetry that participates in the half-space gauging construction of the non-invertible symmetry trivializes in 2d, then the non-invertible symmetry would generally become invertible, much as in our discussion of the 4d non-invertible symmetry on a circle. This is the case for all the examples considered in this work.\footnote{~One might expect that with the right twisted compactification, there are cases for which the non-invertible symmetry survives as a 2d non-invertible symmetry. We leave the exploration of such examples to future work.}

\end{enumerate}

These expectations obtained by integrating the anomaly polynomial can also be motivated otherwise.  We will next discuss different such ways of analyzing the outcome of reducing generalized symmetries on a sphere, and thereby reproduce the expectations presented above in a gauge-independent setting. Moreover, in Section~\ref{sec:2d} we will verify these expectations in the example of  4d $\mathcal{N}=1$ $U(N_c)$ SQCD compactified on a sphere to 2d.

\subsubsection{Reducing the background fields}

Consider the scenario of the second bullet point \ref{item2} above, reducing the 2-group with a non-zero flavor flux $\int_{S^2} c_1(F)= m_F$. Integrating the 2-group transformation \eqref{2grouptr} on the sphere, and denoting $\Phi= \int_{S^2} B_{4d}$, we see that\footnote{~An alternative way to find \eqref{bundle2} is by reducing the 2-group bundle constraint \eqref{eq:2groupbundle}. Denote $\int_{S^2} H = \Sigma$. Integrating both equations in \eqref{eq:2groupbundle} on $S^2$, we get
	$$\Sigma = d \Phi + \kappa m_F A_F, \qquad d\Sigma = 2\pi \kappa m_F c_1(F)\,.$$
	Since $H$, hence $\Sigma$, is invariant under gauge transformations  $A_F\to A_F\to d\lambda$, the transformation of $\Phi$ is given as \eqref{bundle2}.
}
	\begin{eqnarray} \label{bundle2}
\Phi\, \to\, \Phi - \kappa m_F \lambda \ \ \text{ mod } 2\pi\,.
	\end{eqnarray}
In the 2d theory, $\Phi= \int_{S^2} B_{4d}$ is the holonomy of the magnetic 1-form symmetry on the compact surface, which is a constant mod $2\pi$---changing it would change the theory, as per the discussion around \eqref{gamma}. This implies $\lambda \in \frac{2\pi}{\kappa m_F} \bZ$. 
We thus see that $U(1)_F$ is broken to $\mathbb{Z}_{|\kappa m_F|}$, in accordance with the expectation from directly integrating the anomaly polynomial.

We can similarly consider reducing the 2-group with a non-zero gauge flux as per the first bullet point \ref{item1} above, following the similar discussion in \cite{Cordova:2018cvg}. Here we want to consider the 2d theory obtained in a fixed-flux sector. For this, we formally set $c_1 (G) = m_G e_2(S^2) + c_1 (G)_{2d}$, where $e_2(S^2) $ is a unit 2-form on the sphere and $c_1 (G)_{2d}$ is only valued on the 2d spacetime (see Appendix \ref{sec:bott} for our conventions). We note that the coupling $\int_{M_4} B_{4d} c_1 (G)$ between the 2-form background field for the magnetic 1-form symmetry and the magnetic charge now leads to the  counterterm $m_G\int_{M_2} B_{2d}$ in the 2d action expanded around the fixed-flux vacuum. Due to the 2-group transformation law \eqref{2grouptr}, this term leads to an anomalous shift under $U(1)_F$ transformations of the form: $-\frac{m_G}{2}\kappa \int_{M_2} \lambda c_1(F)$.  This term signals the presence of an 't Hooft anomaly of the form $\frac{1}{2}m_G\kappa  c^2_1(F)$, as expected from integrating the anomaly polynomial.  

\subsubsection{Reducing the topological defects}

\begin{figure}
\center
\includegraphics[width=0.66\textwidth]{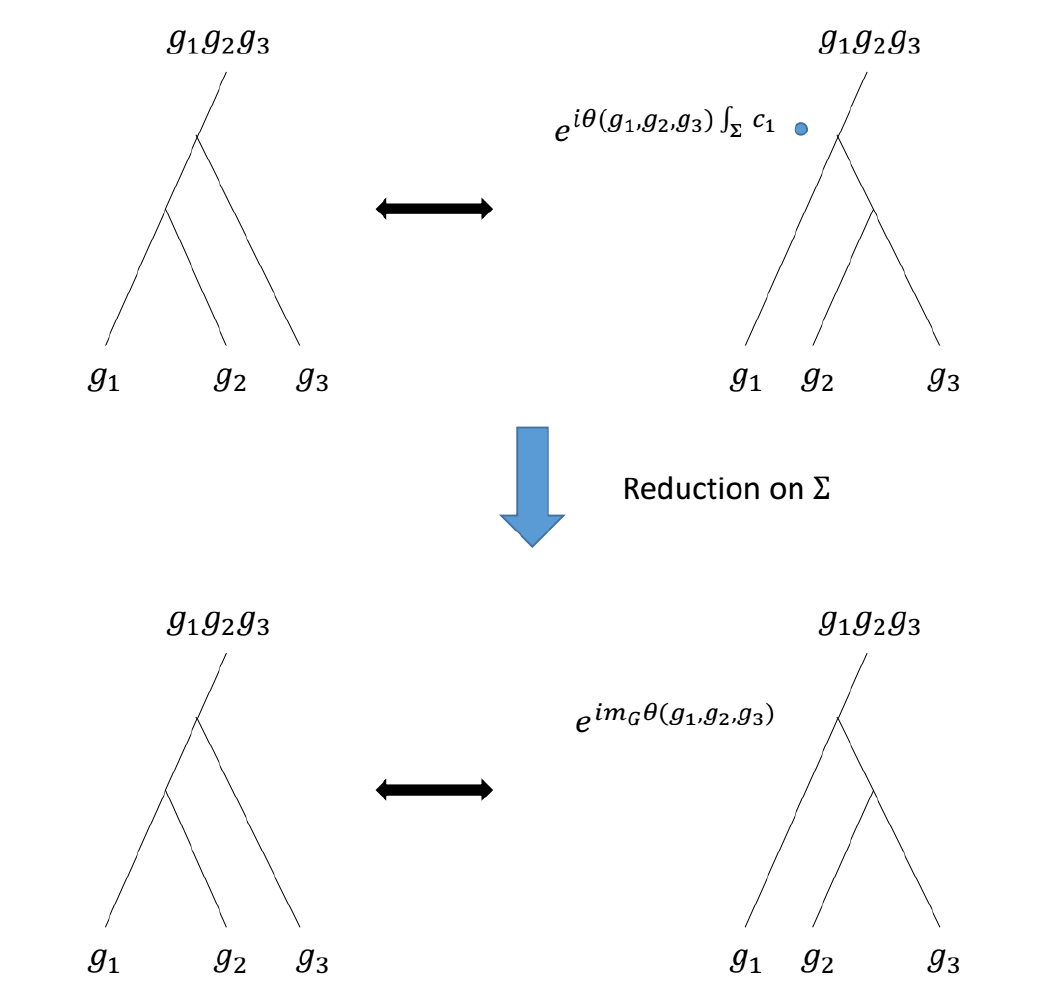} 
\caption{An illustration of the reduction of a 2-group structure in 4d to an 't Hooft anomaly in 2d. The upper picture illustrates the 4d case while the lower one describes the 2d case. The lines represent the topological operators of the 0-form symmetry in the 2d spacetime (in 4d these fill the 2 compact directions that are suppressed in the picture). In 4d, performing the F-move generates the topological operator associated with the 1-form symmetry, illustrated as the blue dot (here again extending in the 2 compact directions). When reducing on the compact surface, $\Sigma$, with fixed gauge flux, this implies that performing the F-move in 2d leads to a phase, which is the manifestation of an 't Hooft anomaly.}
\label{2GrouptoAnom}
\end{figure}

Another approach for investigating the fate of generalized symmetries under compactification is to reduce the defects that generate them. Let us begin by considering the first bullet point \ref{item1} corresponding to reducing a 2-group symmetry with gauge flux, and examine a junction of topological defects associated with $U(1)_F$. We can then consider performing an F-move. That is, we take three of the codimension-one topological defects associated with $U(1)_F$, such that they are parallel in two directions but not in the third. We then consider the transition from merging 1+2 into 3, to merging 1 with 2+3 (see Figure~\ref{2GrouptoAnom}). In a 2-group, this transition results in the generation of the codimension-two topological defect associated with the 1-form symmetry oriented in the two directions shared by the three codimension-one topological defects. Now consider compactifying the system on a 2-surface spanning the two shared directions of all four topological defects. The codimension-one defects would just reduce to the 2d codimension-one defects generating the 2d $U(1)_F$. As such we get a similar F-move junction also in 2d. However, note that the codimension-two defect now fully wraps the 2d surface, and since we consider a gauge flux on the 2d surface it will lead to a phase.\footnote{~Recall that the topological defect is formally defined as $\int_{\Sigma} c_1(G)$ on the surface, where for us $\Sigma=S^2$. As such in the fixed-flux sector, it reduces to a fixed phase.} This just describes the 2d F-symbols expected from an 't Hooft anomaly in the flavor symmetry. As such we see that reducing a 2-group in the presence of gauge fluxes (fixed charge for the 1-form symmetry on the surface) leads to a standard direct product, but with an 't Hooft anomaly for the 0-form symmetry. 

Let us next turn to the third bullet point \ref{item3} above corresponding to reducing a non-invertible symmetry on a sphere with gauge flux, and analyze the fate of the symmetry using the explicit description of the defect in terms of its worldvolume theory. Specifically, we consider the  non-invertible axial defect of a 4d theory with the ABJ anomaly \eqref{fg2}, and its compactification on a sphere, with a fixed gauge flux $m_G$ for the $U(1)_G$ factor of the gauge group on the sphere, $\int_{S^2} c_1(G)= m_G$. The anomaly  \eqref{fg2} leads to the following non-conservation of the axial current $j_{\left(4d\right)}^{A}$,
\begin{equation}
	\label{axial_anom_4d}
	d\star j_{\left(4d\right)}^{A}=\frac{k}{2} c_1(G)^2\,.
\end{equation}
We take the 4d spacetime to be $M_{2}\times S^2$ such that the three-dimensional defect wraps the $S^{2}$ and one of the cycles of $M_2$, \emph{i.e.}~$S^1 \subset M_2$. Everywhere we assume that $M_2$ is much larger than $S^{2}$. For axial rotations by an angle of $\alpha=\frac{2\pi n}{k}$ for $n\in\mathbb{Z}_{k}$, the topological defects,
\begin{equation}
	\label{Axial_defect_inv}
	\mathcal{D}_{\frac{2\pi n}{k}}^{(4d)}=\exp\left[i\int_{S^{2}\times S^{1}}\left(\frac{2\pi n}{k}\star j_{\left(4d\right)}^{A}-\frac{n}{4\pi}a\wedge da\right)\right]
\end{equation}
generate an invertible $\mathbb{Z}_{k}$ subgroup of the anomalously broken $U(1)_A$ axial symmetry (we generally reserve lowercase letters for dynamical gauge fields). However, as shown in \cite{Choi:2022jqy,Cordova:2022ieu}, in addition to the invertible defects in \eqref{Axial_defect_inv} there are non-invertible ones that generate more general axial rotations by angles $\alpha=\frac{2\pi p}{Nk}$ for coprime $p$ and $N$. The explicit form of the non-invertible defect is given by,
\begin{equation}
	\label{Axial_defect_gen}
	\mathcal{D}_{\frac{2\pi p}{Nk}}^{(4d)}=\exp\left[i \int_{S^{2}\times S^{1}}\left(\frac{2\pi p}{Nk}\star j_{\left(4d\right)}^{A}+\mathcal{A}^{N,p}\left[da/N\right]\right)\right]\,,
\end{equation}
where $\mathcal{A}^{N,p}\left[B\right]$ is  any 3d TQFT with a $\mathbb{Z}_{N}^{\left(1\right)}$ 1-form symmetry (with background 2-form $\bZ_N$ gauge field $B$) and an ’t Hooft anomaly labeled by $p$ \cite{Hsin:2018vcg}. 

Let us now analyze the reduction of the 4d axial defects to 2d, starting with the invertible ones in \eqref{Axial_defect_inv}. In this case we can readily see that the resulting defects in 2d obtained by reducing on the sphere with gauge flux $m_G$ are given by 
\begin{equation}
	\label{4d_to_2d_defects}
	\mathcal{D}_{\frac{2\pi n}{k}}^{(4d)}\,\rightarrow\,\exp\left[i\int_{S^{1}}\left(\frac{2\pi n}{k}*j_{\left(2d\right)}^{A}-n m_Ga \right)\right],
\end{equation}
where $j_{\left(2d\right)}^{A} = * \int_{S^2} \star j_{(4d)}^A$ is the 2d axial current and $*$ the 2d Hodge star (in contrast to the 4d one $\star$). In order to identify the group that these defects span, and to compare it with the axial anomaly as viewed from the 2d perspective, let us first recall the form of the defects that generate the discrete subgroup left over by a 2d axial anomaly. The anomaly equation in 2d takes in general the form,
\begin{equation}
	\label{2d_anom}
	d*j_{\left(2d\right)}^{A}=k_{2d} c_1(G)\,,
\end{equation}
and results in the 2d $U(1)_A$ axial symmetry being broken to $\mathbb{Z}_{k_{2d}}$. The defects that realize this invertible discrete symmetry are then given by\footnote{~Although the $U(1)$ Wilson line is not topological, as discussed in Section~\ref{sec:holonomy}, the combination of the two terms in this expression is topologically invariant due to the modified conservation equation \eqref{2d_anom}. }
\begin{equation}
	\label{2d_defects_gen}
	\mathcal{D}_{\frac{2\pi \ell}{k_{2d}}}^{(2d)}=\exp\left[i\int_{S^{1}}\left(\frac{2\pi \ell}{k_{2d}}*j_{\left(2d\right)}^{A}-\ell a\right)\right]\,,\quad \ell\in\mathbb{Z}_{k_{2d}}\,.
\end{equation}
Matching \eqref{2d_defects_gen} with \eqref{4d_to_2d_defects} we find that $k_{2d}=m_Gk$, and that the defects in \eqref{4d_to_2d_defects} resulting from 4d generate only those elements of $\mathbb{Z}_{k_{2d}}$ corresponding to the $\bZ_{\frac{k_{2d}}{m_G}}= \bZ_k \subset\mathbb{Z}_{k_{2d}}$ subgroup. 

In order to obtain the rest of the elements in the $\mathbb{Z}_{k_{2d}}$ group, we need to reduce the 4d non-invertible axial defects in \eqref{Axial_defect_gen} associated with gauging a $\mathbb{Z}_{N}^{(1)}$ subgroup of the magnetic 1-form symmetry. To do so we use the following Lagrangian description of the Hall state TQFT $\mathcal{A}^{N,p}$ (see {\it e.g.}~\cite{Tong:2016kpv} and Appendix C of \cite{vanBeest:2023dbu}) and of the defect  \eqref{Axial_defect_gen},  
\begin{equation}
	\label{Axial_defect_Lag}
\mathcal{D}_{\frac{2\pi p}{Nk}}^{(4d)}=\int D\vec{\mathtt{a}}\,\exp\left[i\int_{S^{2}\times S^{1}}\left(\frac{2\pi p}{Nk}\star j_{\left(4d\right)}^{A}+\frac{1}{4\pi}\vec{\mathtt{a}}^{t}K\land d\vec{\mathtt{a}}+\frac{1}{2\pi}\vec{\mathtt{a}}^{t}\vec{v}\land da\right)\right]
\end{equation}
where the matrix $K$ and the vector $\vec{v}$ are given by
\begin{equation}
	\label{K_v}
K=\left(\begin{array}{cccccc}
	+k_{1} & 1 & 0 & 0 & 0 & \cdots\\
	1 & -k_{2} & 1 & 0 & 0 & \cdots\\
	0 & 1 & +k_{3} & 1 & 0 & \cdots\\
	0 & 0 & 1 & -k_{4} & 1 & \cdots\\
	\vdots & \vdots & \vdots & \vdots & \vdots & \ddots
\end{array}\right)\quad,\quad\vec{v}=\left(1,0,0,\ldots\right)
\end{equation}
and are related to the parameters $p$ and $N$ as follows, 
\begin{equation}
	\label{Relation_pN_Kv}
	p/N=\vec{v}^{\,t}K^{-1}\vec{v}\,.
\end{equation}
Notice also that $\vec{\mathtt{a}}$ in \eqref{Axial_defect_Lag} is a vector of gauge fields living only on the defect, to be distinguished from the dynamical $U(1)$ gauge field $a$ that lives on the entire $M_2\times S^2$. Then, denoting the flux of $\vec{\mathtt{a}}$ over the $S^{2}$ by $\vec{m}$, we find at distances much larger than the size of $S^{2}$, 
\begin{equation*}
\mathcal{D}_{\frac{2\pi p}{Nk}}^{(4d)}\,\rightarrow\,\exp\left(i\frac{2\pi p}{Nk}\int_{S^{1}}*j_{\left(2d\right)}^{A}\right)\sum_{\vec{m}}e^{i\vec{m}^{t}\vec{v}\int_{S^{1}}a}\int D\vec{\mathtt{a}}\,e^{i\left(\vec{m}^{t}K+m_{G}\vec{v}^{t}\right)\int_{S^{1}}\vec{\mathtt{a}}}
\end{equation*}
\begin{equation}
	\label{Axial_defect_red_I}
=\exp\left(i\frac{2\pi p}{Nk}\int_{S^{1}}*j_{\left(2d\right)}^{A}\right)\sum_{\vec{m}}e^{i\vec{m}^{t}\vec{v}\int_{S^{1}}a}\delta_{0,K\vec{m}+m_{G}\vec{v}}
\end{equation}
where $\delta_{0,K\vec{m}+m_{G}\vec{v}}$ is defined as follows,\footnote{~More explicitly, we have $\delta_{0,K\vec{m}+m_{G}\vec{v}}=\delta_{0,k_{1}m_{1}+m_{2}+m_{G}}\delta_{0,m_{1}-k_{2}m_{2}+m_{3}}\delta_{0,m_{2}+k_{3}m_{3}+m_{4}}\cdots$.} 
\begin{equation}
\delta_{0,K\vec{m}+m_{G}\vec{v}}\equiv\delta_{0,(K\vec{m}+m_{G}\vec{v})_{1}}\delta_{0,(K\vec{m}+m_{G}\vec{v})_{2}}\cdots
\end{equation}
and fixes the sum over $\vec{m}$ to $\vec{m}=-m_{G}K^{-1}\vec{v}$. Substituting back to \eqref{Axial_defect_red_I} we find
\begin{equation}
	\label{Axial_defect_red_II}
\mathcal{D}_{\frac{2\pi p}{Nk}}^{(4d)}\,\rightarrow\,\exp\left(i\frac{2\pi p}{Nk}\int_{S^{1}}*j_{\left(2d\right)}^{A}\right)e^{-im_{G}\vec{v}^{t}K^{-1}\vec{v}\int_{S^{1}}a}
\end{equation}
which using \eqref{Relation_pN_Kv} can be put in the form\footnote{~It is tempting to derive \eqref{Axial_defect_red_III} from the presentation of the axial defect using the improperly-quantized Chern-Simons term,  
		\begin{eqnarray*}
		\exp\left[i\int_{S^{2}\times S^{1}}\left(\frac{2\pi p}{Nk}\star j_{(4d)}^{A}-\frac{p}{4\pi N}a\wedge da\right)\right]\,.
		\end{eqnarray*}	
		Upon reduction to 2d on a sphere with gauge flux $\int_{S^2} \frac{da}{2\pi}= m_G$ we still get the same expression for the invertible defect $\exp\left[i\int_{S^{1}}\left(\frac{2\pi p}{Nk}*j_{\left(2d\right)}^{A}-\frac{m_{G}}{N}p\,a\right)\right]$, and the condition of $N|m_G$ in \eqref{Axial_defect_red_final} can be inferred from the invariance under large gauge transformations $\int_{S^{1}}a\to\int_{S^{1}}a+2\pi$.}
\begin{equation}
	\label{Axial_defect_red_III}
\mathcal{D}_{\frac{2\pi p}{Nk}}^{(4d)}\,\rightarrow\,\exp\left[i\int_{S^{1}}\left(\frac{2\pi p}{Nk}*j_{\left(2d\right)}^{A}-\frac{m_{G}}{N}p\,a\right)\right].
\end{equation}
Notice that in order for the fluxes $\vec{m}$ to be integers in this computation, $N$ should divide $m_{G}$. Comparing now \eqref{Axial_defect_red_III} with \eqref{2d_defects_gen} we find again that $k_{2d}=m_Gk$ and obtain the final result,
\begin{equation}
	\label{Axial_defect_red_final}
\mathcal{D}_{\frac{2\pi p}{Nk}}^{(4d)}\,\rightarrow\,\begin{cases}
	\exp\left[i\int_{S^{1}}\left(\frac{2\pi}{k_{2d}}\frac{m_{G}}{N}p*j_{\left(2d\right)}^{A}-\frac{m_{G}}{N}p\,a\right)\right] & N|m_{G}\\
	0 & N\nmid m_{G}
\end{cases}
\end{equation}
in which the elements of $\mathbb{Z}_{k_{2d}}$ that are not captured by \eqref{4d_to_2d_defects} are realized. We therefore reproduce the entire invertible $\mathbb{Z}_{k_{2d}}=\mathbb{Z}_{m_Gk}$ group in 2d. 

In summary, we see that the 4d non-invertible axial defects reduce in the presence of a nonzero gauge flux on the sphere to invertible axial defects in 2d, in such a way that the anomaly we can associate to them (responsible for the breaking of the $U(1)$ group from the 2d perspective) is given by the 4d anomaly coefficient $k$ multiplied by the gauge flux. 

Let us finally comment on the case of a vanishing gauge flux, discussed in bullet point \ref{item4}. In this case, we can substitute $m_G=0$ in \eqref{Axial_defect_red_III} and are left with 2d invertible axial rotations by any rational angle that we had in 4d. Therefore, the $\mathbb{Q}/\mathbb{Z}$ non-invertible symmetry reduces in 2d to a $\mathbb{Q}/\mathbb{Z}$ invertible symmetry.\footnote{~In all the examples considered later the $\mathbb{Q}/\mathbb{Z}$ invertible symmetry is enhanced to a $U(1)$ symmetry in 2d.}

\subsection{Comments on Reducing Dualities}

Another issue we will examine is how the reduction of the generalized symmetry structure interplays with the reduction of 4d dualities to 3d or 2d dualities.
Suppose that two field theories participate in an IR-type duality in four dimensions, meaning that at energies much smaller than their respective strong-coupling scales, they flow (within some conformal window of parameter space) to the same conformal fixed point. Let us denote these theories by $\CT_A$ and $\CT_B$, whose strong coupling scales $\Lambda_{A,B}$ satisfy $\Lambda^{b}=e^{-8\pi^2/g_{4d}^2}$, for $b$ the appropriate 1-loop $\beta$-function coefficient and $g_{4d}$ the running coupling. The statement of duality is that both $\CT_A$ and $\CT_B$ are described by the same CFT $\CT_*$ at energies which satisfy,
	\begin{align}
	E \ll \Lambda_A,\,\Lambda_B\,.
	\end{align}
While the microscopic description may look very different on either side of the duality, in order for the duality to be consistent both $\CT_A$ and $\CT_B$ must share the same global symmetries. In particular, if theory $\CT_A$ enjoys a non-invertible symmetry, then for the duality to hold it must be true that theory $\CT_B$ shares the same non-invertible symmetry. 

The prototypical example of such a duality is Seiberg duality of 4d $\CN=1$ SQCD with gauge group $G=SU(N_c)$ or $U(N_c)$ \cite{Seiberg:1994pq}.  The $U(N_c)$ theory has an ABJ anomaly of the form \eqref{fg2}, which involves the baryonic $U(1)_B$ part of the gauge group, and the $U(1)_R$ symmetry. We will generally consider the pair of dual theories with additional matter fields transforming in the determinant representation of the gauge group, designed to cancel the anomaly involving $U(1)_R$ in favor of a global $U(1)_I$ symmetry under which the determinant matter is charged. Then, the theories enjoy a non-invertible symmetry implementing $U(1)_I$ symmetry rotations by a rational angle, while still preserving a continuous R-symmetry. In Section~\ref{sec:4dnoninv}, we will explicitly demonstrate how the non-invertible symmetry defects of the four-dimensional theories map across the duality. 

Now, we consider placing the 4d theories on a manifold $M_3\times S^1$, denoting the radius of $S^1$ by $R$. 
As was emphasized in \cite{Aharony:2013dha}, in the most naive dimensional-reduction limit whereby the radius is taken to zero with the 3d couplings $g_{3d}^2 = g_{4d}^2/(2\pi R)$ kept finite, we would obtain two 3d theories $\mathcal{T}_{A/B}^{3d}$ that flow to \emph{distinct} SCFTs at low energies $E \ll g_{3d,A}^2,g_{3d,B}^2$. In order to obtain a 3d IR duality in which the two theories flow to the same fixed point, we should instead consider a ``less naive'' compactification limit in which one keeps the radius finite while considering energy scales much smaller than the KK scale $1/R$,
	\begin{align}\label{eq:dimredlimit}
	E \ll \Lambda_A,\,\Lambda_B,\, \frac{1}{R}\,.
	\end{align}
In this limit, the two theories flow to the same effectively three-dimensional SCFT $\CT_*^{3d}$. Another way to see why this is the case is that the effective 3d dual pairs that flow to $\CT_*^{3d}$ must also share the same global symmetries. However, as we will see below, it is not naively obvious in this construction that the global symmetries automatically match. 
The underlying reason is the same one that we have emphasized in Section~\ref{sec:intro3d}: it is only in the limit that the scalar $\sigma = \oint_{S^1} a$ decompactifies---in which case the 1-form symmetry trivializes, and the 2-group structure and non-invertible symmetries that were present in 4d are lost---that the effective 3d theories are dual. 

One can also try to look for two intrinsically three-dimensional theories  that flow to the SCFT $\CT_*^{3d}$ in the IR and are thus dual. As discussed in \cite{Aharony:2013dha,Aharony:2013kma}, these are usually obtained by deforming the theories $\mathcal{T}_{A/B}^{3d}$ coming from the naive dimensional-reduction limit by a suitable monopole superpotential. This deformation has in particular the effect of breaking certain symmetries in the theories $\mathcal{T}_{A/B}^{3d}$ that accidentally arise in the naive reduction of the 4d theories $\mathcal{T}_{A/B}$ and that prevent them from being dual.

Next we consider compactifying the 4d dual theories on $S^2$. These types of compactifications have been discussed \emph{e.g.}~in \cite{Honda:2015yha,Gadde:2015wta} (see also \cite{Dedushenko:2017osi,Tachikawa:2018sae,Sacchi:2020pet}). Again, it is only in the limit \eqref{eq:dimredlimit} that the effective 2d theories are expected to flow to the same SCFT $\mathcal{T}_*^{2d}$, where now $R$ is the radius of the $S^2$. Hence, the 2d theories $\mathcal{T}_{A/B}^{2d}$ obtained from a naive dimensional reduction of $\mathcal{T}_{A/B}$ on $S^2$ above their strong-coupling scales will in general not be dual. Similarly to the reduction to 3d discussed above, this is usually related to the emergence of accidental symmetries in the dimensional reduction. Unfortunately, in this case it is less clear how the theories $\mathcal{T}_{A/B}^{2d}$ can be deformed to obtain two genuinely dual 2d theories that flow to the same SCFT $\mathcal{T}_*^{2d}$ to which also the effective 2d theories arising from the limit \eqref{eq:dimredlimit} flow. However, it is often the case that in the reduction there are no accidental symmetries, and the theories $\mathcal{T}_{A/B}^{2d}$ obtained from the naive dimensional reduction are still dual.\footnote{~One should  be careful when the 2d theories have non-compact directions in the target space, see \emph{e.g.}~\cite{Aharony:2016jki,Aharony:2017adm}, however we do not discuss this here. This issue can be avoided by considering the duality between the massively deformed theories, where most of the vacua are lifted so that we are left with discrete isolated vacua.}

\section{Reducing the Symmetry Topological Field Theory}
\label{sec:symtft}

In Section~\ref{sec:discussion}, we described several approaches for analyzing the reduction of generalized symmetries on $S^1$ and $S^2$,  including the reduction of the 5d anomaly inflow action or anomaly polynomials, and making use of the background fields and the topological defects corresponding to a certain generalized symmetry. One loophole of the discussion so far is that the 2-group structure and non-invertible symmetry are discussed separately, \emph{i.e.}~either $k=0$ or $\kappa=0$, and their interplay has not been addressed. When both $k$ and $\kappa$ are non-vanishing, not only does the expression of the non-invertible defect \eqref{Axial_defect_gen} gets modified, but also the junction between three such defects, as well as higher-codimensional junctions between four such defects (associated with the $F$-move), are also modified depending on $\kappa$ and the cubic self anomaly, $k_{F^3}$. The detailed discussion of these higher structures can be very complicated, see \cite{Copetti:2023mcq} for extensive discussions. In this section we take an alternative route, by taking advantage of the recently developed Symmetry TFT (SymTFT) for continuous symmetries \cite{Gaiotto:2020iye,Apruzzi:2021nmk,Freed:2022qnc, Bonetti:2024cjk, Apruzzi:2024htg}, as a convenient tool for studying the higher structure of the non-invertible symmetry.

As mentioned in the introduction, a natural generalization of the relation between anomaly theories under compactification is that their SymTFTs should be related by a similar dimensional reduction. Here we shall show how one can indeed recover the previous statements by reducing the 5d SymTFT, describing the 2-group structure or non-invertible symmetry and their anomalies, on $S^2$.\footnote{~As alluded to in Section~\ref{sec:intro3d}, in the strict 3d limit the interesting generalized symmetries decouple. We therefore focus on the $S^2$ compactification only.}

\subsection{Symmetry TFT and Boundary Conditions}
\label{sec:symbc}

We begin by specifying the SymTFT and its topological boundary conditions for the 4d theory with $\mathbb{Q}/\mathbb{Z}$ axial symmetry. Such a SymTFT was recently found in \cite{Antinucci:2024zjp,Brennan:2024fgj} (see also \cite{Apruzzi:2024htg}), and here we will follow the description in \cite{Antinucci:2024zjp}.

We start with the SymTFT of a $U(1)_G\times U(1)_F$ 0-form \emph{global} symmetry, where the 6d anomaly polynomial and 5d anomaly inflow action obtained by descent are
\begin{align}
\CI_6 &= \frac{k}{2} c_1(G)^2 c_1(F) + \frac{\kappa}{2} c_1(G)c_1(F)^2 + \frac{k_{A^3}}{6} c_1(F)^3\,,\nonumber\\
\mathcal{A}_5^{\text{inf}}&= \int_{X_5} \left(\frac{k}{2} A_F c_1(G)^2 + \frac{\kappa}{2} A_F c_1(F) c_1(G)+ \frac{k_{F^3}}{6} A_Fc_1(F)^2\right)\,,
\end{align}
where $A_F$ is the background gauge field for $U(1)_F$. We have introduced the self 't Hooft anomaly coefficient $k_{F^3}$ for the classical $U(1)_F$ symmetry. The SymTFT is obtained by gauging the $U(1)_G\times U(1)_F$ global symmetry, leading to,
\begin{equation}
	\begin{split}
		\mathcal{S}_5= \int_{X_5} \bigg(\frac{1}{2\pi} \sfb_3 \wedge  da_1 &+ \frac{1}{2\pi} \sfb_3' \wedge da_1' + \frac{k}{8\pi^2} a_1 \wedge da_1' \wedge da_1' \\&+ \frac{\kappa}{8
	\pi^2} a_1 \wedge d a_1 \wedge d a_1' + \frac{k_{F^3}}{24\pi^2} a_1 \wedge d a_1 \wedge d a_1 \bigg)\,.
	\end{split}
\end{equation}
Since there are many fields with various degrees, in this section we introduce subscripts to denote the form degree. The fields labeled by standard fonts such as $a_i, a_i'$ are $U(1)$-valued gauge fields, while those labeled by mathsf fonts such as $\sfb_i, \sfb_i'$ are $\mathbb{R}$-valued fields. The $U(1)_G\times U(1)_F$ global symmetry is obtained by specifying the Dirichlet boundary conditions for $a_1$ and $a_1'$.

We then gauge the $U(1)_G$ symmetry in 4d. Unlike the finite symmetry case where it is sufficient to change the topological boundary condition while keeping the SymTFT unchanged, for the continuous symmetry, the SymTFT needs to be modified as well. Following the discussion in  \cite{Antinucci:2024zjp}, we simply need to change 
\begin{eqnarray}
	da_1' \to \sff_2, \qquad \sfb_3' \to db_2\,.
\end{eqnarray}
The SymTFT of the $\mathbb{Q}/\mathbb{Z}$ non-invertible symmetry and 1-form symmetry is then,
\begin{align}\label{SymTFT_4d}
	\begin{split}
		\CS_5= \int_{X_5} \bigg(\frac{1}{2\pi} \sfb_3 \wedge  da_1 &+ \frac{1}{2\pi} \sff_2 \wedge db_2 + \frac{k}{8\pi^2} a_1 \wedge \sff_2 \wedge \sff_2 \\&+ \frac{\kappa}{8
			\pi^2} a_1 \wedge d a_1 \wedge \sff_2 + \frac{k_{F^3}}{24\pi^2} a_1 \wedge d a_1 \wedge d a_1 \bigg)\,.
	\end{split}
\end{align}
The boundary condition is modified to the Dirichlet boundary condition for $a_1, b_2$.

To see the global symmetry after gauging $U(1)_G$, we study the topological operators in the SymTFT \eqref{SymTFT_4d}. Those that are not trivialized by the Dirichlet boundary condition when pushed to the boundary become the topological defects of the global symmetry. One way to determine the topological operators is to impose gauge invariance. We first find the gauge transformation of $a_1, b_2, \sfb_3, \sff_2$ under which the action \eqref{SymTFT_4d} is gauge invariant, and then identify the operators invariant under gauge transformations. Another easier, and equivalent, way to determine the topological operators is to use the equations of motion, 
\begin{align}
&da_1 = 0\,,\qquad db_2 + \frac{k}{2\pi} a_1 \wedge \sff_2 + \frac{\kappa}{4\pi} a_1 \wedge d a_1 =0\,,\nonumber\\
&d \sff_2 = 0\,,\qquad d \sfb_3 + \frac{k}{4\pi} \sff_2 \wedge \sff_2 + \frac{\kappa}{2\pi} d a_1 \wedge \sff_2 + \frac{k_{F^3}}{4\pi} da_1 \wedge da_1 =0\,.
\end{align}
We thus derive the topological defects 
\begin{align}
&W_n[\gamma_1] = \exp\left(i n \int_{\gamma_1} a_1\right)\,,\qquad \CT_n[\gamma_2] = \exp\left(i n \int_{\gamma_2} b_2\right) \CB_{n}[\gamma_2;a_1, \sff_2]\,,\nonumber\\
&V_{\alpha}[\gamma_2] = \exp\left(i \alpha \int_{\gamma_2} \sff_2\right)\,,\qquad U_{p,N}[\gamma_3] = \exp\left(i \frac{p }{Nk}\int_{\gamma_3} \sfb_3\right) \CA^{p,N}[\gamma_3; \sff_2]\,,
\end{align}
where $n, p, N\in \bZ$ and $\alpha\in U(1)$. 

Let us comment on the properties of these operators. First, both $W_n$ and $V_\alpha$ are invertible operators. For $\CT_n$, the first factor $e^{i n \int b_2}$ is invertible, while the second factor $\CB_{n}[\gamma_2;a_1, \sff_2]$ is a 2d TQFT coupled to $a_1$ and $\sff_2$ fields with the anomaly $ \frac{kn}{2\pi} a_1 \wedge \sff_2$, so that its anomaly cancels the contribution of the $b_2$ field coming from its equation of motion. A representative example is  $\bZ_{\kappa n}$ gauge theory. Finally, for the operators $U_{p,N}$, the second factor $\CA^{p,N}[\gamma_3; \sff_2]$ is a 3d TQFT coupled to $\sff_2$ field with the anomaly $\frac{p}{4\pi N} \sff_2 \wedge \sff_2$. The fact that these operators are topological follows from the equations of motion.

With the chosen boundary conditions the operators $W_{n}$ and $\mathcal{T}_{n}$ are trivialized when taken to lie on the boundary, and thus do not survive as topological operators in the 4d field theory. They can, however, end on the boundary, giving rise to a local operator charged under the axial symmetry and to an 't Hooft line charged under the magnetic 1-form symmetry, respectively. Meanwhile, the operators $V_{\alpha}$ and $U_{p,N}$ can be pushed to the boundary and result in the 1-form and $\mathbb{Q}/\bZ$ axial non-invertible topological  defects. The properties of the topological defects, \emph{e.g.}~fusion rules and F-symbols, can be analyzed in principle, but they are beyond the scope of this paper. We will be mainly interested in how the symmetries reduce upon compactification, which will be explained in the subsections below.

\subsection{Reducing on \texorpdfstring{$S^2$}{S2} with Gauge Flux}

We would next like to compactify the SymTFT \eqref{SymTFT_4d} to 2d on a sphere with gauge flux $m_G$ for the gauge field of the boundary theory. Since the bulk operator $V_{\alpha}$ corresponds to the magnetic 1-form symmetry operator on the boundary, the gauge flux $m_G$ implies a corresponding holonomy for $\sff_{2}$ on the sphere,
\begin{equation}
\int_{S^{2}}\sff_{2}=2\pi m_G\,.
\end{equation}
Taking $X_{5}=X_{3}\times S^{2}$, we therefore obtain (assuming a vanishing flux for $a_1$ since we assumed zero flavor flux),
\begin{equation}\label{red_SymTFT}
\mathcal{S}_{3}=\int_{X_{3}}\left(\frac{1}{2\pi}\sfb_{1}\wedge da_{1}+\frac{1}{2\pi} \sff_{2}\wedge db_{0}+m_Gdb_{2}+\frac{k m_G}{2\pi}a_{1}\wedge \sff_{2}+ \frac{\kappa m_G}{4\pi} a_1 \wedge d a_1\right)\,,
\end{equation}
where 
\begin{equation}
\sfb_{1}=\int_{S^{2}}\sfb_{3}\quad,\quad b_{0}=\int_{S^{2}}b_{2}\,.
\end{equation}
The boundary condition is the Dirichlet boundary condition for $a_1, b_0, b_2$. Note that the third term in \eqref{red_SymTFT} is a total derivative and hence can be dropped.

In order to interpret \eqref{red_SymTFT} as the SymTFT of a 2d theory and match it with our previous analysis, let us first find the SymTFT of a 2d theory with an anomalous axial symmetry. Following \cite{Antinucci:2024zjp}, we start from the 3d SymTFT for two 0-form global symmetries $U(1)_{G}\times U(1)_{F}$ with a mixed anomaly, and a self anomaly of $U(1)_F$,
\begin{equation}
\mathcal{S}_{3}'=\int_{X_{3}}\left(\frac{1}{2\pi}\sfb_{1}\wedge da_{1}+\frac{1}{2\pi} \sfc_{1}\wedge dv_{1}+\frac{k_{2d}}{2\pi}a_{1}\wedge dv_1+ \frac{k_{F^2,2d} }{4\pi} a_1 \wedge d a_1\right)\,,
\end{equation}
and then gauge $U(1)_{G}$, which amounts to replacing $dv_{1}\rightarrow \sff_{2}$ and $\sfc_{1}\rightarrow db_{0}$ for $\sff_{2}$  an $\mathbb{R}$ 2-form field and $b_{0}$ a compact scalar (see \cite{Antinucci:2024zjp} for more details). We then obtain the SymTFT, 
\begin{equation}
\label{SymTFT_2d}
\mathcal{S}_{3}=\int_{X_{3}}\left(\frac{1}{2\pi}\sfb_{1}\wedge da_{1}+\frac{1}{2\pi} \sff_{2}\wedge db_{0}+\frac{k_{2d}}{2\pi}a_{1}\wedge \sff_{2}+ \frac{k_{F^2,2d}}{4\pi} a_1 \wedge d a_1\right)
\end{equation}
with Dirichlet boundary conditions for $a_1, b_0$. The mixed anomaly implies that the global symmetry $U(1)_F$ is anomalously broken to $\bZ_{|k_{2d}|}$. Moreover, the self anomaly of $U(1)_F$ implies the self anomaly for $\bZ_{|k_{2d}|}$ valued in $k_{F^2,2d}$ mod $k_{2d}$.

Comparing \eqref{red_SymTFT} with \eqref{SymTFT_2d}, we see that the SymTFTs indeed match under the compactification, with the identification
\begin{eqnarray}
	k_{2d}=km_G, \qquad k_{F^2, 2d}= \kappa m_G\,.
\end{eqnarray}
The 4d non-invertible symmetry reduces to a $\bZ_{|k_{2d}|} = \bZ_{|k m_G|}$ invertible symmetry  in 2d. Moreover, the 2-group structure of the non-invertible symmetry, which depends on $\kappa$, reduces to the self anomaly of $\bZ_{|k m_G|}$, valued in $\kappa m_G $ mod $k m_G$. The self-anomaly of the non-invertible symmetry in 4d does not influence the properties of the symmetries in 2d. There is also a $(-1)$-form symmetry generated by $e^{i \alpha \int \sff_2}$ which does not mix with the discrete 0-form symmetry.

The above result automatically contains the special cases discussed in Section~\ref{sec:intro2d}. When $k=0$ and $\kappa \neq 0$, the $U(1)_F$ symmetry in 4d reduces to a $U(1)_F$ symmetry in 2d, and the 2-group structure in 4d reduces to a self anomaly of $U(1)_F$ in 2d. When $\kappa=0$ and $k\neq 0$ instead, the non-invertible symmetry in 4d reduces to an invertible symmetry in 2d, again as expected.

\subsection{Reducing on \texorpdfstring{$S^2$}{S2} with Flavor Flux}

We finally compactify the SymTFT \eqref{SymTFT_4d} to 2d on a sphere with flavor flux $m_F$. Since the bulk field $a_1$ reduces to the background field for the flavor symmetry $A_F$ by imposition of the Dirichlet boundary conditions, the flavor flux $m_F$ implies a corresponding flux for $a_1$ on the sphere,
\begin{equation}
	\int_{S^{2}}da_{1}=2\pi m_F\,.
\end{equation}
Taking $X_{5}=X_{3}\times S^{2}$, we therefore obtain (considering a vanishing holonomy for $\sff_2$ since we assumed zero gauge flux),
	\begin{align}\label{eq:S3flavor}
	\begin{split}
	\CS_3= \int_{X_3} \Bigg(m_F \sfb_3 + \frac{1}{2\pi} \sfb_1 \wedge  d a_1 &+ \frac{1}{2\pi} \sff_2  \wedge d b_0 + \frac{k m_F}{4\pi} \text{CS}(\sff_2)  \\
	& +  \frac{\kappa m_F}{2\pi} a_1 \wedge \sff_2 + \frac{k_{F^3} m_F}{4\pi} a_1 \wedge d a_1\Bigg)
	\end{split}
	\end{align}
where
\begin{eqnarray}
	\sfb_1 = \int_{S^2} \sfb_3, \qquad b_0 = \int_{S^2} b_2\,.
\end{eqnarray}
We have defined $\text{CS}(\sff_2)$ by $d \text{CS}(\sff_2) = \sff_2 \wedge \sff_2$, which is possible since $\sff_2$ is an $\mathbb{R}$-valued field. Since we are interested in the case with Dirichlet boundary conditions for $a_1, b_0, a_1$, it is only possible for $\sff_2$ to have Neumann boundary condition. The action \eqref{eq:S3flavor} is compatible with this boundary condition only when $k m_F=0$, since otherwise the term $\text{CS}(\sff_2)$ would indicate that the gauge symmetry is anomalous in 2d, which is inconsistent with the observation in bullet point \ref{item3}. Below, we thus assume $k=0$, that is we are only interested in an invertible 2-group symmetry. Further noting that the first term in \eqref{eq:S3flavor} decouples from the rest of the terms, the SymTFT \eqref{eq:S3flavor} reduces to 
\begin{eqnarray}\label{eq:S3flavor2}
	\CS_3= \int_{X_3} \left(\frac{1}{2\pi} \sfb_1 \wedge  d a_1 + \frac{1}{2\pi} \sff_2  \wedge d b_0+ \frac{\kappa m_F}{2\pi} a_1 \wedge \sff_2 + \frac{k_{F^3} m_F}{4\pi} a_1 \wedge d a_1\right)\,.
\end{eqnarray}

Comparing \eqref{eq:S3flavor2} with \eqref{SymTFT_2d}, we obtain,
\begin{eqnarray}
	k_{2d} = \kappa m_F\,, \qquad k_{F^2, 2d} = k_{F^3} m_F\,.
\end{eqnarray}
This result means that the invertible continuous 2-group symmetry whose Postnikov class is related to $\kappa$ and whose self anomaly is related to $k_{F^3}$ in 4d, reduces to a discrete $\bZ_{|\kappa m_F|}$ symmetry with self anomaly $k_{F^3} m_F $ mod $\kappa m_F$ in 2d. There is also a $(-1)$-form symmetry generated by $e^{i \alpha \int \sff_2}$ which does not mix with the discrete 0-form  symmetry. 
These results are again consistent with the special case $k_{F^3}=0$ and $\kappa\neq 0$ discussed in Section~\ref{sec:intro2d}.

\section{4d \texorpdfstring{$\mathcal{N}=1$}{N=1}  Supersymmetric QCD with \texorpdfstring{$U(N_c)$}{U(Nc)} Gauge Group }
\label{sec:4d}

In the rest of this work we will focus on a series of specific examples of 4d gauge theories and their dimensional reductions that exhibit  the general features discussed in Sections \ref{sec:discussion} and \ref{sec:symtft}.
In this section we begin by reviewing the symmetries of 4d $\mathcal{N}=1$ supersymmetric QCD with $U(N_c)$ gauge group, focusing on aspects of its  generalized symmetry structure.

\subsection{Generalized Symmetries}
\label{sec:4dUN}

We consider 4d $\mathcal{N}=1$  SQCD in four dimensions, with gauge group $U(N_c)$ and $N_f$ flavors of fundamental and antifundamental chiral superfields $Q,\widetilde{Q}$. At the classical level, the global 0-form symmetries are, 
	\begin{align} \label{unglobal}
	\frac{SU(N_f)_L\times SU(N_f)_R\times U(1)_A\times U(1)_R}{(\mathbb{Z}_{N_f})_L \times (\mathbb{Z}_{N_f})_R\times \mathbb{Z}_2}\,.
	\end{align}
Here, $U(1)_R$ is the R-symmetry under which the gauginos have unit charge; the $SU(N_f)_L\times SU(N_f)_R$ global symmetry corresponds to separate rotations of the $Q$ and $\widetilde{Q}$; and $U(1)_A$ is the classical chiral axial symmetry. The charges of the fields under these  symmetries are listed in Table~\ref{tab:UNdetcharges}. The quotient by $(\mathbb{Z}_{N_f})_L \times (\mathbb{Z}_{N_f})_R$ is as the center of the two $SU(N_f)$ groups can be mimicked by rotations in $U(1)_A$ and the $U(1)$ gauge part. The quotient by the $\mathbb{Z}_2$ is as the $\mathbb{Z}_2$ part of $U(1)_A$ acts identically to the $\mathbb{Z}_2$ part of the $U(1)$ gauge group. 
This theory is obtained from the one with $SU(N_c)$ gauge group by gauging the baryonic vector-like symmetry.
The model with $U(1)_B=U(1)_V/\mathbb{Z}_{N_c}$ gauged possesses a magnetic $U(1)^{(1)}_M$ 1-form symmetry that acts on the 't Hooft lines associated with $U(1)_B$.

There are several anomalies involving the gauge symmetry, captured by the following terms in the six-form anomaly polynomial,
	\begin{align}
	\label{I6UN}
	\begin{split}
	\CI_6 
	\, \supset \, 
	&2N_f c_1(A) c_2(U(N_c))  + N_f c_1(A) c_1(B)^2 -c_1(R)  c_1(B)^2\\
	&+ \left[ c_2(SU(N_f)_L) - c_2(SU(N_f)_R) \right] c_1(B)\,.
	\end{split}
	\end{align}
In this expression, $c_1(A)$ denotes the first Chern class for the classical $U(1)_A$ symmetry, $c_1(R)$ for the $U(1)_R$ symmetry, and $c_1(B)$ for the $U(1)_B$ part of the gauge group---see Appendix \ref{sec:appconventions} for more details on our conventions. 
Let us discuss each of these anomalies in turn. 
Because the second Chern class $c_2(U(N_c))$ is integer-valued on spin manifolds, the first term in \eqref{I6UN} implies that the classical $U(1)_A$ axial symmetry is broken to $\mathbb{Z}_{2N_f}$. Then, the background gauge field $A_A$ for this symmetry should be understood as a $\mathbb{Z}_{2N_f}$ gauge field, regarded as a 1-cochain $\hat{A}_A$ valued in $\mathbb{Z}_{2N_f}$ and related to the $U(1)_A$ uplift as $A_A\to \frac{2\pi}{2N_f} \hat{A}_A$ (so that $2\pi  \hat{A}_A$ is pure gauge). Accordingly, the curvature $F_A=dA_A$ is to be understood as the Bockstein homomorphism $\beta(\hat{A}_A)\in H^2(M_4,\mathbb{Z})$, so we should replace $c_1(A)=\frac{dA_A}{2\pi} \to \beta(\hat{A}_A)$. As was emphasized in \cite{Csaki:1997aw}, this discrete $\mathbb{Z}_{2N_f}$ subgroup of the axial symmetry can be absorbed into the (gauged) vector-like $U(1)_V$ symmetry, which assigns charges $\pm1$ to the quark superfields. This is because the composition of a $U(1)_V$ rotation  with the action of the $(\mathbb{Z}_{N_f})_L\times (\mathbb{Z}_{N_f})_R$ center of the non-abelian flavor symmetries  can always be chosen to cancel an overall $\mathbb{Z}_{2N_f}$-valued phase on $Q$ and $\widetilde{Q}$. 

 	\def\arraystretch{1.2} 
	\setlength\tabcolsep{5pt} 
	\begin{table}
	\centering
	\begin{tabular}{|c||c|c||c|c|c|c|c|c|}
	\hline
	 & $SU(N_c)$ & $U(1)_V$  & $SU(N_f)_L$ & $SU(N_f)_R$ & $U(1)_A$ & $U(1)_R$ & $U(1)_I$ & $U(1)_t$ \\ 
	 \hline \hline
 	$Q$ & $\Box$ & $1$ & $\Box$ & $1$ & $1$ & $\frac{N_f-N_c}{N_f}$ & $0$ & $0$ \\ 
	\hline
	  $\widetilde{Q}$ & $\overline{\Box}$ & $-1$ & $1$ & $\overline{\Box}$ & $1$ & $\frac{N_f-N_c}{N_f}$ & $0$ & $0$ \\ 
	  \hline
 	 $\lambda$ & adj & $0$ & $1$ & $1$  & $0$ & $1$ & $0$ & $0$ \\ 
	 \hline
  	$\Omega^\pm$ & $1$ & $\pm N_c$ & $1$ & $1$ & $0$ & $2$ & $1$ & $\pm 1$ \\ 
	\hline
	\end{tabular}
	\caption{ The charges of the fields under the classical 0-form symmetries of $U(N_c)$ SQCD, including the fields $\Omega^\pm$ in the determinant representation discussed in Section~\ref{sec:4dUNdet}. $U(1)_R$ is the classical anomaly-free R-symmetry. The baryonic gauge symmetry is $U(1)_B=U(1)_V/\mathbb{Z}_{N_c}$, so that the charges of the fields under $U(1)_B$ are divided by a factor of $N_c$. The  $U(1)_I$ and $U(1)_t$ symmetries are only present in the theory with the determinant matter included.  \label{tab:UNdetcharges}}
	\vspace{0.35cm}
	\begin{tabular}{|c||c|c||c|c|c|c|c|c|}
	\hline
	 & $SU(\widetilde{N}_c)$ & $U(1)_{\widetilde{V}}$  & $SU(N_f)_L$ & $SU(N_f)_R$ & $U(1)_{\widetilde{A}}$ & $U(1)_R$ & $U(1)_I$ & $U(1)_t$ \\ 
	 \hline \hline
	 $q$ & $\Box$ & $1$ & $\overline{\Box}$ & $1$ & $1$ & $\frac{N_c}{N_f}$ & $0$ & $0$ \\ 
	 \hline
 	 $\widetilde{q}$ & $\overline{\Box}$ & $-1$ & $1$ & ${\Box}$ & $1$ & $\frac{N_c}{N_f}$ & $0$ & $0$ \\ 
	 \hline
  	$\widetilde{M}$ & $1$ & $0$ & $\Box$ & $\overline{\Box}$ & $-2$ & $\frac{2(N_f-N_c)}{N_f}$ & $0$ &  $0$ \\ 
	\hline
  	$\widetilde{\lambda}$ & adj & $0$ & $1$ & $1$  & $0$ & $1$ & $0$ & $0$ \\ 
	\hline
  	$\widetilde{\Omega}^\pm$ & $1$ & $\pm \widetilde{N}_c$ & $1$ & $1$ & $0$ & $2$ 	& $1$ & $\pm 1$ \\ 
	\hline
	\end{tabular}
	\caption{ The charges of the fields under the symmetries of the magnetic dual of $U(N_c)$ SQCD, whose gauge group is $U(\widetilde{N}_c = N_f-N_c)$. The generators of the dual $U(1)_{\widetilde{V}}$ and $U(1)_{\widetilde{A}}$ groups are related to the electric versions according to \eqref{chargemap}.  \label{tab:UNdualdetcharges}}
	\end{table}
	\setlength\tabcolsep{6pt} 
	\def\arraystretch{1}

	The remaining global-gauge-gauge anomalies in the first line of \eqref{I6UN} involve the instanton number $n_B$ for the $U(1)_B$ gauge symmetry, which is an integer class on spin manifolds,   written in terms of the $U(1)_B$ gauge field $a_B$ as,
	\begin{align}
	n_B = \frac{(da_B)^2}{8\pi^2} = \frac{1}{2} c_1(B)^2\,,\qquad \int_{M_4} n_B\in \mathbb{Z}\,.
	\end{align}
Therefore the axial anomaly is trivialized by the previously determined breaking of $U(1)_A\to \mathbb{Z}_{2N_f}$, while the  ABJ anomaly involving the R-symmetry implies that $U(1)_R$ is partially broken, with discrete invertible remainder $(\mathbb{Z}_2)_R$. In our notation from  \eqref{fg2}, $k_A=2N_f$ for the gauge-axial anomaly in \eqref{fg2}, and $k_R = -2$ for the gauge-R symmetry anomaly.  In fact, following the arguments of \cite{Choi:2022jqy,Cordova:2022ieu} one concludes that the $U(1)_A$ symmetry is still explicitly broken to $\bZ_{2N_f}$ as it participates in the mixed anomaly with the non-abelian gauge symmetry $U(N_c)$, while an R-symmetry rotation by any rational angle $\pi p/q$ for co-prime integers  $p$ and $q$ is preserved and enacted by a non-invertible topological defect, one argument for which goes as follows.\footnote{~While previously (around \eqref{Axial_defect_gen}) we used $p$ and $N$ for the coprime integers, here we switch to $p$ and $q$ so as to not confuse with $N_c,N_f$.}

By itself, the codimension one defect $U_\alpha(\CM_3)=\exp[ i  \alpha \int_{\CM_3} \star j_R]$ that enacts an R-symmetry rotation by angle $\alpha = \frac{2 \pi p}{|k_R|q} = \frac{\pi p}{q}$ is not topological due to the ABJ anomaly, but the composition,
	\begin{align}\label{upq}
\exp \left[ \frac{ i\pi p}{q}  \left( \int_{M_3} \star j_R -2\int_{M_4} n_B \right)\right] \,,\qquad 
	\text{gcd}(p,q)=1\,,\qquad M_3=\partial M_4 
	\end{align}
is both gauge-invariant and topological.
 In order to promote \eqref{upq} into a genuinely three-dimensional defect that does not depend on the 4d bulk, we stack it with an abelian 3d TQFT with $\mathbb{Z}_q^{(1)}$ 1-form symmetry, whose anomaly is labeled by the integer $p\in \mathbb{Z}_q$. Every such TQFT can be decomposed into a minimal one with $M_4$ bulk dependence,\footnote{~A presentation of the Lagrangian density $\CA^{q,p}$ in terms of an abelian CS theory was given in Appendix C of \cite{vanBeest:2023dbu}. We have already encountered it in Section~\ref{sec:intro2d} in analyzing the $S^2$ reduction of the non-invertible axial defect.}
	\begin{align}\label{aqp}
	\exp\left[i  \left(\int_{M_3}\CA^{q,p}(\hat{B})  +\frac{ 2\pi p}{q} \int_{M_4} \frac{\CP(\hat{B})}{2}   \right) \right]\,,
	\end{align}
and one that is neutral under $\mathbb{Z}_q^{(1)}$ \cite{Hsin:2018vcg}.  Here, $\CP(\hat{B})$ is the Pontryagin square operation, with $\hat{B} \in H^2(M_4,\mathbb{Z}_q)$ the background gauge field for the $\mathbb{Z}_q^{(1)}$ 1-form symmetry.
The four-dimensional bulk magnetic 1-form symmetry couples with the internal 1-form symmetry of the degrees of freedom living on the defect, so that by identifying $\hat{B}$ with $c_1(B)$ mod $q$, the bulk dependence will precisely cancel that in \eqref{upq}. There are thus an infinite number of non-invertible 0-form defects in the $U(N_c)$ SQCD labeled by the rational numbers $p/q$,  given by the following,
\begin{align}\label{SQCDdefect}
	\CD_{\frac{\pi p}{q}} (M_3) = \exp \left[ i  \int_{M_3} \left( \frac{\pi p}{q} \star j_R + \CA^{q,p}(c_1(B)) \right)  \right]\,.
	\end{align}
As discussed in \cite{Choi:2022jqy,Cordova:2022ieu}, this non-invertible symmetry can be obtained equivalently by gauging a $\mathbb{Z}^{(1)}_q$ subgroup of the $U(1)_M^{(1)}$ magnetic 1-form symmetry, performing the chiral rotation by the angle $\pi p/q$, and then gauging the dual $\mathbb{Z}_q^{(1)}$ electric 1-form symmetry.

We comment that one must be careful to account for the non-canonical quantization of the background R-symmetry gauge field $A_R$ in drawing the conclusions of the previous paragraphs. Since the anomaly-free $U(1)_R$ symmetry in Table~\ref{tab:UNdualdetcharges} assigns non-integer charges to the fields, $A_R$ should be regarded as a $U(1)/\mathbb{Z}_{N_f / \text{gcd}(N_c,N_f)}$ gauge field, with flux satisfying,
	\begin{align}
	\int_{S^2} c_1(R) \, \in \, \frac{N_f}{\text{gcd}(N_c,N_f)}\mathbb{Z}\,.
	\end{align}
This subtlety can be addressed by changing bases to one in which all fields have integral charges. In such a basis, the canonically normalized background fields $A_{R'}\sim A_{R'} + 2\pi$ and $A_{A'}\sim A_{A'} + 2\pi$ are related to the original ones as,\footnote{~The now-integral $U(1)_{R'}$ and $U(1)_{A'}$ charges of an operator $\CO$ are related to the charges in Table~\ref{tab:UNdetcharges} as,
	\begin{align} \label{unprimeR}
	R'(\CO) = R(\CO) - \frac{N_f-N_c}{N_f} A(\CO)\,,\qquad A'(\CO) = A(\CO)\,.
	\end{align}} 
	\begin{align} \label{unprimeA}
	A_{A'} = A_{A} + \frac{N_f-N_c}{N_f} A_R\ \  \  \text{mod} \ \ 2\pi\,,\qquad A_{R'} = A_R\ \ \  \text{mod}\ \ \frac{2\pi}{N_f/\text{gcd}(N_c,N_f)}\,.
	\end{align}
A more careful analysis of the anomalies \eqref{I6UN} in terms of these integer quantized gauge fields yields the same conclusions that we have presented above, with pertinent anomaly terms  written in \eqref{ael0}.

Moving on to the second line of \eqref{I6UN}, the presence of  anomalies linear in the $U(1)_B$ gauge symmetry implies an extension of the 0-form symmetry participating the mixed anomaly by $U(1)_M^{(1)}$, as pointed out in \cite{Cordova:2018cvg}. This model then exhibits a 2-group involving the non-abelian $SU(N_f)_{L/R}$ flavor symmetries
\begin{equation} \label{2groupSQCD}
\begin{split}
	H &= dB +\frac{1}{4\pi} \left(\mathrm{CS}(A_{SU(N_f)_L}) - \mathrm{CS}(A_{SU(N_f)_R})\right)\,,\\
	 dH &= \frac{1}{4\pi} \left(\tr F_{SU(N_f)_L}^2 -\tr F_{SU(N_f)_R}^2\right) \,.
	\end{split}
\end{equation}
Note that the 2-group structure does not involve the R-symmetry and hence does not mix with the non-invertible symmetry.

We emphasize that the theory with $U(N_c)$ gauge group has several important features that are new relative to the case of $SU(N_c)$ gauge group, related to the presence of the $U(1)^{(1)}_M$ magnetic 1-form symmetry. Principally, the  theory exhibits the non-invertible defects $\CD_{\frac{\pi p}{q}}$, which implement a $\mathbb{Q}/\mathbb{Z}$ symmetry which is not present in the $SU(N_c)$ theory. By contrast, the $SU(N_c)$ ABJ anomaly $\CI_6\supset 2N_f c_2({SU(N_c)}) c_1(A)$ cannot be trivialized by stacking with a TQFT  associated to the axial symmetry defect, essentially since $\pi^1(SU(N_c))$ is trivial and so the second Chern class cannot be written as a Pontryagin square. Furthermore, the $U(N_c)$ theory enjoys the 2-group \eqref{2groupSQCD}.
 
\subsubsection*{Consequences of the Non-Invertible Symmetry}

Let us briefly comment on the implication of the non-invertible R-symmetry.  
For the $SU(N_c)$ case in the conformal window, we have a UV gauge theory that flows to an interacting SCFT at low energies. After gauging the $U(1)_B$ symmetry, will the same conclusions hold also for the $U(N_c)$ theory in the conformal window? On the one hand, the issue with the R-symmetry is quite dramatic, since gauging $U(1)_B$ breaks the continuous R-symmetry to the $\mathbb{Q}/\mathbb{Z}$ non-invertible symmetry. 
However, the anomaly which leads to the R-symmetry breaking comes entirely from the $U(1)_B$ part of the gauge group, which we expect to become weakly coupled in the IR. Then, we would expect that the $U(N_c)$ case should flow to a weakly gauged version of the SCFT of the $SU(N_c)$ case. Indeed, normally in 4d $\CN=1$ theories we do not expect phase transitions to occur as scales cross one another, leading us to expect that we can perform the gauging of the SCFT in the IR, and this should be the same regardless of the size of the couplings, as long as they are not strictly taken to zero or infinity.    

This is subtle, however, since the $SU(N_c)$ part of the gauge group flows to strong coupling, and in principle the $SU(N_c)$ strong dynamics might generate new superpotential terms now that the R-symmetry is broken. If said superpotential terms are relevant in the IR then they might drive the theory to a new fixed point. In said scenario, gauging $U(1)_B$ in the IR would not commute with gauging it at the UV point, as the latter gauging corresponds to taking the coupling of such terms to be strictly zero. For this reason it might be difficult to rule out (for example) the generation of new superpotential terms that only violate the continuous R-symmetry, if all we had to go on was the remaining invertible discrete part of the R-symmetry. However, note that the non-invertible part allows us to closely mimic any continuous R-symmetry rotation, as the rationals are dense in $\mathbb{R}$. Therefore, any such putative superpotential terms would be forbidden by the non-invertible symmetry and we can indeed argue that the $U(N_c)$ case should just flow to the SCFT associated with the $SU(N_c)$ case, but with its $U(1)_B$ symmetry weakly gauged.

\subsection{Mapping Symmetries Across Seiberg Duality}
\label{sec:4dnoninv}

Having identified a large class of global symmetries of SQCD implemented by the non-invertible defects \eqref{SQCDdefect}, one should next verify that these symmetries are matched across duality.  Both the $SU(N_c)$ and the $U(N_c)$ SQCD indeed participate in electric-magnetic Seiberg dualities \cite{Seiberg:1994pq}, where the magnetic dual theory has gauge group $SU(N_f-N_c)$ or $U(N_f-N_c)$ accordingly, with $N_f$ chiral superfields $q,\widetilde{q}$ in the fundamental and antifundamental representations of the gauge group, and a gauge singlet $\widetilde{M}$ which couples via the superpotential $W_{\text{mag}} = q \widetilde{M} \widetilde{q}$ (Throughout a tilde over a letter will denote the magnetic analogue of an electric variable, with the exception that $\widetilde{Q},\widetilde{q}$ denote anti-quarks and $\widetilde{B},\widetilde{b}$ anti-baryons). The charges of the fields in the magnetic dual under the classical 0-form symmetries are listed in Table~\ref{tab:UNdualdetcharges}. The magnetic description is useful for values of $N_f$ greater than $N_c+1$: in the range $N_c+2 \leq N_f \leq 3/2\, N_c$ the dual is IR free, indicating that the theory is in a free magnetic phase consisting of massless magnetically charged fields; meanwhile, in the conformal window of $3/2 \,N_c < N_f < 3 N_c$, both the electric and magnetic theories are asymptotically free and flow at low energies to the same nontrivial superconformal fixed point.

We begin by considering how the (invertible) global symmetries are mapped across duality for the case of $SU(N_c)$ gauge group. The first step is to note that the baryonic $U(1)_B=U(1)_V/\mathbb{Z}_{N_c}$ symmetry of $SU(N_c)$ SQCD is mapped to the baryonic $U(1)_{\widetilde{B}}=U(1)_{\widetilde{V}} / \mathbb{Z}_{N_f-N_c}$ symmetry of its magnetic dual up to a shift with the discrete $\mathbb{Z}_{2N_f}$ remnant of the axial symmetry. Regarding this discrete subgroup as embedded in its $U(1)_A$ predecessor---as per the comments below \eqref{I6UN}, practically this amounts to the replacement $F_A = dA_A\to 2\pi \beta(\hat{A}_A)$---the background field strengths for the axial and vector symmetries are related across the duality as follows,
	\begin{align}\label{curvemap}
	F_{\widetilde{A}} = - F_A\,,\qquad 
	(N_f-N_c)F_{\widetilde{V}} = N_c F_V - N_f F_A\,.
	\end{align} 
An equivalent expression to \eqref{curvemap} can be given in terms of the symmetry generators of these groups; denoting the generators of $(\mathbb{Z}_{2N_f})_A$ and $U(1)_V$ by $a$ and $v$ respectively, and similarly for the generators $\tilde{a}$ and $\tilde{v}$ of the magnetic  $(\mathbb{Z}_{2N_f})_{\widetilde{A}}$ and $U(1)_{\widetilde{V}}$ symmetries, these satisfy,\footnote{~These can also be understood as the fugacities in the supersymmetric index or $S^3\times S^1$ partition function.}
	\begin{align}
	\label{chargemap}
	\tilde{a} \to a^{-1}\,,\qquad\quad 
	\tilde{v} \to v^{\frac{N_c}{N_f-N_c}} \,a^{-\frac{N_f}{N_f-N_c}}\,.
	\end{align}
The relations \eqref{chargemap} are required by matching the charges of the gauge-invariant operators. In particular, using the charges in Table~\ref{tab:UNdetcharges} one may verify that the electric mesons $M^i_j = Q^{ai}\widetilde{Q}_{aj}$, baryons $B^{i_1\dots i_{N_c}} = \epsilon_{a_1\dots a_{N_c}} Q^{a_1 i_1}\dots Q^{a_{N_c} i_{N_c}}$, and antibaryons $\widetilde{B}_{{i_1}\dots i_{N_c}}= \epsilon^{a_1 \dots a_{N_c}} \widetilde{Q}_{a_1 i_1}\dots \widetilde{Q}_{a_{N_c} i_{N_c}}$ carry the following charges under $(\mathbb{Z}_{2N_f})_A \times U(1)_R \times U(1)_V$\,,
	\begin{align}
	M\sim a^2 \, r^{\frac{2(N_f-N_c)}{N_c}} \,,\qquad 
	B \sim \left( a \, v \right)^{N_c} \,r^{\frac{N_c(N_f-N_c)}{N_f}}\,,\qquad 
	\widetilde{B} \sim \left( \frac{a}{v}\right)^{N_c}  \, r^{\frac{N_c(N_f-N_c)}{N_f}}\,.
	\end{align}
The electric mesons $M$ map to the magnetic gauge singlets $\widetilde{M}$, while the electric baryons $B,\widetilde{B}$ map to the magnetic baryons $b,\widetilde{b}$ up to contraction with the $\epsilon$ symbol for $SU(N_f)$. The charges of the gauge-invariant operators in the magnetic theory are then computed from Table~\ref{tab:UNdualdetcharges} as,
	\begin{align}
	\widetilde{M}\sim \tilde{a}^{-2} \, r^{\frac{2 (N_f-N_c)}{N_f}} \,,\qquad 
	b\sim \left(\tilde{a} \,\tilde{v} \right)^{N_f-N_c} \, r^{\frac{N_c(N_f-N_c)}{N_f}}\,,\qquad 
	\widetilde{b} \sim \left( \frac{\tilde{a}}{\tilde{v}}\right)^{N_f-N_c}\,   r^{\frac{N_c(N_f-N_c)}{N_f}}\,.
	\end{align}
Using \eqref{chargemap} and  $a^{2N_f}=\tilde{a}^{2N_f}=1$, the dual pairs of gauge-invariant operators evidently carry the same charge under all global symmetries. 

In order to examine the dual theories with unitary gauge group, we gauge the baryonic $U(1)_B$ symmetry on the electric side of the duality, and $U(1)_{\widetilde{B}}$ on the magnetic side of the duality, keeping in mind that the gauge fields are related as in \eqref{curvemap} (where now $F_V\to f_V$ is the curvature of a dynamical gauge field, and similarly for $F_{\widetilde{V}}\to f_V$ in the dual). The two partition functions will still be dual, differing only up to a counterterm $\exp(\frac{iN_f}{2\pi} \int_{M_4 }B \wedge da_A)$ that arises from coupling to the 1-form magnetic $U(1)^{(1)}_M$ symmetry background field. One should also keep in mind that after gauging $U(1)_B$ the baryons are no longer individually gauge-invariant operators, as only the combinations $B\widetilde{B}$ and $b\widetilde{b}$ are gauge invariant. As we show explicitly in Appendix \ref{sec:appumatch}, one can  verify that all 't Hooft anomalies---including those of the discrete $\mathbb{Z}_{2N_f}$ symmetries---match across duality after applying the identification \eqref{curvemap}.\footnote{~As we noted below \eqref{I6UN}, the discrete $\mathbb{Z}_{2N_f}$ chiral symmetry is contained in the continuous symmetries, and therefore the $\mathbb{Z}_{2N_f}$ anomalies are automatically matched and do not present a new check of the duality. However, we find that it is useful to present the global symmetries as in \eqref{unglobal} in order to describe the mapping of the non-invertible topological defects of the theory that arise after gauging $U(1)_B$.}

The matching of the non-invertible defects can then be accomplished as follows.
From our discussion in the previous subsection, the non-invertible defects on the electric side of the duality follow from the $\tr\, U(1)_B^2 U(1)_R = -2$ anomaly in \eqref{I6UN}. The corresponding term in the anomaly polynomial of the magnetic dual leads to $\tr\, U(1)_{\widetilde{B}}^2 U(1)_R = -2$, and so by the same logic the dual has an infinite set of  $\mathbb{Q}/\mathbb{R}$ symmetries enacted by non-invertible defects which are the analogues of \eqref{SQCDdefect}. Because the R-symmetry matches on either side of the duality,  the defects $\CD_{\frac{\pi p}{q}}$ enacting an R-symmetry rotation by rational angle $\pi p/q$ are mapped in the dual to defects $\widetilde{\CD}_{\frac{\pi p}{q}}$, which are labeled by the same co-prime integers $p$ and $q$.

\subsection{Adding Determinant Matter}
\label{sec:4dUNdet}

As we have seen, the R-symmetry of $U(N_c)$ SQCD has a mixed anomaly with the gauge symmetry $U(1)_B$, with the non-anomalous part being the non-invertible $(\mathbb{Q}/\mathbb{Z})_R$ symmetry. Since we would like to compactify the 4d theory on $S^2$ (which will be elaborated on in Section~\ref{sec:2d}), generically all supersymmetries are broken due to the non-trivial curvature of $S^2$, and we lose control over the compactified theory. On the other hand, it is possible to preserve half of the supercharges via an R-symmetry twist, which requires $U(1)_R$ to be unbroken. This motivates us to consider a modified theory by introducing two chiral fields $\Omega^\pm$ in the determinant of the fundamental and  antifundamental representations respectively of the $U(N_c)$ gauge symmetry, whose charges  are designed to cancel off the $\tr\, U(1)_B^2 U(1)_R$ anomaly \cite{Gadde:2015wta}.  This theory classically possesses two additional global symmetries $U(1)_I$ and $U(1)_t$ that act on $\Omega^\pm$. The charges of the fields under these symmetries are listed in Table~\ref{tab:UNdetcharges}.  
	
We may now repeat the analysis from the previous subsections to determine the symmetries of this model. 
The anomalies involving the dynamical gauge fields are now captured by the following terms in the anomaly polynomial,
	\begin{align}
	\label{I6UNdet}
	\begin{split}
	 \CI_6
	\, \supset \, 
	&2 N_f c_1(A) c_2({U(N_c)})  +  N_f c_1(A)  c_1(B)^2+  c_1(I) c_1(B)^2\\
	&+ \left[ 2c_1(t)\left(  c_1(R) + c_1(I)\right)+  c_2(SU(N_f)_L) - c_2(SU(N_f)_R) \right] c_1(B)\,.
	\end{split}
	\end{align}
The $\tr\, U(N_c) U(1)_A$ anomaly implies that $U(1)_A$ is again broken to $\mathbb{Z}_{2N_f}$, which can be absorbed by a combination of a $U(1)_V$ and $SU(N_f)_L\times SU(N_f)_R$ rotation. 
The $\tr\, U(1)_B^2 U(1)_I$ anomaly implies that $U(1)_I$ will be broken to a  non-invertible $\mathbb{Q}/\mathbb{Z}$ symmetry, associated to defects $\CD_{\frac{\pi p}{q}}$ that perform a $U(1)_I$ rotation by a rational angle $\pi p/q$. The defect is of the same form as \eqref{SQCDdefect}, which we reproduce here,  
\begin{eqnarray}\label{eq:Idefect}
	\CD_{\frac{\pi p}{q}}(M_3) = \exp\left[i \int_{M_3} \left(\frac{\pi p}{q} \star j_I + \CA^{q,p}(c_1(B))\right)\right]\,.
\end{eqnarray}

The terms on the second line of \eqref{I6UNdet} lead to a 2-group  involving the non-abelian flavor symmetries, in addition to an extension of  the $U(1)_R$ and $U(1)_t$ global symmetries by $U(1)_M^{(1)}$. The background fields are then related by 
\begin{eqnarray}\label{eq:2group111}
	\begin{split}
		H&= dB + \frac{1}{2\pi} A_{R}dA_{t} + \frac{1}{4\pi} \mathrm{CS}(A_{SU(N_f)_L}) - \frac{1}{4\pi} \mathrm{CS}(A_{SU(N_f)_R}) \,, \\
		dH &= \frac{1}{2\pi} F_R \wedge F_t + \frac{1}{4\pi} \tr F_{SU(N_f)_L}^2 -  \frac{1}{4\pi} \tr F_{SU(N_f)_R}^2\,.
	\end{split}
\end{eqnarray}
However, unlike in the case without determinant matter where the 0-form symmetries participating in the 2-group and the non-invertible symmetries do not mix, here the situation is more involved. In particular, the $U(1)_I$ symmetry participates in both the  anomaly quadratic in $c_1(B)$, breaking $U(1)_I$ to a non-invertible symmetry, and the  anomaly linear in $c_1(B)$, leading to higher structure of the non-invertible defects that is encoded in the junctions among these defects \cite{Copetti:2023mcq}. Here instead of discussing such higher structure of non-invertible defects in detail, we point out that it is conveniently packaged into the SymTFT and its boundary conditions, thanks to the recent developments of SymTFT for continuous symmetries \cite{Gaiotto:2020iye,Apruzzi:2021nmk,Freed:2022qnc, Bonetti:2024cjk, Apruzzi:2024htg}, which we have discussed in Section~\ref{sec:symtft}. Moreover, we have also seen  that after compactifying the 4d theory on $S^2$ with a non-trivial gauge flux in $c_1(B)$, the resulting anomaly of the 0-form symmetries of the 2d theory depends sensitively on the $c_1(B)$ linear anomaly in \eqref{I6UNdet}. As such, further analysis of the SymTFT in this example can provide a useful probe for the presence of higher structure for non-invertible symmetries in 4d.

\section{From Four to Three Dimensions}
\label{sec:3d}

In this section we will consider a series of examples of 4d gauge theories compactified on a circle, which illustrate the  features discussed in Section~\ref{sec:intro3d}.

\subsection{Maxwell Theory on a Circle}
\label{sec:3dmaxwell}

Let us first discuss the case without matter, and consider free Maxwell theory in four spacetime dimensions. The 4d gauge connection $a_\mu$, $\mu=\{0,1,2,3\}$ reduces to the 3d vector field $a_i$, $i=\{0,1,2\}$, and a compact scalar $\sigma$ coming from the holonomy of the 4d vector field along the circle direction \eqref{sigma},
	\begin{align}\label{sigma2}
	\sigma = \int_{S^1}a\,,\qquad  \sigma \sim \sigma + 2\pi\,.
	\end{align}
The pure 4d gauge theory  possesses a $U(1)^{(1)}_{E}$ electric 1-form symmetry under which the Wilson line changes by a phase, and a magnetic $U(1)_{M}^{(1)}$ 1-form symmetry under which the 't Hooft line changes by a phase.  On general grounds, each of these 1-form symmetries reduces to a 1-form and 0-form symmetry when compactified on a circle, which can be seen as follows  \cite{Gaiotto:2014kfa}. The 4d Wilson line that does not wrap the circle reduces to the 3d Wilson line, which is acted upon by the 3d 1-form symmetry. Meanwhile, the Wilson line wrapping the circle becomes the exponentiated scalar $e^{i\sigma}$, whose rotation by a phase is equivalent to a shift of the scalar field, $\sigma \to \sigma + \alpha$. Therefore, the $U(1)^{(1)}_E$ electric 1-form symmetry reduces to a 3d electric 1-form symmetry, and a 0-form symmetry that acts by shifting the scalar.  

We can similarly consider what happens to the 't Hooft lines. The 't Hooft line not wrapping the circle gives rise to an 't Hooft line in 3d, which has the property that the Wilson line around them transforms by phase which is an integer multiple of $2\pi$, so that in particular the scalar $\sigma$ undergoes a multiple of $2\pi$ shift when going around them. Meanwhile, the 't Hooft lines wrapping the circle become 3d monopole operators. In this way the magnetic 1-form symmetry reduces to a 3d magnetic 1-form symmetry that acts on the winding defects, and a 0-form symmetry (sometimes also called the topological symmetry) acting on the monopole operators. 

Let us next explore the reduction of non-invertible symmetry in this example.
Free Maxwell theory is known to enjoy electric-magnetic duality exchanging $f$ and $\star f$. This transformation, in addition with $2\pi$ shifts of the $\theta$ angle, generates the $SL(2,\mathbb{Z})$ duality group of  free Maxwell theory. Here we shall for simplicity set $\theta=0$, and concentrate only on the electric-magnetic duality. The action of the theory is then given by,
	\begin{align} \label{Action:fMt}
	S^{4d} \, = \, -\frac{1}{2 e_{4d}^2} \int da_{4d} \wedge \star da_{4d}\,. 
	\end{align}
Under a duality transformation the coupling constant changes as $e_{4d}^2 \rightarrow \frac{(2\pi)^2}{e_{4d}^2}$, so that for generic values of the coupling, the duality relates Maxwell theories with different values of $e_{4d}$. At the special value of $e^2_{4d} = 2\pi$, however, the theory is invariant under electric-magnetic duality and this transformation becomes a symmetry.  More generally, whenever $\frac{2\pi}{e^2_{4d}}$ is rational, Maxwell theory possesses a non-invertible electric-magnetic duality symmetry---we refer the reader to \cite{Niro:2022ctq} for details about this structure. Here we shall be concerned with the fate of this symmetry upon a circle reduction to three dimensions.

For this, we begin by reducing the action \eqref{Action:fMt} on the circle to three dimensions, obtaining,
	\begin{align}
S^{4d}\,\rightarrow \,S^{3d}  = & - \frac{1}{2e^2_{3d}} \int d a_{3d} \wedge \star d a_{3d} - \frac{1}{2 e^2_{3d}(2\pi R)^2} \int d \sigma \wedge \star d \sigma \,.
	\end{align}
Here we have identified the gauge couplings as $e_{4d}^2 = (2\pi R)e_{3d}^2$, and used that $\sigma \simeq 2\pi R a_3$.\footnote{~Since $\sigma$ is taken to be independent of $x^3$, we can effectively rewrite $\sigma = \frac{1}{2\pi R}\int_{S^1} dx^3\sigma$ and identify $\sigma \simeq 2\pi R a_3$. }
Next, we can dualize the 3d gauge field to a scalar using $d a_{3d} = \star d\varphi$. The scalar $\varphi$ is periodic, which can be seen  by performing an integral $\int_L \varphi$ over a line $L$ starting at a point, expanding to some circle, and then contracting back to a point. In particular, we have that $\int_L \varphi = \int_{\Sigma} d \varphi$, where $\Sigma$ is a surface bounded by $L$. It is possible to choose a continuous deformation of $L$ so that $\Sigma$ will trace a sphere, in which case $\int_{L_{\text{initial}}} \varphi - \int_{L_{\text{final}}} \varphi = \int_{\Sigma} d \varphi = \int_{\Sigma} \star d a_{3d}$. However, $L_{\text{initial}}$ and $L_{\text{final}}$ are both a point, so we just have that $\varphi  = \varphi + \int_{\Sigma} \star d a_{3d}$, which requires $\varphi$ to be periodic with period $\int_{\Sigma} \star d a_{3d} = e^2_{3d}$.\footnote{~The more common convention is that $\int \star da$ gives the electric charge $e$ rather than $e^2$, which is related to our convention here by absorbing a factor of $e$ into the gauge field.} 
We thus conclude that $\varphi$ is a periodic scalar with period $e^2_{3d}$.

Now that there are two periodic scalars, $\sigma$ and $\varphi$,  one might envision a symmetry exchanging the two. For this to actually be a symmetry of the action, we must have that (a) both scalars have the same period, and (b) the kinetic terms of the two scalars agree. As to the first point, we are clearly free to redefine the scalars to ensure that they have the same period. For instance, we can define $\varphi' = \frac{2\pi}{e^2_{3d}}\varphi$, so that $\varphi'$ is $2\pi$ periodic. In terms of these fields, the action takes the form,
	\begin{align}
		\label{3dMaxwell_scalar}
	S^{3d} = -\frac{e^2_{3d}}{2 (2\pi)^2 } \int d \varphi' \wedge \star d \varphi' - \frac{1}{ 2e^2_{3d}(2\pi R)^2} \int d \sigma \wedge \star d \sigma \,.
	\end{align}
The exchange of $\varphi'$ and $\sigma$ is evidently a symmetry  when $e_{3d}^2 = 1/R$, or equivalently when  $e^2_{4d} = 2\pi$, which is precisely the condition for the presence of  the duality symmetry in 4d. We now see that this invertible symmetry of Maxwell theory reduces to an invertible symmetry of  the theory on the circle, which acts by exchanging the two periodic scalars $\varphi'$ and $\sigma$.

What about other values of $e_{4d}$? For this let us first redefine $\sigma$ such that the kinetic terms are equal. This requires setting $\sigma = \frac{e^2_{4d}}{2\pi}\sigma'$, where now $\sigma'$ satisfies $\sigma' = \sigma' + \frac{(2\pi)^2}{e^2_{4d}}$. The only thing now impeding the symmetry is the different periods of $\sigma'$ and $\varphi'$. 
However, note that we can change the period of the scalars by gauging various subgroups of the zero- or 1-form symmetries. Specifically, by gauging a $\mathbb{Z}_k$ subgroup of the shift symmetry of $\sigma'$, we can change its period such that it satisfies $\sigma' = \sigma' +\frac{(2\pi)^2}{k e^2_{4d}}$. Similarly, by gauging a $\mathbb{Z}_k$ subgroup of the 1-form winding symmetry of $\varphi'$, we can change its period such that $\varphi' = \varphi' +2\pi k$. As such we see that if $\frac{2\pi}{e^2_{4d}}=N$ for integer $N$, we can still obtain a symmetry as follows. First, we gauge a $\mathbb{Z}_N$ subgroup of the 0-form shift symmetry of $\sigma'$ (or a $\mathbb{Z}_N$ subgroup of the 1-form winding symmetry of $\varphi'$). After this, the two scalars have the same period and there is a symmetry exchanging them. We can then perform said symmetry transformation, and return back to the original period by gauging the dual $\mathbb{Z}_N$. The combined operation becomes a non-invertible symmetry, which is the 3d reduction of the non-invertible symmetry of Maxwell theory at $e^2_{4d}=\frac{2\pi}{N}$. More generally, we expect to be able to define non-invertible symmetries in a similar manner whenever $\frac{2\pi}{e^2_{4d}}\in \mathbb{Q}$.

In conclusion, we see that the non-invertible electric-magnetic duality symmetry of 4d Maxwell theory reduces to a non-invertible symmetry of 3d Maxwell theory plus a periodic scalar, which arises by compactifying the four-dimensional theory on a circle. Said symmetry can be interpreted as exchanging the periodic scalar and the dual photon. Notably, the presence of the periodic scalar was necessary for the operation of the symmetry. Furthermore, gauging discrete subgroups of the shift and winding symmetries was required to recover the non-invertible symmetry for any rational value of $\frac{2\pi}{e_{4d}^2}$. 

\subsubsection{Reducing the topological defects}

Let us next expand more on the reduction of non-invertible symmetries in 4d Maxwell theory  from the perspective of reducing the corresponding defects themselves, and investigate in more detail their action on operators in the resulting 3d theory.  We focus for simplicity on two of the many non-invertible defects in the theory: the (electric) condensation defects, and the $S$-duality defects we discussed above. The former can be obtained by higher gauging of a $\mathbb{Z}_{N}^{\left(1\right)}$ subgroup of the electric 1-form symmetry on a codimension one submanifold, and can be described by the (Euclidean) action,
\begin{equation}
		\label{Cond_defect_Lag}
	\frac{1}{2e_{4d}^{2}}\int_{x<0}da_{4d}^{\left(L\right)}\wedge\star da_{4d}^{\left(L\right)}+\frac{1}{2e_{4d}^{2}}\int_{x>0}da_{4d}^{\left(R\right)}\wedge\star da_{4d}^{\left(R\right)}+\frac{iN}{2\pi}\int_{x=0}\tilde{a}\wedge\left(da_{4d}^{\left(L\right)}-da_{4d}^{\left(R\right)}\right)\,,
\end{equation}
where we consider a planar defect placed at $x=0$ and where $\tilde{a}$ is a gauge field living on it. The latter, as previously discussed, exist at $\frac{2\pi}{e_{4d}^2}=N$ (for integer $N$)\footnote{~While defects with the same action on local operators (which hence can also be called $S$-defects) exist for other values of $e_{4d}^2$ as well,  for simplicity we will only consider the ones at $\frac{2\pi}{e_{4d}^2}=N$. }  and correspond to the action,
\begin{equation}
	\label{S_defect_Lag}
\frac{N}{4\pi}\int_{x<0}da_{4d}^{\left(L\right)}\wedge\star da_{4d}^{\left(L\right)}+\frac{N}{4\pi}\int_{x>0}da_{4d}^{\left(R\right)}\wedge\star da_{4d}^{\left(R\right)}+\frac{iN}{2\pi}\int_{x=0}a_{4d}^{\left(L\right)}\wedge da_{4d}^{\left(R\right)}\,.
\end{equation}
Notice that fusing two duality defects \eqref{S_defect_Lag} with opposite orientation results in the condensation defect \eqref{Cond_defect_Lag}. 

We will next explore what happens to these defects upon reduction on a circle to 3d. For the condensation defect \eqref{Cond_defect_Lag} we obtain 
\begin{equation}
			\label{Cond_defect_Lag_red}
\frac{iN}{2\pi}\int_{x=0}\tilde{a}\wedge\left(da_{4d}^{\left(L\right)}-da_{4d}^{\left(R\right)}\right)\rightarrow\frac{iN}{2\pi} \left[\int_{x=0}\tilde{\phi}\left(da_{3d}^{\left(L\right)}-da_{3d}^{\left(R\right)}\right)+\tilde{a}\wedge\left(d\sigma^{\left(L\right)}-d\sigma^{\left(R\right)}\right)\right]
\end{equation}
where we have defined the defect compact scalar $\tilde{\phi}$ as,
\begin{equation}
\tilde{\phi}=\int_{S^{1}}\tilde{a}\quad,\quad\tilde{\phi}\sim\tilde{\phi}+2\pi\,.
\end{equation}
We see that the non-invertible condensation defect of 4d Maxwell theory reduces to that of 3d Maxwell theory (given by the first term on the RHS of \eqref{Cond_defect_Lag_red}), and to a defect in the free scalar theory which we now address. One can show (see \emph{e.g.}~Appendix E of \cite{Niro:2022ctq}) that this latter defect can be rewritten as,
\begin{align} \label{Cond_discrete}
\begin{split}
\int D\tilde{a}&\exp\left[ \frac{iN}{2\pi}\int_{x=0}\tilde{a}\wedge\left(d\sigma^{\left(L\right)}-d\sigma^{\left(R\right)}\right)\right]\\
&\quad\to\, \sum_{\tilde{\eta}=0}^{N-1}\int D\tilde{a}\exp\left[-\frac{i}{2\pi}\int_{x=0}d\tilde{a}\left(\sigma^{\left(L\right)}-\sigma^{\left(R\right)}+\frac{2\pi}{N}\tilde{\eta}\right)\right]\,,
\end{split}
\end{align}
where $\tilde{\eta}=0,1,\ldots,N-1$ is an integer-valued field, which in turn evaluates to the following sum of 0-form defects,
\begin{equation}
\sum_{\tilde{\eta}=0}^{N-1}\exp\left(\frac{2\pi\tilde{\eta}}{N}\frac{1}{e_{3d}^{2}\left(2\pi R\right)^{2}}\int_{x=0}\star d\sigma\right).
\end{equation}
In other words, the reduction of the 4d condensation defect to the 3d scalar theory results in a non-simple defect, which is given by the projection operator associated with the $\mathbb{Z}_{N}^{\left(0\right)}$ 0-form symmetry of this 3d theory (coming from a $\mathbb{Z}_{N}^{\left(1\right)}$ subgroup of the 4d electric 1-form symmetry). In particular, it annihilates all the local operators $\exp\left(ik\sigma\right)$ for $k\in\mathbb{Z}$, except for those with $k=N\mathbb{Z}$. However it is not a genuine non-invertible symmetry, but just the sum of familiar 0-form defects. 

Let us present an alternative argument for the non-simplicity of this defect. The 4d condensation defect \eqref{Cond_defect_Lag} results from 1-gauging the $\mathbb{Z}_{N}^{\left(1\right)}$ subgroup of the electric 1-form symmetry along a 3d submanifold (say $x=0$). When reducing to the 3d scalar theory, this submanifold becomes two dimensional, but the electric defects remain of the same dimension (that is two dimensional), and so we obtain the 1-gauging of a 0-form symmetry which is simply a projection operator. 

Before turning to the reduction of the duality defect \eqref{S_defect_Lag}, let us briefly discuss another type of condensation defect in 4d Maxwell theory,  corresponding to a higher gauging of a $\mathbb{Z}_{N}^{\left(1\right)}$ subgroup of the magnetic 1-form symmetry. In this case the defect is (see \emph{e.g.}~\cite{Choi:2022jqy,Choi:2022rfe}) 
\begin{equation}
\int D\tilde{a}D\tilde{\tilde{a}}\, \exp\left[i\int_{x=0}\left(\frac{N}{2\pi}\tilde{a}\wedge d\tilde{\tilde{a}}+\frac{1}{2\pi}\tilde{a}\wedge da_{4d}\right)\right]\,,
\end{equation}
where $\tilde{a}$, $\tilde{\tilde{a}}$ are gauge fields living on the defect, and its reduction on the circle is given by 
\begin{equation}
	\label{Mag_Cond_red}
	\begin{split}
\int D\tilde{a}D\tilde{\tilde{a}}D\tilde{\phi}D\tilde{\tilde{\phi}}\,& \exp\left[i\int_{x=0}\left(\frac{N}{2\pi}\tilde{\phi}\,d\tilde{\tilde{a}}+\frac{1}{2\pi}\tilde{\phi}\,da_{3d}\right) \right. \\
&\left.\qquad\qquad\quad+i\int_{x=0}\left(\frac{N}{2\pi}\tilde{a}\wedge d\tilde{\tilde{\phi}}+\frac{1}{2\pi}\tilde{a}\wedge d\sigma\right)\right].
\end{split}
\end{equation}
The  defect in the first line of \eqref{Mag_Cond_red} exists in 3d Maxwell theory, and integrating out $\tilde{\tilde{a}}$ sets $\tilde{\phi}=2\pi\tilde{\eta}/N$ where $\tilde{\eta}$ is the integer-valued field we discussed above in \eqref{Cond_discrete}. One can similarly see that the defect is non-simple and given by the sum, 
\begin{equation}
\sum_{\tilde{\eta}=0}^{N-1}\exp\left(\frac{2\pi i\tilde{\eta}}{N}\int_{x=0}\frac{da_{3d}}{2\pi}\right).
\end{equation}
The second defect in \eqref{Mag_Cond_red} is a genuine condensation defect in the 3d scalar theory, given by higher gauging its 1-form symmetry. As can clearly be seen, this discussion of the 4d magnetic condensation defect is dual to that of the electric one. 

Let us finally reduce the duality defect \eqref{S_defect_Lag}. Here we have, 
\begin{equation}
	\label{SDefect_red}
\frac{iN}{2\pi}\int_{x=0}a_{4d}^{\left(L\right)}\wedge da_{4d}^{\left(R\right)}\,\rightarrow\,\frac{iN}{2\pi}\int_{x=0}\sigma^{\left(L\right)}da_{3d}^{\left(R\right)}+\frac{iN}{2\pi}\int_{x=0}a_{3d}^{\left(L\right)}\wedge d\sigma^{\left(R\right)}
\end{equation}
and the vanishing of the field variations on the defect (when also taking into account the bulk contributions) yields,
\begin{equation}
	\label{Sdefect_EOM_1}
x=0:\qquad\star da_{3d}^{\left(L\right)}=-\frac{i}{2\pi R}d\sigma^{\left(R\right)}\quad,\quad\star da_{3d}^{\left(R\right)}=\frac{i}{2\pi R}d\sigma^{\left(L\right)}\,.
\end{equation}
Since at $e_{4d}^{2}=\frac{2\pi}{N}$ (where we consider the defect) we have $e_{3d}^{2}=\frac{e_{4d}^{2}}{2\pi R}=\frac{1}{NR}$, we can rewrite \eqref{Sdefect_EOM_1} as,
\begin{equation}
	\label{Sdefect_EOM_2}
x=0:\qquad da_{3d}^{\left(L\right)}=-N\frac{ie_{3d}^{2}}{2\pi}\star d\sigma^{\left(R\right)}\quad,\quad\frac{ie_{\left(3d\right)}^{2}}{2\pi}d\sigma^{\left(L\right)}=\frac{1}{N}\star da_{3d}^{\left(R\right)}\,.
\end{equation}
For $N=1$ these are exactly the familiar relations mentioned above between the free 3d photon and its dual compact scalar with periodicity $2\pi$, and we conclude that the 4d invertible duality defect reduces to the sum of two topological 3d duality interfaces in 3d Maxwell theory plus a compact scalar. Alternatively, this 4d defect reduces to the invertible symmetry exchanging the scalar obtained from the holonomy of the 4d gauge field around the circle, with the scalar which is the dual of the 3d photon, as discussed below Eq.~\eqref{3dMaxwell_scalar}. For $N>1$, the extra factors of $N$ in \eqref{Sdefect_EOM_2} correspond to supplementing the sum of duality interfaces for $N=1$ with gauging a $\mathbb{Z}_{N}^{\left(1\right)}$ subgroup of the 1-form symmetry of 3d Maxwell theory and a $\mathbb{Z}_{N}^{\left(0\right)}$ subgroup of the the 0-form symmetry of the scalar theory (which are both equivalent to rescalings of the corresponding fields) in the half space to the right of the interfaces, for the present choice of orientation. This is indeed clearly a reduction of supplementing (for $N>1$) the $N=1$ duality defect of 4d Maxwell theory,  with the gauging of a $\mathbb{Z}_{N}^{\left(1\right)}$ subgroup of its electric 1-form symmetry in the half space to the right of the defect. 

Let us examine how the sum of interfaces \eqref{SDefect_red} acts on operators in the theory, which as discussed is given by 3d Maxwell theory and a free compact scalar. We begin by identifying the symmetries and some operators that are acted upon non-trivially by them, which will be useful for describing the action of the sum of interfaces. First, in addition to the two $\mathbb{Z}_{2}$ 0-form symmetries which act as $a_{3d}\rightarrow-a_{3d}$ and $ \sigma\rightarrow-\sigma$, there are two $U(1)$ 0-form symmetries with charges,
\begin{equation}
	\label{Charges_0-form}
Q_{a}^{\left(0\right)}=\oint\frac{da_{3d}}{2\pi}\quad,\quad Q_{\sigma}^{\left(0\right)}=-\frac{ie_{3d}^{2}}{4\pi^{2}}\oint\star d\sigma\,,
\end{equation}
under which the flux $n$ monopole of $a_{3d}$ and the operator $\exp\left(in\sigma\right)$ have charge $n$, respectively. Second, there are two $U(1)$ 1-form symmetries with charges, 
\begin{equation}
	\label{Charges_1-form}
Q_{a}^{\left(1\right)}=-\frac{i}{e_{3d}^{2}}\oint\star da_{3d}\quad,\quad Q_{\sigma}^{\left(1\right)}=\oint\frac{d\sigma}{2\pi}\,,
\end{equation}
that act non-trivially on the Wilson line $\exp\left(in\int a_{3d}\right)$ and on the disorder $\sigma$-line around which $\sigma$ has a nontrivial winding $\sigma\rightarrow\sigma+2\pi n$. Then, using these charges \eqref{Charges_0-form} and \eqref{Charges_1-form} we can rewrite the matching conditions \eqref{Sdefect_EOM_2} across the interfaces as follows, 
\begin{equation}
	\label{Charge_matching_3dMaxwell}
Q_{a}^{\left(0\right)L}=NQ_{\sigma}^{\left(0\right)R}\ \ \,,\ \ \ Q_{a}^{\left(1\right)L}=-NQ_{\sigma}^{\left(1\right)R}\ \ \,,\ \ \ Q_{\sigma}^{\left(1\right)L}=\frac{1}{N}Q_{a}^{\left(1\right)R}\ \ \,,\ \ \ Q_{\sigma}^{\left(0\right)L}=-\frac{1}{N}Q_{a}^{\left(0\right)R}\,.
\end{equation}
This form makes the action of the sum of interfaces on the operators of the theory quite manifest. For example, bringing from the left a flux $n$ monopole with $Q_{a}^{\left(0\right)L}=n$ and passing it across the interfaces results in $Q_{\sigma}^{\left(0\right)R}=-n/N$ which corresponds to the operator $\exp\left(-i\frac{n}{N}\sigma^{\left(R\right)}\right)$. For $n\neq N\mathbb{Z}$ this operator breaks the periodicity of $\sigma^{\left(R\right)}$, and is therefore not a genuine local operator but in fact the line operator $\exp\left(i\frac{n}{N}\int d\sigma^{\left(R\right)}\right)$ stretching between the defect (or interface) and the point under consideration. Similarly, moving a parallel Wilson loop $\exp\left(in\oint a_{3d}^{\left(L\right)}\right)$ with $Q_{a}^{\left(1\right)L}=n$ across the interfaces from the left results in a $\sigma^{\left(R\right)}$ winding loop with $Q_{\sigma}^{\left(1\right)R}=-n/N$ which for $n\neq N\mathbb{Z}$ corresponds to the disc operator $\exp\left(-\frac{2\pi n}{N}\frac{e_{3d}^{2}}{4\pi^{2}}\int_{D}\star d\sigma^{\left(R\right)}\right)$, where the boundary of $D$ is the loop of the $\sigma^{\left(R\right)}$ winding operator.

\subsection{Adding Matter}
\label{sec:3dmatter}

The previous discussion demonstrates that the compactness of the scalar, as well as the presence of shift and winding symmetries, can play an important role in the reduction of non-invertible symmetries.  
In the pure gauge theory case, the scalar $\sigma$ remains compact even at energies $E \ll 1/R$, and it only decompactifies in the dimensional reduction limit in which $R$ is taken to zero. Adding matter qualitatively changes this conclusion, as we will now illustrate.

Let us consider the case of a four-dimensional $U(1)$ gauge theory with a Dirac fermion of integer gauge charge $q$, with action
	\begin{align}
	\label{oneweyl}
	S_{4d}\, =  \, - \frac{1}{2 e_{4d}^2}  \int_{M_4} da_{4d} \wedge \star da_{4d} +\int_{M_4} d^4x\,  i \bar\psi \slashed{D} \psi \,.
	\end{align}
The covariant derivative is $D_\mu = \partial_\mu - i q a_\mu$ and $\slashed{D}= \gamma^\mu D_\mu$. The charged fermion breaks the electric 1-form symmetry to $\mathbb{Z}_q^{(1)}$, while the magnetic $U(1)_M^{(1)}$ 1-form symmetry remains unbroken. 
Taking spacetime to decompose as $M_4 = M_3\times S^1$ and integrating over the circle of radius $R$, we naively obtain, 
	\begin{align}
	\label{dimreds}
	\begin{split}
	S_{3d} \, =  \, & - \frac{1}{2e_{3d}^2} \int_{M_3} da_{3d} \wedge \star da_{3d}  - \frac{1}{2  e_{3d}^2(2\pi R)^2} \int_{M_3} d\sigma \wedge \star d \sigma      \\
	&+ (2\pi R)\sum_{j=1,2} \int_{M_3} d^3x\, i \bar\psi_{j,3d} \slashed{D}_{3d} \psi_{j,3d} + q\sum_{j=1,2} \int_{M_3} d^3x\, (-1)^j \sigma ( i \bar\psi_{j,3d}  \psi_{j,3d})\,,
	\end{split}
	\end{align}
where $\bar{\psi}_{j,3d}= \psi_{j,3d}^\dagger \sigma^3$ for $\sigma^3$  the third Pauli matrix, and we have again identified the gauge couplings as $e_{4d}^2 = (2\pi R)e_{3d}^2$, and used \eqref{sigma2} to identify $\sigma$ in terms of $(a_{4d})_3$. In the reduction, the 4d Dirac spinor (with four complex components) splits into two independent Dirac spinors with two complex components each.

We can now ask, what becomes of the 4d 1-form symmetries in the 3d model? Due to the last term in \eqref{dimreds}, the action is no longer invariant under shifts of $\sigma \to \sigma + 2\pi/q$. This would imply that $\sigma$ is no longer periodic, so that the shift symmetry (coming from the 4d electric 1-form symmetry) is completely broken, and the winding lines accounting for the magnetic 1-form symmetry are not present. However, we would have expected to retain at least some 0-form shift symmetry---and what happened to the winding symmetry?

This puzzle is resolved by recalling that in the compactification we expand the fields in KK modes, and that the action \eqref{dimreds} only keeps the tree-level terms of the zero modes in the limit $E \ll 1/R$. To keep track of the KK tower for the fermion (we shall not need to track the KK tower for the gauge fields), we expand
	\begin{align}
	\label{psikk}
	\psi(\vec{x},z) = \sum_{n\in \mathbb{Z}} \begin{pmatrix}
	\psi_{1,n}(\vec{x})\\
	\psi_{2,n}(\vec{x})
	\end{pmatrix} e^{\frac{i n z}{R}}\,,
	\end{align}
so that the terms on the bottom line of \eqref{dimreds} become,
	\begin{align}\label{skk}
	\int_{M_4}d^4x\,  i\bar\psi \slashed{D} \psi 
	\,  \to \,
	\sum_{n\in \bZ,\, j=1,2} \int_{M_3}d^3x\, i \bar\psi_{j,n} \left[ (2\pi R) \slashed{D}_{3d}    +   (-1)^j  \left(q\sigma - 2\pi n\right)  \right] \psi_{j,n}\,.
	\end{align}
Evidently this action is invariant under a shift of $\sigma \to \sigma+ 2\pi$, which mixes the KK modes as $n\to n-q$. We identify the shift symmetry $\sigma \to \sigma + 2\pi$ as a gauge symmetry, and the part $\sigma \to \sigma + 2\pi/q$ as the 0-form symmetry coming from the $\mathbb{Z}_q^{(1)}$ 1-form symmetry, which essentially shifts the entire KK tower by one.  We conclude that the compactness of the scalar is only retained in the effective theory with its full KK tower intact, and lost in the restriction to zero modes.

We can further ask what happens once we integrate out the KK tower---is the shift symmetry broken explicitly, spontaneously, or does it just act trivially? It appears to be spontaneously broken, for the following reason. When we compactify, we have the freedom of introducing holonomies on the circle for both background and dynamical gauge fields. Such a holonomy sets the background around which we expand in the IR, as $\int_{S^1} a = H_0 + \sigma$,  so that the low-energy theory depends on the choice of vacuum labeled by $H_0$. 
Since the shift symmetry maps one choice of vacuum to another, this symmetry is spontaneously broken in the IR. In accordance with this discussion, we propose that the $\sigma \to \sigma + 2\pi/q$ 0-form shift symmetry is spontaneously broken, and that in the limit $E\ll 1/R$ the dynamical scalar $\sigma$ can be regarded as non-compact.  

It remains to determine what happens to the winding symmetry in this limit. The natural proposal is that said symmetry just acts trivially in the IR. This is because the winding defects still exist, but now lead to a change in vacua as we go around them. This is similar to the vortex solutions in spontaneously broken $U(1)$ gauge theories, which are massive. As such we expect the winding defects to become massive, leading to them being integrated out and not present in the IR theory.

\subsection{\texorpdfstring{$\mathcal{N}=1$}{N=1}  \texorpdfstring{$U(N_c)$}{U(Nc)} SQCD with Determinant Matter}
\label{sec:3dsqcd}

As we emphasized in Section~\ref{sec:intro3d}, the survival of the 1-form winding symmetry in 3d depends crucially on the compactness of the scalar, which does not survive the IR limit in the theory with matter. We will next explore how this affects the generalized symmetries of $\CN=1$ supersymmetric QCD, when compactified on a circle to a three-dimensional $\CN=2$ theory.

We consider  SQCD with $U(N_c)$ gauge group on $\mathbb{R}^3\times S^1$, leading to an effective 3d $\CN=2$ theory with the same gauge group and $N_f$ chiral multiplets, and monopole superpotential (to be reviewed below). The charges of the fields present in the 3d theory under the classical 0-form global symmetries are given in Table~\ref{tab:UN3ddet}.
The features of this theory will depend on the energy scale at which we examine it. We will first consider the effective theory with the circle radius $R$ small but finite, at energies above the scale $1/R$ so that we have not yet integrated out the KK tower, and then note how the conclusions change as we lower the energy.  The 4d and 3d gauge couplings are related as $g_4^2 = 2\pi R g_3^2$, so that the 4d strong coupling scale $\Lambda$ satisfies,
	\begin{align}
	\eta = \Lambda^{b_1} = e^{-8\pi^2/g_4^2} = e^{-4\pi / (R g_3^2)}\,,
	\end{align}
where $b_1=3N_c-N_f$ is the one-loop $\beta$-function coefficient.
Note that compared with the same theory with $SU(N_c)$ gauge group, the $U(N_c)$ theory has a dynamical compact scalar $\sigma$ associated with the holonomy  of the $SU(N_c)$ vector field on the circle, as well as a compact scalar associated with the $U(1)_B$ gauge field which we will denote by $\sigma_B$.\footnote{~We expect that gauging $U(1)_B$ in the $SU(N_c)$ theory only commutes with the compactification/reduction limit in the deep IR limit $E\ll 1/R$, since reducing $U(N_c)$ leads to the additional compact scalar $\sigma_B$ which is not present when gauging baryon number directly in three dimensions.}

As reviewed in Section~\ref{sec:4dUN}, the $\tr \,U(1)_B^2 U(1)_R$ ABJ anomaly \eqref{I6UN} leads to a breaking of $U(1)_R\to \mathbb{Q}/\mathbb{Z}$ enacted by non-invertible defects. The fact that this theory does not preserve a continuous $U(1)_R$ leads to subtletites in the 3d reduction, and for this reason we will instead consider the theory with the additional chiral multiplets $\Omega^\pm$ in the determinant representation of $U(N_c)$, whose charges, listed in Table~\ref{tab:UNdetcharges}, are designed to cancel the offending ABJ anomaly and thus preserve a continuous $U(1)_R$. Recall from Section~\ref{sec:4dUNdet}  that this theory has a non-invertible symmetry coming from the classical $U(1)_I$ symmetry that acts on the determinant matter, whose discrete invertible part is $\mathbb{Z}_2$, and also exhibits a 2-group that extends the various 0-form symmetries by the $U(1)_M^{(1)}$ magnetic 1-form symmetry.

The theory on a circle  exhibits the following features. 

\subsubsection*{1-form symmetry} Firstly,  the magnetic 1-form symmetry $U(1)_{M}^{(1)}$ reduces to the topological 0-form symmetry of the 3d theory which we will denote by $U(1)_J$, and whose conserved current is related to the reduction of the $U(1)_B$ field strength to 3d, $j\sim \star (f_{B})_{3d}$ ;  and a magnetic 1-form winding symmetry, whose conserved current is $j_w\sim \star d\sigma_B$. Furthermore, the 2-group structure \eqref{eq:2group111}, reduces to a 2-group in the effective 3d theory involving this 1-form symmetry, and the various 0-form symmetries. As we have discussed, the 3d 1-form symmetry trivializes in the limit that the scalar decompactifies. 

\subsubsection*{Monopole superpotential}  
The  KK monopole along the circle direction plays an important role in the compactification, leading to a non-perturbative monopole superpotential $W_\eta$ in the effective 3d theory, which explicitly breaks the axial symmetry $U(1)_A$ to a discrete subgroup. Let us briefly review the origin of this superpotential.

At a generic point on the Coulomb branch of the 3d theory, the $U(N_c)$ gauge group is broken to $U(1)^{N_c}$. We denote the scalars dual to these abelian gauge fields as $\varphi_1,\dots,\varphi_{N_c}$, which are paired with $N_c$ scalars $\sigma_i$ coming from the eigenvalues of the compact scalars $\sigma$ and $\sigma_B$. Weyl transformations can be used to arrange these eigenvalues in descending order, as $\sigma_1\geq\dots\geq \sigma_{N_c}$. The scalars $\sigma_i$ and dual photons $\varphi_i$ can be organized into chiral multiplets $V_i$ that classically parametrize the Coulomb branch, 
	\begin{align}
	\label{vi}
	V_i \sim \exp \left[ 4 \pi / g_3^2\, \sigma_i + i \varphi_i\right] \,,\qquad i = 1,\dots,N_c\,.
	\end{align}
Each $V_i$ is associated to a 3d instanton that is obtained from compactifying the $N_c$ independent 4d 't Hooft Polyakov monopoles. Moreover, the reason for the $\sim$ symbol is that the coordinates $V_i$ are only a semi-classical description of the monopole operators that parametrize the quantum Coulomb branch (see {\it e.g.}~\cite{Intriligator:2013lca} for further discussion).  Quantum mechanically, they acquire charges due to zero modes in the monopole background, as we review in Appendix~\ref{sec:appzero}.  

Most of the Coulomb branch is lifted by instanton effects via Affleck--Harvey--Witten type superpotentials \cite{Aharony:1997bx,deBoer:1997kr}, except for two coordinates $V_+$ and $V_- $ which are oppositely charged under the topological $U(1)_J$ which shifts the dual photons $\varphi_i$ in the exponent. There is a KK monopole wrapping the circle direction, leading to an additional superpotential,
	\begin{align}
	\label{weta}
	W_\eta = \eta V\,.
	\end{align}
$V$ can be represented semi-classically as
\begin{align}
V\sim\exp \left[ 4 \pi / g_3^2\, (\sigma_1-\sigma_{N_c}) + i (\varphi_1-\varphi_{N_c})\right]\,,
\end{align}
however it should  really be understood as the quantum monopole operator. As reviewed in Appendix \ref{sec:appzero},  this is the only monopole operator with $U(1)_R$ charge 2 and $U(1)_A$ charge $0~\text{mod}~2N_f$ that preserves the topological $U(1)_J$ symmetry, and  it manifestly breaks $U(1)_A\to \mathbb{Z}_{2N_f}$.\footnote{~Much as is the case in 4d, for some values of $N_f$ non-perturbative effects generate a further contribution to the effective superpotential, which can depend on the monopole operators as well as the other gauge invariant moduli. For simplicity we focus on large enough values of $N_f$ where this is not the case.} Moreover, it is a monopole only for the $SU(N_c)$ part of the gauge group, and so the superpotential \eqref{weta} only lifts the corresponding direction of the Coulomb branch, while it does not lift the direction associated with the $U(1)_B$ part. This superpotential \eqref{weta} manifestly vanishes in the dimensional reduction limit of $R\to 0$, when $\eta=\Lambda^{b_1}$ goes to zero, but is present in the effective theory at small $R$.

 	\def\arraystretch{1.2} 
	\setlength\tabcolsep{4.5pt} 
	\begin{table}
	\centering
	\begin{tabular}{|c||c|c||c|c|c|c|c|c|c|}
	\hline
	 & $SU(N_c)$ & $U(1)_V$  & $SU(N_f)_L$ & $SU(N_f)_R$ & $U(1)_A$ & $U(1)_R$ & $U(1)_I$ & $U(1)_t$ & $U(1)_J$  \\ 
	 \hline \hline
 	$Q$ & $\Box$ & $1$ & $\Box$ & $1$ & $1$ & $\frac{N_f-N_c}{N_f}$ & $0$ & $0$ & $0$ \\ 
	\hline
	  $\widetilde{Q}$ & $\overline{\Box}$ & $-1$ & $1$ & $\overline{\Box}$ & $1$ & $\frac{N_f-N_c}{N_f}$ & $0$ & $0$ & $0$ \\ 
	  \hline
 	 $\lambda$ & adj & $0$ & $1$ & $1$  & $0$ & $1$ & $0$ & $0$ & $0$ \\ 
	 \hline
  	$\Omega^\pm$ & $1$ & $\pm N_c$ & $1$ & $1$ & $0$ & $2$ & $1$ & $\pm 1$ & $0$ \\ \hline
	$V$ & $1$ & $0$ & $1$ & $1$ & $-2N_f$ & $2$ & $0$ & $0$ & $0$ \\ \hline
	\end{tabular}
	\caption{ The charges of the fields under the classical 0-form global symmetries of 3d $U(N_c)$ SQCD, including $\Omega^\pm$ in the determinant representation. We have also listed the charges of the monopole operator $V$.  \label{tab:UN3ddet}}
	\end{table}
	\setlength\tabcolsep{6pt} 
	\def\arraystretch{1}

\subsubsection*{Reduction of the 4d anomaly action}

The axial symmetry is not preserved in the effective 3d theory on $S^1$, being explicitly broken by the monopole superpotential \eqref{weta}. Another perspective on this symmetry breaking arises from the reduction of the anomaly polynomial $\CI_6$ of the 4d theory on the circle, given in \eqref{I6UNdet}.
Taking the anomaly inflow action $\CA_{5}^{\text{inf}}$ satisfying $d\CA_{5}^{\text{inf}}=2\pi \int_{M_6} \CI_6$, decomposing the gauge fields into their three-dimensional components which are taken as usual to be independent of the circle direction, and then reducing on $M_5=M_4\times S^1$, leads to the following inflow action for the 3d theory involving the compact scalars,
		\begin{align}
		\label{anomtheory}
		\begin{split}
		\CA_{4}^{\text{inf}}\,  \supset \,  &\frac{1}{4\pi^2}  \int_{M_4} \bigg[  2 N_f A_A \tr \,f_{3d}\wedge D\sigma +2 A_I (f_B)_{3d} \wedge d\sigma_B   \\
		&\quad + \left(2 (A_I + A_R) d A_t + \mathrm{CS}(A_{SU(N_f)_L}) - \mathrm{CS}(A_{SU(N_f)_R})  \right)\wedge d \sigma_B\bigg]\,.
		\end{split}
	\end{align}
	Here $f_{3d}$ is the $U(N_c)$ field strength of the three-dimensional theory, and $(f_B)_{3d}$ is the 3d $U(1)_B$ field strength, namely $(f_B)_{\mu\nu}\to \{  (f_B)_{ij},(f_B)_{i3}=\frac{1}{2\pi R} \partial_i\sigma_B \}$. 
	To derive this expression we used in particular the following reductions of 4d characteristic classes,
	\begin{align}
		\int_{S^1_R} c_2(U(N_c))^{4d} \, &= \, \frac{1}{4\pi^2} \left[ \tr f_{3d}\wedge D\sigma - (f_B)_{3d} \wedge d\sigma_B \right] \,,\\
		\int_{S^1_R} (n_B)^{4d} \, &=  \, \frac{1}{4\pi^2} (f_B)_{{3d}} \wedge d\sigma_B\,,
	\end{align}
where recall that the integer-valued  instanton number for the $U(1)_B$ part of the gauge group is $n_B = \frac{1}{2} c_1(B)^2$.  For simplicity, we also restricted ourselves to ordinary compactification (as opposed to twisted compactification) for which the holonomy of the background fields are assumed to be trivial. 
	
The terms in \eqref{anomtheory} can be interpreted as anomalies involving the dynamical gauge symmetry, the $(-1)$-form $U(1)$ gauge symmetries associated to the dynamical compact scalars, and the global 0-form $U(1)_A$ and $U(1)_I$ symmetries. These anomalies are of course trivialized if we demand that $\sigma$ and $\sigma_B$ are constants. However, allowing for configurations in which the scalars have non-trivial winding on 3-cycles  in spacetime, the inflow action \eqref{anomtheory} can have a fractional part that is not cancellable by counterterms,  implying that the partition function shifts by a non-trivial phase. Then, the first term in \eqref{anomtheory} reproduces the fact that $U(1)_A$ is broken to $\mathbb{Z}_{2N_f}$ in the effective theory on the circle, consistent with the subgroup of $U(1)_A$ that is unbroken by the monopole superpotential \eqref{weta}. 
We emphasize that this effect explicitly trivializes in {either the $R\to 0$ limit or the decompactification limit}, and so should be viewed exclusively as a feature of the effective 4d theory on a circle---of course, there are no ABJ anomalies in the usual sense in 3d QFTs.

\subsubsection*{Non-invertible symmetry}

As discussed in Section~\ref{sec:4d}, the mixed gauge-global anomaly \eqref{I6UNdet} implies that the $U(1)_A$ axial symmetry is explicitly broken to $\bZ_{2N_f}$, and the $U(1)_I$ symmetry becomes non-invertible. Here, we discuss the fate of the non-invertible defect in 4d under compactification on $S^1$ in several approaches. For the discussion to be interesting, we keep the radius $R$ of $S^1$ finite and we do not integrate out the KK modes.

In the first approach, we start with the ``ABJ" anomaly \eqref{anomtheory} after the compactification. The first term in \eqref{anomtheory} means that $U(1)_A$ is explicitly broken to $\bZ_{2N_f}$ because it involves the non-abelian gauge symmetry. The second term in \eqref{anomtheory} means that $U(1)_I$ is explicitly broken to the $\mathbb{Q}/\bZ$ non-invertible symmetry, with $\bZ_2$ being its invertible part. The reasoning for the non-invertible symmetry is precisely in parallel with that in 4d: the anomaly means that the worldvolume of the $U(1)_I$ defect supports degrees of freedom that have gauge anomaly $2\alpha (f_B)_{3d} \wedge d\sigma_B$, and to cancel this gauge anomaly we introduce a TQFT localized on the defect worldvolume. Such a TQFT exists only when  $2\alpha$ is a rational number multiple of $2\pi$, \emph{i.e.}~$\alpha = \frac{\pi p}{q}$. The last term in \eqref{anomtheory} implies the 2-group involving $U(1)_R, U(1)_t, SU(N_f)_{L,R}$ and $U(1)^{(1)}_M$. The two-group structure is captured precisely by  \eqref{eq:2group111}. Finally, the anomaly $\frac{1}{4\pi^2} 2 A_I d A_t d\sigma_B$ implies the higher structure of the non-invertible symmetry. This approach has a conceptual issue that gauge symmetry is not a symmetry, and as such it would be nice to have an alternative derivation of the non-invertible symmetry that keeps track of the global symmetries only.  

In the second approach, we directly reduce the non-invertible defect \eqref{eq:Idefect} (similarly to the discussion around \eqref{Axial_defect_Lag} reducing the defect to 2d). For example, consider $(p,q)=(1,N)$, and taking the TQFT $\CA^{1, N}$ to be the minimal TQFT $U(1)_N$, such that the defect is,
\begin{eqnarray}
	\int D \ra \, \exp\left[i \int_{M_3} \left(\frac{\pi}{N} \star j_I + \frac{N}{4\pi} \ra \wedge d \ra + \frac{1}{2\pi} \ra \wedge d a_B\right)\right].
\end{eqnarray}
On $S^1$, we have $\sigma_B = \int_{S^1} a_B$ and $\phi = \int_{S^1} \ra$, and the defect reduces to, 
\begin{eqnarray}
\int D\ra\,  D\phi\, \exp\left[i\int_{M_{3}}\left(\frac{\pi}{N}\star(j_{I})_{2d}+\frac{N}{2\pi}\phi\,d\ra+\frac{1}{2\pi}\phi\,da_{B}+\frac{1}{2\pi}\ra\wedge d\sigma_{B}\right)\right].
\end{eqnarray}
The last three terms are precisely the TQFT mentioned in the previous approach.


	\newcommand{\specialitem}[3][black]{%
	  \item[%
	    \colorbox{#2}{\textcolor{#1}{\makebox[1em]{#3}}}%
	  ]
	}
	\def\xA{3.5}
	\def\yA{2}

	\def\yB{-1.3}

	\def\xCA{7}
	\def\xCB{4.8}
	\def\xCC{2.2}
	\def\yC{-4.2}

	\def\xD{2.5}
	\def\yD{-5.75}

	\def\yE{-8.2}

	\def\arrowscale{0.85}

	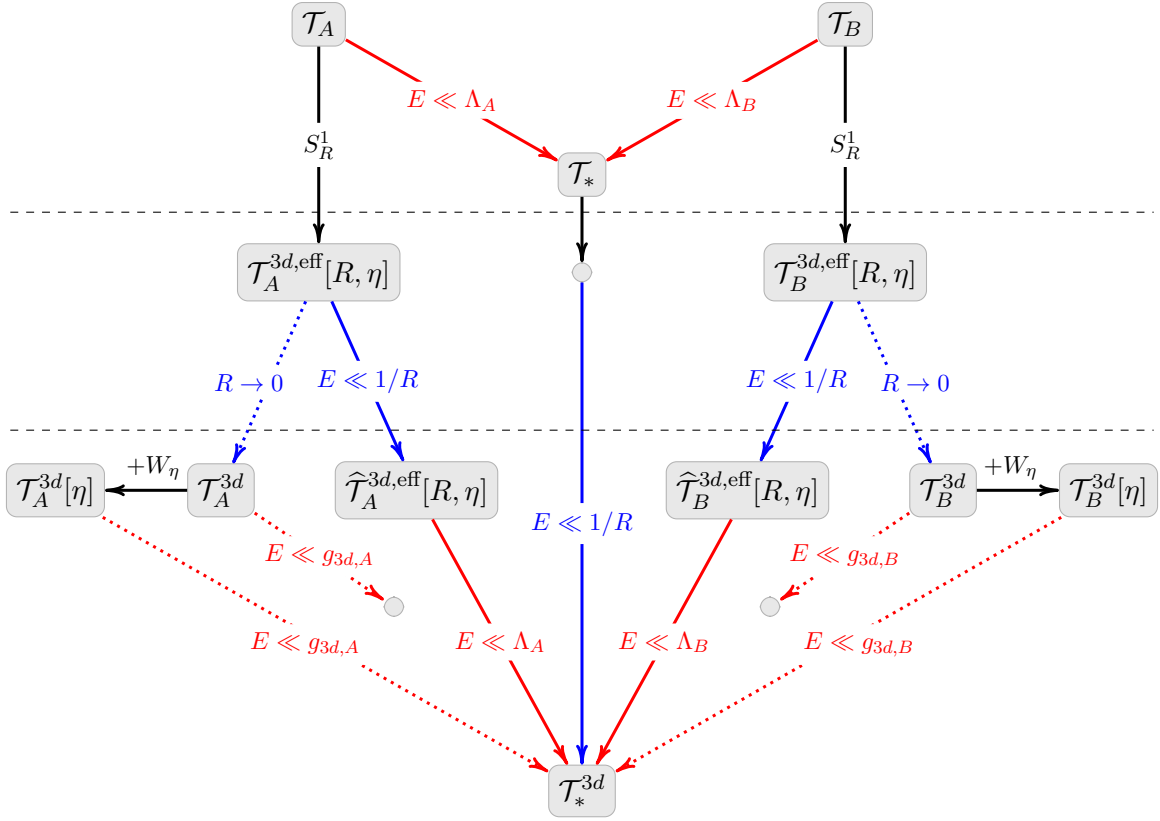
\begin{figure}[h!]
	\begin{center}
	\begin{tikzpicture}[boxstyle/.style={rectangle,draw=gray!60,fill=gray!18,rounded corners}]

	\node at (-\xA,\yA) (A4d) [boxstyle] {$\CT_{A}$};
	\node at (\xA,\yA) (B4d) [boxstyle] {$\CT_{B}$};
	\node at (0,0) (CFT4d) [boxstyle]  {$\CT_{*}$};

	\node at (-\xA,\yB) (A3deff) [boxstyle]  {$\CT_A^{3d,\text{eff}}[R,\eta]$};
	\node at (0,\yB) (dotUV) [boxstyle]  {};
	\node at (\xA,\yB) (B3deff) [boxstyle]  {$\CT_B^{3d,\text{eff}}[R,{\eta}]$};

	\node at (-\xCA,\yC) (AdW) [boxstyle]  {${\CT}_A^{3d}[\eta]$};
	\node at (-\xCB,\yC) (A3d) [boxstyle]  {${\CT}_A^{3d}$};
	\node at (-\xCC,\yC) [boxstyle]  (Ahat3deff) {$\widehat{\CT}_A^{3d,\text{eff}}[R,\eta]$};
	\node at (\xCC,\yC) (Bhat3deff) [boxstyle]  {$\widehat{\CT}_B^{3d,\text{eff}}[R,{\eta}]$};
	\node at (\xCB,\yC) (B3d)  [boxstyle]  {$\CT_B^{3d}$};
	\node at (\xCA,\yC) (BdW) [boxstyle]  {${\CT}_B^{3d}[{\eta}]$};

	\node at (-\xD,\yD) (AdotIR) [boxstyle]  {};
	\node at (\xD,\yD) (BdotIR) [boxstyle]  {};

	\node at (0,\yE) (C) [boxstyle]  {$\CT_*^{3d}$};

	\draw [-{>[scale=1.5, length=5, width=3,flex]},line width=0.4mm,black] (A4d) to  (A3deff);
	\path (A4d) -- node[rectangle,fill=white,rounded corners,scale=\arrowscale]  {$S^1_R$} (A3deff);
	\draw [-{>[scale=1.5, length=5, width=3,flex]},line width=0.4mm,black] (CFT4d) to  (dotUV);
	\draw [-{>[scale=1.5, length=5, width=3,flex]},line width=0.4mm,black] (B4d) to  (B3deff);
	\path (B4d) -- node[rectangle,fill=white,rounded corners,scale=\arrowscale]  {$S^1_R$} (B3deff);

	\draw [-{>[scale=1.5, length=5, width=3,flex]},line width=0.4mm,red] (A4d) to  (CFT4d);
	\path (A4d) -- node[rectangle,fill=white,rounded corners,scale=\arrowscale]  {$\textcolor{red}{E\ll \Lambda_A}$} (CFT4d);
	\draw [-{>[scale=1.5, length=5, width=3,flex]},line width=0.4mm,red] (B4d) to  (CFT4d);
	\path (B4d) -- node[rectangle,fill=white,rounded corners,scale=\arrowscale] {$\textcolor{red}{E\ll {\Lambda_B}}$} (CFT4d);

	\draw [-{>[scale=1.5, length=5, width=3,flex]},line width=0.4mm,red] (Ahat3deff) to  (C);
	\path (Ahat3deff) -- node[rectangle,fill=white,rounded corners,scale=\arrowscale]  {$\textcolor{red}{E\ll \Lambda_A}$} (C);
	\draw [-{>[scale=1.5, length=5, width=3,flex]},line width=0.4mm,red] (Bhat3deff) to  (C);
	\path (Bhat3deff) -- node[rectangle,fill=white,rounded corners,scale=\arrowscale] {$\textcolor{red}{E\ll {\Lambda}_B}$} (C);

	\draw [-{>[scale=1.5, length=5, width=3,flex]},line width=0.4mm,red,dotted] (A3d) to  (AdotIR);
	\path (A3d) -- node[rectangle,fill=white,rounded corners,scale=\arrowscale]  {$\textcolor{red}{E\ll g_{3d,A}}$} (AdotIR);
	\draw [-{>[scale=1.5, length=5, width=3,flex]},line width=0.4mm,red, dotted] (B3d) to  (BdotIR);
	\path (B3d) -- node[rectangle,fill=white,rounded corners,scale=\arrowscale]  {$\textcolor{red}{E\ll {g}_{3d,B}}$} (BdotIR);

	\draw [-{>[scale=1.5, length=5, width=3,flex]},line width=0.4mm,red, dotted] (AdW) to  (C);
	\path (AdW) -- node[rectangle,fill=white,rounded corners,scale=\arrowscale]  {$\textcolor{red}{E\ll {g}_{3d,A}}\ \ \ \ \ $} (C);
	\draw [-{>[scale=1.5, length=5, width=3,flex]},line width=0.4mm,red, dotted] (BdW) to  (C);
	\path (BdW) -- node[rectangle,fill=white,rounded corners,scale=\arrowscale]  {$\ \ \ \ \ \textcolor{red}{E\ll {g}_{3d,B}}$} (C);

	\draw [-{>[scale=1.5, length=5, width=3,flex]},line width=0.4mm,blue, dotted] (A3deff) to  (A3d);
	\path (A3deff) -- node[rectangle,fill=white,rounded corners,scale=\arrowscale]  {$\textcolor{blue}{R\to 0}\ \ \ \ \ $} (A3d);
	\draw [-{>[scale=1.5, length=5, width=3,flex]},line width=0.4mm,blue, dotted] (B3deff) to  (B3d);
	\path (B3deff) -- node[rectangle,fill=white,rounded corners,scale=\arrowscale]  {$\textcolor{blue}{\ \ \ \ \ R\to 0}$} (B3d);

	\draw [-{>[scale=1.5, length=5, width=3,flex]},line width=0.4mm,blue] (A3deff) to  (Ahat3deff);
	\path (A3deff) -- node[rectangle,fill=white,rounded corners,scale=\arrowscale]  {$\textcolor{blue}{E\ll 1/R}$} (Ahat3deff);
	\draw [-{>[scale=1.5, length=5, width=3,flex]},line width=0.4mm,blue] (B3deff) to  (Bhat3deff);
	\path (B3deff) -- node[rectangle,fill=white,rounded corners,scale=\arrowscale]  {$\textcolor{blue}{E\ll 1/R}$} (Bhat3deff);
	\draw [-{>[scale=1.5, length=5, width=3,flex]},line width=0.4mm,blue] (dotUV) to  (C);
	\path (dotUV) -- node[rectangle,fill=white,rounded corners,scale=\arrowscale]  {$\textcolor{blue}{E\ll 1/R}$} (C);

	\draw [-{>[scale=1.5, length=5, width=3,flex]},line width=0.4mm,black] (A3d) to  (AdW);
	\draw [-{>[scale=1.5, length=5, width=3,flex]},line width=0.4mm,black] (B3d) to  (BdW);

	\node at (-\xCA+1.3,\yC+0.3) (Weta1) [scale=\arrowscale] {$+ W_\eta$};
	\node at (\xCA-1.3,\yC+0.3) (Weta2) [scale=\arrowscale] {$+ W_{{\eta}}$};
	
	\draw [dashed]  (-\xCA-0.6,-0.5) -- (\xCA+0.6,-0.5);
	\draw [dashed]  (-\xCA-0.6,-3.4) -- (\xCA+0.6,-3.4);

	\end{tikzpicture}
	\end{center}
	\caption{ Compactifying 4d SQCD with $SU(N_c)$ or $U(N_c)$ gauge group ($\CT_{A}$) and its magnetic dual $(\CT_{B})$ on a circle, in the conformal window. Red arrows indicate strong coupling limits, blue arrows indicate small radius limits, and dotted arrows are used for limits that specifically involve dimensional reduction $R\to 0$.  
	 The theories above the top dashed horizontal line are defined in 4d; the intermediate level consists of the effective 3d theories above the KK scale; and the bottom level theories are defined after integrating out the KK towers. This figure may be viewed as a refinement of Figure 1 in \cite{Aharony:2013dha}.
	\label{fig:3dpic}}
	\end{figure}

\subsection{Comments on Infrared Dualities}
\label{sec:3ddualities}

Let us now comment on how the generalized symmetry structure interplays with the  reduction of 4d dualities to 3d dualities. For concreteness we will focus on 4d $U(N_c)$ SQCD as our starting point, although the structure of the diagram in Figure~\ref{fig:3dpic} is the same for the $SU(N_c)$ case. 
In the UV we begin with  4d $\CN=1$ $U(N_c)$ gauge theory with $N_f$ fundamentals and anti-fundamental chirals, denoted $\CT_{A}$ in Figure~\ref{fig:3dpic}. The Seiberg dual with $U(N_f-N_c)$ gauge group, $N_f$ fundamental and anti-fundamental chirals $q,\tilde{q}$ under the new gauge group, and singlets $M$ that couple via a superpotential $W\sim Mq\tilde{q}$, is denoted  $\CT_{B}$.  The strong coupling scales $\eta = \Lambda^{b_1}$ of the dual theories are related as, 
	\begin{align}
	\eta_A {\eta}_B = (-1)^{N_f-N_c} \,.
	\end{align}
In the conformal window, theories $\CT_{A}$ and $\CT_{B}$ flow to the same interacting SCFT  at energies satisfying $E \ll \Lambda_A,{\Lambda}_B$.

Compactifying each of these 4d theories on a circle with small but finite radius $R$, and including the full KK tower, results in theories we denote by $\CT_A^{3d,\text{eff}}[R,\eta]$ and $\CT_B^{3d,\text{eff}}[R,{\eta}]$, where the arguments are meant to indicate the explicit dependence of these theories on the radius $R$ and on the strong coupling scale $\eta$ through their superpotential. The 4d and 3d gauge couplings are related as $g_{4d}^2 = 2\pi R g_{3d}^2$. As per our discussion in Section~\ref{sec:3dsqcd}, the theories $\CT_A^{3d,\text{eff}}[R,\eta]$ and $\CT_B^{3d,\text{eff}}[R,{\eta}]$ include the following features. Firstly, they possess compact scalars from the holonomies of the gauge fields on the circle, leading to a compact moduli space of the effective 3d $\CN=2$ theory, and preserving a 0-form shift symmetry. The $U(1)^{(1)}$ 1-form winding symmetry is preserved, leading to the non-invertible symmetry and 2-group structure. They generate a superpotential $W_\eta$ coming from the KK monopole on the circle, which also explicitly only preserves a $ \mathbb{Z}_{2N_f}$ subgroup of the classical $U(1)_A$ axial symmetry. Finally, the anomaly \eqref{anomtheory} implies that the $U(1)_A$ symmetry is non-perturbatively broken to the $\mathbb{Z}_{2N_f}$ subgroup.

Integrating out the KK tower and flowing to energies $E \ll 1/R$, the shift 0-form and winding 1-form symmetries are lost. We denote the resulting low-energy theories by   $\widehat{\CT}_A^{3d,\text{eff}}[R,\eta]$, and $\widehat{\CT}_B^{3d,\text{eff}}[R,\eta]$. For these theories, the scalar decompactifies, so that the 0-form shift symmetry that acts on it is spontaneously broken and the moduli space is no longer compact.  The 1-form winding symmetry now acts trivially, so that both the non-invertible symmetry and 2-group trivialize. Furthermore the $U(1)_A$ axial anomaly trivializes, although the $W_\eta$ superpotential still explicitly breaks $U(1)_A\to \mathbb{Z}_{2N_f}$.
It is in this limit that the  effective 3d $\CN=2$ theories  are dual to each other, so that at energies $E \ll \Lambda_A,\Lambda_B$  they are expected to flow to the same 3d SCFT, $\CT_*^{3d}$.  As emphasized in \cite{Aharony:2013dha}, this duality---which involves the additional monopole superpotential---is  not the same as 3d $\CN=2$ Aharony duality \cite{Aharony:1997gp}, although the latter can be recovered from it by considering real mass deformations. 

The 3d SCFT $\CT_*^{3d}$ is expected to match onto the theory obtained by first putting the 4d SCFT $\CT_*$ on the circle, and then taking the $E\ll 1/R$ limit; {\it i.e.}~the $E\ll \Lambda_A,{\Lambda}_B$ and $E \ll 1/R$ limits commute. However, we emphasize that this might not hold for other theories or choices of gauge groups, for the following reason. 
Our conclusion that the shift 0-form and winding 1-form symmetries are lost in the IR is based on our use of the Lagrangian gauge-theory descriptions of the theories. We have been considering a scenario in which a 4d UV gauge theory with a 1-form symmetry flows to an interacting SCFT in the IR. Compactifying the UV gauge theory to 3d, we have argued that we do not expect to get both the 0- and 1-form symmetries. However, compactifying the IR SCFT may lead to a different 3d theory, and it is in principle possible for the resulting 3d theory to retain both the 0- and 1-form symmetries. 

For the theories at hand that are based on unitary gauge groups, this possibility does not seem to be realized, due to the following. Consider first  a $U(1)$ gauge theory. Then, the IR theory is the same as the UV gauge theory, so we expect the result to be the same for both. The non-abelian $U(N_c)$ case is more subtle due to the $SU(N_c)$ part, which can flow to strong coupling. However, the $SU(N_c)$ sector by itself has no 1-form symmetry---all the generalized symmetry structure comes from the $U(1)$ part (as we also emphasized at the end of Section~\ref{sec:4dUN}). Now, say we gauge the $U(1)$ baryon symmetry of  $SU(N_c)$ SQCD. Since the 2-group and non-invertible structure stem from mixed anomalies involving only the $U(1)$, we expect them to be the same for both the UV and IR theories. 
Compactifying these theories to 3d, the UV $U(N_c)$ gauge theory just turns to a 3d UV gauge theory, where we expect the previous results regarding the fate of the 4d 1-form symmetry to apply. In the 4d IR, however, we expect to get a $U(1)$ gauge theory that weakly gauges whatever 4d SCFT the $SU(N_c)$ gauge theory flows to. Upon compactification to 3d, this should then give a 3d $U(1)$ gauge theory that weakly gauges whatever 3d SCFT is obtained from compactifying the 4d SCFT.  While we may not know the precise details of the 3d SCFT resulting from the compactification of the 4d $SU(N_c)$ part, we do know what happens to the $U(1)$. And since the 1-form symmetry only involves the $U(1)$ part, our previous arguments should still hold, and we expect the generalized symmetry structure mentioned so far in the $U(N_c)$ theories to trivialize upon compactification to 3d. 
By contrast, if the part of the gauge group that is engaged in the 1-form symmetry {\it also} has non-trivial IR dynamics---as for instance is the case for $SO(N_c)$ gauge group---we might be led to a different conclusion. For this reason it would be especially interesting to examine $SO(N_c)$ gauge theories, although we leave this to future work.  

Returning to Figure~\ref{fig:3dpic}, it remains to consider the dimensional reduction limit $R\to 0$. As was emphasized in \cite{Aharony:2013dha}, the $R\to 0$ limit does {\it not} commute with the $E\ll \Lambda_A,{\Lambda}_B$ limit, and in general the theories obtained from dimensional reduction are not themselves dual in 3d. In the diagram, these theories obtained by dimensionally reducing the two 4d Seiberg-dual theories are denoted by $\CT_A^{3d}$ and $\CT_B^{3d}$. In the strong-coupling limits where $E\ll g_{3d}^A,g_{3d}^B $, these theories do {not} flow to the same fixed point. 
However, upon deforming the dimensionally-reduced theories $\CT_A^{3d}$ and $\CT_B^{3d}$ by the appropriate relevant monopole superpotential, we can indeed obtain 3d theories which {\it are} dual, and are expected to flow to the same 3d SCFT $\CT_*^{3d}$.

\section{From Four to Two Dimensions}
\label{sec:2d}

We shall next consider the compactification of 4d field theories on a sphere, focusing on the cases of four-dimensional $\CN=1$ SQCD with $SU(N_c)$ or $U(N_c)$ gauge group compactified to $\CN=(0,2)$ models in two dimensions.

\subsection{Generalities of the \texorpdfstring{$S^2$}{S2} Compactification}
\label{sec:gen2d}

We begin with some general considerations for compactifying four-dimensional $SU(N_c)$ or $U(N_c)$ SQCD on $S^2$. This type of compactification was studied in \cite{Honda:2015yha,Gadde:2015wta} (see also \cite{Dedushenko:2017osi,Tachikawa:2018sae,Sacchi:2020pet}), and we shall review some of their results here. The curvature of the $S^2$ breaks all supersymmetry, but it is possible to preserve half of the supercharges via an R-symmetry twist, by turning on a non-trivial R-symmetry background that cancels against the spin connection. This amounts to introducing an R-symmetry magnetic flux $m_R=-1$, such that 
\begin{align}\label{eq:Rsymflux}
\int_{S^2}c_1(R) = -1\,.
\end{align}
One complication is that as we turn on a magnetic flux, we should ensure that the Dirac quantization condition is obeyed. Since the R-symmetry flux on $S^2$ should be $-1$, all R-charges must therefore be integers.  While this is obviously not obeyed by the superconformal R-symmetry of SQCD (see Table~\ref{tab:UNdetcharges}), it is sufficient to twist using any R-symmetry and not necessarily the one at the superconformal point. The solution adopted in \cite{Gadde:2015wta} is to mix the R-symmetry with a $U(1)$ subgroup of the $SU(N_f)_L$ non-abelian flavor symmetry, in such a way that all R-charges become integer. We shall implement such a shift in the examples we study below. 

Then, the topological twist that preserves $\CN=(0,2)$ supersymmetry in two dimensions amounts to shifting the first Chern class of the R-symmetry bundle  that is associated to a $U(1)_R$ symmetry under which all fields have integer charges, as
	\begin{align}
	\label{twist}
	c_1({R}) = - \frac{1}{2} t + \hat{c}_1({R})\,.
	\end{align}
Here we have denoted by $t$ the Chern root of the tangent bundle to the sphere, related to the global angular form of the $SU(2)_{\text{ISO}}$ isometry of the sphere by $t=2e_2(S^2)$,  and which integrates to the Euler characteristic as,\footnote{~Equation \eqref{twist} holds more generally for an $n$-punctured Riemann surface of genus $g$ upon substituting $\chi(\Sigma_{g,n})=2-2g-n$, although we focus on $g=n=0$.} 
	\begin{align}\label{eq:euler}
	\int_{S^2} t = \chi(S^2) = 2\,.
	\end{align}
We denote by $\hat{c}_1(R)$ the first Chern class of the R-symmetry bundle of the 2d theory, so as to not burden the characteristic classes with (4d) versus (2d) subscripts. Our conventions are spelled out in Appendix \ref{sec:bott}. In this way, \eqref{twist} ensures that \eqref{eq:Rsymflux} is automatically satisfied.

The compactification on $S^2$ was thereby carried out in \cite{Gadde:2015wta} based on the analysis of the $T^2\times S^2$ partition function \cite{Closset:2013sxa,Benini:2015noa,Honda:2015yha}. By supersymmetric localization, this partition function takes the schematic form,
	\begin{align}
	\mathcal{I}=\sum_{\vec{m}_G\in\Lambda^\vee_w(G)}\oint_{\text{JK}}\prod_{a=1}^{\mathrm{rk}\,G}\frac{\udl{z_a}}{2\pi iz_a}\mathcal{Z}(\vec{z})\,,
	\end{align}
where $\vec{m}_G$ is the gauge flux which takes values in the co-weight lattice of the gauge group $G$, and $\vec{z}$ are gauge fugacities over which we have to integrate with a specific choice of contour known as the Jeffrey--Kirwan (JK) countour \cite{1993alg.geom..7001J} (see also \cite{Benini:2013xpa} for a pedagogical explanation). The key observation of \cite{Gadde:2015wta} is that for each value of $\vec{m}_G$, the corresponding integral over $\vec{z}$ takes the form of the elliptic genus, or $T^2$ partition function, of a 2d $\CN=(0,2)$ theory \cite{Gadde:2013wq,Gadde:2013dda,Benini:2013nda,Benini:2013xpa}. The expression for the $T^2\times S^2$ partition function of the original 4d theory can indeed be re-interpreted as the sum of the partition functions of the 2d theories obtained by compactification on the $S^2$ (see Section~\ref{sec:undetsneg} for more details), since there is no dependence on the radius of the sphere thanks to the topological twist. This observation thus tells us that from the topologically-twisted compactification of a 4d theory on $S^2$ we can get a direct sum of different 2d theories, in agreement with our discussion in Section~\ref{sec:holonomy}. 

In \cite{Gadde:2015wta} it was also pointed out that if we choose an R-symmetry for the topological twist such that all chiral fields have a non-negative integer R-charge, then the summation over $\vec{m}_G$ actually truncates to the single $\vec{m}_G=\vec{0}$ sector. In this case we obtain a single 2d theory from the compactification, whose field content can be deduced from that of the original 4d theory by re-interpreting the $T^2\times S^2$ partition function as a 2d elliptic genus. The resulting rule is as follows:
\begin{itemize}
\item a 4d $\mathcal{N}=1$ chiral multiplet of R-charge $r$ gives rise in 2d to $r-1$ Fermi multiplets if $r>1$, $1-r$ chiral multiplets if $r<1$, and no field if $r=1$;
\item a 4d $\mathcal{N}=1$ vector multiplet gives rise to a 2d $\mathcal{N}=(0,2)$ vector multiplet.
\end{itemize}

In the following we will both consider examples in which the sum over magnetic fluxes truncates to the zero sector, and cases in which it does not. When there is no truncation, then the above rule gets modified in a way that depends on the specific value of $\vec{m}_G$ and on the representation of the fields under the gauge group. In particular, since the gauge flux $\vec{m}_G$ breaks the gauge group $G$ to some residual subgroup $H$, then the 4d $\mathcal{N}=1$ vector multiplet in the adjoint representation of $G$ will decompose into a 2d $\mathcal{N}=(0,2)$ vector multiplet in the adjoint of $H$ and Fermi multiplets, according to the branching rule for the adjoint representation of $G$ with respect to the subgroup $H$. We will see this in more detail later in examples. In particular, we will see that each term in the sum of the $T^2\times S^2$ partition function can be understood as associated to a 2d theory obtained by $S^2$ compactification with a topological twist and a flux for the gauge symmetry.

\subsection{\texorpdfstring{$\mathcal{N}=1$}{N=1} \texorpdfstring{$U(N_c)$}{U(Nc)} SQCD with Determinant Matter} 
\label{sec:undets2}

 	\def\arraystretch{1.1} 
	\setlength\tabcolsep{2pt} 
	\begin{table}[h!]
	\centering
	\resizebox{\columnwidth}{!}{%
	\begin{tabular}{|c||c|c||c|c|c|c|c|c|c|c|c|c|}
	\hline
	 & $SU({N}_c)$ & $U(1)_V$ & $SU(N_1)$ & $SU(2n)$  & $SU(N_3)$  & $SU(N_f)_R$ & $U(1)_x$  & $U(1)_y$ & $U(1)_{{R_1}}$  & $U(1)_A$ & $U(1)_{I}$  & $U(1)_{t}$ \\ \hline \hline
	$Q_1$ & $\Box$ & 1 & $\Box$ &  $\bf 1$ & $\bf 1$ & $\bf 1$ & $0$ & $-2n$ & 0 & 1 & 0 & 0 \\ \hline
	$Q_2$ & $\Box$ & 1 & $\bf 1$ &  $\Box$  & $\bf 1$ & $\bf 1$ & $-N_3$ & $N_1$ & 1 & 1 & 0 & 0 \\ \hline
	$Q_3$ & $\Box$ & 1 & $\bf 1$ &  $\bf 1$ & $\Box$ & $\bf 1$ & $2n$ & $0$ & 2 & 1 & 0 & 0 \\ \hline
	$\widetilde{Q}$ & $\overline{\Box}$ & $-1$ & $\bf 1$ &  $\bf 1$ & $\bf 1$ & $\overline{\Box}$ & $0$ & $0$ & 0 & 1 & 0 & 0 \\ \hline
	$\Omega^\pm$ & $\bf 1$ & $\pm N_c$ & $\bf 1$ &  $\bf 1$ & $\bf 1$ & $\bf 1$ & $0$ & $0$ & $2$  & 0 & $1$ & $\pm 1$ \\ \hline 
	\end{tabular} 
	}%
	\caption{The charges of the fields for 4d $U(N_c)$ SQCD, including the determinant matter $\Omega^\pm$, upon twisting $U(1)_{x,y}\subset SU(N_f)_L$ with the original R-symmetry to obtain the charges listed in the table ({\it twist 1}). Here $N_1+2n+N_3=N_f$, and $n+N_3 = N_f-N_c$. 
\label{tab:s2red}
	}
	\vspace{0.21cm}
	\resizebox{\columnwidth}{!}{%
	\begin{tabular}{|c||c|c||c|c|c|c|c|c|c|c|c|c|}
	\hline
	 & $SU(\widetilde{N}_c)$ & $U(1)_{\widetilde{V}}$ & $SU(N_1)$ & $SU(2n)$  & $SU(N_3)$  & $SU(N_f)_R$ & $U(1)_{\widetilde{x}}$  & $U(1)_{\widetilde{y}}$ & $U(1)_{{R_1}}$  & $U(1)_{\widetilde{A}}$ & $U(1)_{I}$  & $U(1)_{t}$ \\ \hline \hline
	$q_1$ & $\Box$ & 1 & $\overline{\Box}$ &  $\bf 1$ & $\bf 1$ & $\bf 1$ & $2n$  & $0$ & 2 & 1 & 0 & 0 \\ \hline
	$q_2$ & $\Box$ & 1 & $\bf 1$ &  $\overline{\Box}$  & $\bf 1$ & $\bf 1$  & $- N_1$ & $ N_3$ & 1 & 1 & 0 & 0 \\ \hline
	$q_3$ & $\Box$ & 1 & $\bf 1$ &  $\bf 1$ & $\overline{\Box}$ & $\bf 1$ &  $0$ & $-2n$  & 0 & 1 & 0 & 0 \\ \hline
	$\widetilde{q}$ & $\overline{\Box}$ & $-1$ & $\bf 1$ &  $\bf 1$ & $\bf 1$ & ${\Box}$ & $0$ & $0$ & 0 & 1 & 0 & 0 \\ \hline
	$\widetilde{M}_1$ & $\bf 1$ & 0 & $\Box$  & $\bf 1$ & $\bf 1$  & $\overline{\Box}$ & $-2n$ & 0  & 0 & $-2$ & 0 & 0 \\ \hline
	$\widetilde{M}_2$ & $\bf 1$ & 0 & $\bf 1$  & $\Box$ & $\bf 1$  & $\overline{\Box}$ & $N_1$ & $-N_3$ & 1 & $-2$ & 0 & 0 \\ \hline
	$\widetilde{M}_3$ & $\bf 1$ & 0 & $\bf 1$ & $\bf 1$ &  $\Box$   & $\overline{\Box}$ & 0 & $2n$  & 2 & $-2$ & 0 & 0 \\ \hline
	$\widetilde{\Omega}^\pm$ & $\bf 1$ & $\pm\widetilde{N}_c$ & $\bf 1$ &  $\bf 1$ & $\bf 1$ & $\bf 1$ & $0$ & $0$ & $2$  & 0 & $1$ & $\pm 1$ \\ \hline 
	\end{tabular}
	}%
	\caption{The $U(\widetilde{N}_c=N_f-N_c)$ dual of the theory described in Table~\ref{tab:s2red}.
	\label{tab:dual1}}
	\vspace{0.2cm}
	\resizebox{0.85\columnwidth}{!}{%
	\begin{tabular}{|c||c|c||c|c|c|c|c|c|c|c|}
	\hline
	 & $SU(N_c)$ & $U(1)_V$ & $SU(N_1)$   & $SU(N_3)$  & $SU(N_f)_R$ & $U(1)_z$   & $U(1)_{{R_1}}$  & $U(1)_A$ & $U(1)_{I}$  & $U(1)_{t}$ \\ \hline \hline
	$Q_1$ & $\Box$ & 1 & $\Box$  & $\bf 1$ & $\bf 1$ & $-N_3$ & 0 & 1 & 0 & 0 \\ \hline
	$Q_3$ & $\Box$ & 1 & $\bf 1$  & $\Box$ & $\bf 1$ & $N_1$  & 2 & 1 & 0 & 0 \\ \hline
	$\widetilde{Q}$ & $\overline{\Box}$ & $-1$ & $\bf 1$  & $\bf 1$ & $\overline{\Box}$ & $0$ & 0 & 1 & 0 & 0 \\ \hline
	$\Omega^\pm$ & $\bf 1$ & $\pm N_c$ & $\bf 1$  & $\bf 1$& $\bf 1$ & $0$  & $2$  & 0 & $1$ & $\pm 1$ \\ \hline
	\end{tabular}
	}%
	\caption{Taking $n=0$ in Table~\ref{tab:s2red} and twisting the R-symmetry with $U(1)_z\subset SU(N_f)_L$.
	\label{tab:s2red2}
	}
	\vspace{0.2cm}
	\resizebox{0.88\columnwidth}{!}{%
	\begin{tabular}{|c||c|c||c|c|c|c|c|c|c|c|}
	\hline
	 & $SU(\widetilde{N}_c)$ & $U(1)_{\widetilde{V}}$ & $SU(N_1)$   & $SU(N_3)$  & $SU(N_f)_R$ & $U(1)_{\widetilde{z}}$   & $U(1)_{{R_1}}$  & $U(1)_{\widetilde{A}}$ 	& $U(1)_{I}$  & $U(1)_{t}$ \\ \hline \hline
	$q_1$ & $\Box$ & 1 & $\overline{\Box}$  & $\bf 1$ &  $\bf 1$ & $N_3$ & 2 & 1 & 0 & 0 \\ \hline
	$q_3$ & $\Box$ & 1 &  $\bf 1$  & $\overline{\Box}$ &  $\bf 1$ & $-N_1$  & 0 & 1 & 0 & 0 \\ \hline
	$\widetilde{q}$ & $\overline{\Box}$ & $-1$ &  $\bf 1$  &  $\bf 1$ & ${\Box}$ & $0$ & 0 & 1 & 0 & 0 \\ \hline
	$\widetilde{M}_1$ &  $\bf 1$ & 0 & $\Box$ &  $\bf 1$ & $\overline{\Box}$ & $-N_3$ & 0 & $-2$ & 0 & 0 \\ \hline
	$\widetilde{M}_3$ &  $\bf 1$ & 0 &  $\bf 1$ & $\Box$  & $\overline{\Box}$ & $N_1$ & 2 & $-2$ & 0 & 0 \\ \hline
	$\widetilde{\Omega}^\pm$ &  $\bf 1$ & $\pm \widetilde{N}_c$ &  $\bf 1$  &  $\bf 1$ &  $\bf 1$ & $0$  & $2$  & 0 & $1$ & $\pm 1$ \\ \hline
	\end{tabular}
	}%
	\caption{The $U(\widetilde{N}_c=N_f-N_c)$  dual of the theory described in Table~\ref{tab:s2red2}.
	\label{tab:dual2}}
	\end{table}
	\def\arraystretch{1}

We begin with the compactification of  $U(N_c)$ SQCD on $S^2$, including the determinant matter. As per the previous discussion, we mix the R-symmetry with an abelian subgroup of the $SU(N_f)_L$ flavor symmetry, so that all R-charges become non-negative integers. We therefore consider the breaking pattern,
	\begin{align} \label{SUNfLbreakfirst}
	SU(N_f)_L\to \left\{
	\begin{array}{cc}
	 SU(N_1) \times SU(2n) \times SU(N_3) \times U(1)_x \times U(1)_y & \quad n > 0 \\
	  SU(N_1) \times SU(N_3) \times U(1)_z & \quad n = 0
	  \end{array}
	 \right.
	\end{align}
where the integers $N_1$, $n$, and $N_3$ are related as,
	\begin{align}
	\label{constraintfirst}
	N_1+2n+N_3=N_f\,,\qquad n+  N_3 =  N_f-N_c\,.
	\end{align}
The latter constraint is for gauge-anomaly cancellation.  
We separately consider the cases $n>0$ and $n=0$, as their compactifications exhibit qualitatively different features. The charges of the four-dimensional fields under the classical 0-form symmetries are given in Table~\ref{tab:s2red} for the $n>0$ case, and Table~\ref{tab:s2red2} for the $n=0$ case. The Seiberg-like duals of these theories are given in Tables~\ref{tab:dual1} and \ref{tab:dual2}. We shall refer to the breaking pattern \eqref{SUNfLbreakfirst} leading to the non-negative integer R-charges written in these tables as {\it twist 1} to differentiate it from the two other R-symmetry twists we will consider in subsequent subsections, and accordingly we denote the twisted R-symmetry with a subscript, $R_1$. 

The perturbative anomalies of these 4d theories are computed in Appendix \ref{sec:unngtr0}, and lead to the following features.\footnote{~The anomalies for the case of $SU(N_c)$ gauge group also appear in Appendix \ref{sec:sunngtr0}.} The gauge-gauge-flavor anomalies, written in \eqref{gaugeu}, are   the same for any $n\geq 0$ and given as follows,
	\begin{align}
	\label{gaugeufirst}
	\CI_6\big|_{\text{gauge}} &= 2 N_f  c_1(A)  \left( c_2({U(N_c)})   +   \frac{1}{2}c_1(B)^2 \right)   + c_1(B)^2  c_1(I) \,.
	\end{align}
	These are the same terms that were discussed in Section~\ref{sec:4dUNdet}.
The last term implies that $U(1)_I$ is broken to a non-invertible symmetry labeled by the rational numbers $\mathbb{Q}/\mathbb{Z}$, whose invertible part is $\mathbb{Z}_2$---namely, $k = 2$ in \eqref{fg2}. Furthermore, the first term implies that the classical axial symmetry is broken to the discrete subgroup $U(1)_A\to (\mathbb{Z}_{2N_f})_A$. 

Much as was the case in Section~\ref{sec:4dUNdet}, there are also anomalies linear in $U(1)_B$ that lead to a 2-group structure. These are written in full in equations \eqref{2groupan} and \eqref{2groupan0} in Appendix \ref{sec:unngtr0}, and here we  highlight the terms involving the R-symmetry for which we turn on the flux \eqref{eq:Rsymflux}, and so are  relevant for the compactification, 
	\begin{align} \label{2groupa}
	\begin{split}
	 \CI_6\big|_{2\text{-group}} \, \supset\,   &c_1(B) c_1(R_1) \left(  2 (N_f-N_c) c_1(A) + 2 c_1(t)  \right) \\
	  + & c_1(B) c_1(R_1)   \cdot\left\{ \begin{array}{cc} - n c_1(R_1) + 2n (N_3 c_1(x) + N_1 c_1(y))  & \ \ n>0 \\ 2 N_c (N_f-N_c) c_1(z) & \ \ n=0\end{array}  \right\} 
	  \end{split}
	\end{align}
	For simplicity, in this expression we treat the discrete axial symmetry as embedded in the classical continuous $U(1)_A$.
As was also previously discussed, the non-invertible symmetry participates in the 2-group  as signaled by a term $2 c_1(B) c_1(I) c_1(t)$ in $\CI_6$, although we will not discuss this feature further. 

We proceed to compactify these theories on the sphere, threaded with flavor flux $m_R = -1$ for the R-symmetry as in \eqref{twist}. 
The compactification to 2d results in the field content given in Table~\ref{tab:2dtheories}.
The fundamental and anti-fundamental chiral fields in 4d lead to $N_1$ fundamental chiral multiplets $P$,  $N_3$ fundamental Fermi multiplets $\Psi$, and $N_f$ anti-fundamental chiral multiplets $\Phi$; the 4d $\CN=1$ vector multiplet leads to a $(0,2)$ vector multiplet; and  the determinant matter leads to two Fermi multiplets $\Gamma^{\pm}$. Defining  the supercharge to have positive chirality, then the chiral fields contain positive chirality fermions, and the $(0,2)$ vector and Fermi multiplets contain negative chirality fermions.  The perturbative anomalies of the 2d theories as computed from Table~\ref{tab:2dtheories}---including the gauge anomalies---precisely match those obtained from reducing the 4d anomaly polynomial on $S^2$, and are written in full in \eqref{2dun1} for the case $n>0$, and in  \eqref{2dun2} for $n=0$.\footnote{~{In Table~\ref{tab:2dtheories} we have defined  $(\mathbb{Z}_{2N_f})_A$ as the discrete subgroup of the axial symmetry that is non-anomalous in 4d. However from the purely 2d perspective, there is a priori no reason to expect that the corresponding axial symmetry is broken to $(\mathbb{Z}_{2N_f})_A$. As we comment below \eqref{map2ops}, one hint in favor of the occurrence of this breaking can be inferred from the 2d duality. One option is that this is due to some mechanism happening in the compactification, similar to the monopole superpotential that is dynamically generated in the compactification from 4d to 3d  reviewed in Section~\ref{sec:3d} (and see {\it e.g.}~\cite{Gadde:2015wta} for a related discussion). We leave further investigation of this phenomenon for future work.
}}

  \def\arraystretch{1.3}
\setlength\tabcolsep{0.19em}
\begin{table}[t!]
\centering
	\resizebox{\columnwidth}{!}{
\begin{tabular}{|c||c|c||c|c|c|c|c|c|c|c|c|c|}
\hline
 & $SU(N_c)$ & $U(1)_V$ & $SU(N_1)$   & $SU(N_3)$  & $SU(N_f)_R$ & $(\mathbb{Z}_{2N_f})_{A}$  & $U(1)_{{R_1}}$   & $U(1)_{I}$  & $U(1)_{t}$  & $U(1)_x$  & $U(1)_y$ & $U(1)_z$  \\
 [-1.7ex] 
& & & & & & & & & &  \multicolumn{2}{c|}{  \footnotesize{$n>0$} }    &     \footnotesize{$n=0$} \vspace{-0.13cm}    \\ \hline \hline
$P$ & $\Box$  &  1 & $\Box$ & ${\bf 1}$ & ${\bf 1}$ & $1$ & $0$ & 0& 0  & 0 & $-2n$ & $-N_3$  \\ \hline
$\Psi$ & $\Box$ & 1 &  ${\bf 1}$ & $\Box$ & ${\bf 1}$  & 1 & 1  & 0 & 0 & $2n$ & $0$ & $N_1$ \\ \hline
$\Phi$ & $\overline{\Box}$ & $-1$ & ${\bf 1}$  & ${\bf 1}$ & $\overline{\Box}$ & 1 &  $0$ & $0$ & 0  & 0 & 0 & 0 \\ \hline
$\lambda$ & adj & $0$ & ${\bf 1}$ &  ${\bf 1}$ & ${\bf 1}$ & 0  & 1 & 0 &  0 & $0$ & $0$  & 0 \\ \hline
$\Gamma^\pm$ & ${\bf 1}$ & $\pm N_c$ & ${\bf 1}$  & ${\bf 1}$ & ${\bf 1}$ & 0 & $1$   & $1$ & $\pm 1$ & $0$ & $0$ & 0 \\ \hline
\end{tabular}
}
\caption{The 2d matter content that results from compactifying 4d $U(N_c)$ SQCD with determinant matter on the sphere, using {\it twist 1} in Tables~\ref{tab:s2red} and \ref{tab:s2red2}, and where $N_1+2n+N_3=N_f$, and $2n+2N_3 = 2(N_f-N_c)$. For $n=0$, the $U(1)_{x,y}$ symmetries reduce to a single $U(1)_z$. For $n>0$, the $(\mathbb{Z}_{2N_f})_{A}$ symmetry can be reabsorbed by a combination of the other abelian symmetries and the gauge symmetry.
\label{tab:2dtheories}
} 
\vspace{0.35cm}
	\resizebox{\columnwidth}{!}{
\begin{tabular}{|c||c|c||c|c|c|c|c|c|c|c|c|c|}
\hline
 & $SU(\widetilde{N}_c)$ & $U(1)_{\widetilde{V}}$ & $SU(N_1)$   & $SU(N_3)$  & $SU(N_f)_R$ & $(\mathbb{Z}_{2N_f})_{\widetilde{A}}$  & $U(1)_{{R_1}}$   & $U(1)_{I}$  & $U(1)_{t}$  & $U(1)_{\widetilde{x}}$  & $U(1)_{\widetilde{y}}$ & $U(1)_{\widetilde{z}}$  \\
 [-1.7ex] 
& & & & & & & & & &  \multicolumn{2}{c|}{  \footnotesize{$n>0$} }    &     \footnotesize{$n=0$} \vspace{-0.13cm}    \\ \hline \hline
$\widetilde{\Psi}$ & $\Box$  &  1 & $\overline{\Box}$ & ${\bf 1}$ & ${\bf 1}$ & $1$ & $1$ & 0& 0  & $2n$ & $0$ & $N_3$  \\ \hline
$\widetilde{P}$ & $\Box$ & 1 &  ${\bf 1}$ & $\overline{\Box}$ & ${\bf 1}$  & 1 & 0  & 0 & 0 & $0$ & $-2n$ & $-N_1$ \\ \hline
$\widetilde{\Phi}$ & $\overline{\Box}$ & $-1$ & ${\bf 1}$  & ${\bf 1}$ & $\overline{\Box}$ & 1 &  $0$ & $0$ & 0  & 0 & 0 & 0 \\ \hline
$\widetilde{M}_1$ & ${\bf 1}$ & $0$ & $\Box$ & ${\bf 1}$ & $\overline{\Box}$ & $-2$ & $0$ & $0$ & $0$ & $-2n$ & $0$ & $-N_3$ \\ \hline
$\widetilde{M}_3$ & ${\bf 1}$ & $0$ & ${\bf 1}$ & $\Box$ & $\overline{\Box}$ & $-2$ & {$1$} & $0$ & $0$ & $0$ & $2n$ & $N_1$ \\ \hline
$\widetilde{\lambda}$ & adj & $0$ & ${\bf 1}$ &  ${\bf 1}$ & ${\bf 1}$ & 0  & 1 & 0 &  0 & $0$ & $0$  & 0 \\ \hline
$\widetilde{\Gamma}^\pm$ & ${\bf 1}$ & $\pm \widetilde{N}_c$ & ${\bf 1}$  & ${\bf 1}$ & ${\bf 1}$ & 0 & $1$   & $1$ & $\pm 1$ & $0$ & $0$ & 0 \\ \hline
\end{tabular}
}
\caption{The 2d matter content that results from compactifying the duals in Tables~\ref{tab:dual1}, \ref{tab:dual2}. 
\label{tab:2dtheoriesdual}
}
\end{table}

The ABJ anomalies of the 2d theories are captured by the following terms,
	\begin{align} \label{abj2d}
	\begin{split}
	\CI_4 \,\supset\, &-2 c_1(B)\left(  c_1(t) +  (N_f-N_c) c_1(A) \right) \\
	&
	+2c_1(B) \cdot \left\{ \begin{array}{cc} 
n \hat{c}_1(R_1)- n \left(N_3 c_1(x) + N_1 c_1(y) \right) & \ \ n>0 \\ 
  - N_c(N_f-N_c) c_1(z)   & \ \ n=0 \end{array} \right\}
  	\end{split}
	\end{align}
These terms match the integration of the 2-group terms \eqref{2groupa} on the sphere and imply a symmetry breaking pattern in the 2d theory in accordance with our general discussions in Section~\ref{sec:intro2d} and Section~\ref{sec:symtft}. 
Indeed, the $U(1)_{R_1}$ symmetry inherited from 4d is anomalous in the $n>0$ theory, resulting in the breaking $U(1)_{R_1}\to \mathbb{Z}_{2n}$. This is precisely in accordance with the discussion in point \ref{item1} of Section~\ref{sec:intro2d}, since the pertinent 2-group coefficient from \eqref{2groupa} is $\kappa = 2n$, so that  $U(1)_{R_1}\to \mathbb{Z}_{|\kappa m_R |=2n}$. More precisely, the symmetry that is broken to a discrete subgroup is a particular combination of all the abelian symmetries in the theory, while the remaining independent combinations survive as non-anomalous symmetries in 2d. For example, we can redefine the field strength for $U(1)_A$ as,
\begin{align}\label{eq:newU1A}
F_{A'}=F_A+\left\{ \begin{array}{cc} 
\frac{1}{N_f-N_c}\left(F_t-{n}F_{R_1}+{n}\left(N_3F_x+N_1F_y\right)  \right) & \ \ n>0 \\ 
\frac{1}{N_f-N_c}F_t+N_cF_z   & \ \ n=0 \end{array} \right\}
\end{align}
so that the anomaly \eqref{abj2d} becomes (for both $n>0$ and $n=0$)
\begin{align}
\CI_4 \,\supset\, &-2 c_1(B)(N_f-N_c) c_1(A')\,.
\end{align}
In other words, the redefinition \eqref{eq:newU1A} changes the charges of the fields under the  abelian symmetries so  that only the axial symmetry is anomalous. The anomaly indicates that the new $U(1)_{A'}$ symmetry is broken to $U(1)_{A'}\to\mathbb{Z}_{2(N_f-N_c)}$. However, we should remember that for $n=0$ the axial symmetry was actually $(\mathbb{Z}_{2N_f})_{A'}$ (for $n>0$ this symmetry can be reabsorbed with a combination of the other abelian symmetries and the gauge symmetry), so that overall we find the symmetry breaking pattern,
\begin{align}\label{eq:U1Abreaking}
U(1)_{A'}\to\begin{cases}\mathbb{Z}_{2(N_f-N_c)} & n>0\,, \\ \mathbb{Z}_{2\gcd{(N_f,N_c)}} & n=0\,.\end{cases}
\end{align}

\subsubsection{Reduction of the 4d Duality}

Let us now consider the 4d duality in the new parameterization of the global symmetries that we used to perform {\it twist 1}. First consider the duality for the $U(N_c)$ theory with $n>0$, with the field content and symmetries given in Table~\ref{tab:s2red}. The pertinent composite chiral operators in the original theory include the gauge invariant mesons $M_i=Q_i\widetilde{Q}$ for $i=1,2,3$; the baryons $B^{(\alpha_1,\alpha_2,\alpha_3)} = (Q_1)^{\alpha_1} (Q_2)^{\alpha_2} (Q_3)^{\alpha_3}$; and the antibaryons $\widetilde{B}=\widetilde{Q}^{N_c}$. The baryon indices satisfy $\alpha_1+\alpha_2+\alpha_3=N_c$, so that all gauge indices are contracted by an epsilon tensor with $N_c$ indices, and  $\alpha_i \leq N_i$ (defining $N_2=2n$) so that the flavor indices are not over-saturated. 
These operators carry the following charges under $(\mathbb{Z}_{2N_f})_A\times U(1)_{R_1}\times U(1)_V \times U(1)_x\times U(1)_y$,
	\begin{align}
	\begin{split}
	M_1\sim a^2 \,  y^{-2n}\,,\qquad M_2\sim r^1 \, a^2  \,  x^{-N_3} \, y^{N_1}\,,\qquad M_3\sim r^2\,  a^2 \,    x^{2n}\,, \\
	B^{(\alpha_1,\alpha_2,\alpha_3)} \sim  r^{\alpha_2 + 2\alpha_3}\, a^{N_c}\,  v^{N_c}\,  x^{-N_3 \alpha_2 + 2n\alpha_3}\,  y^{-2n\alpha_1 + N_1\alpha_2}\,,\qquad \widetilde{B}\sim \left(a/v \right)^{N_c}\,,
	\end{split}
	\end{align}
where here we use the same notation as in Section~\ref{sec:4dnoninv}, so that $a,r,v,x,y$ can also be understood as fugacities in the supersymmetric index.
Since $U(1)_B$ is gauged, only baryon/anti-baryon pairs are gauge invariant, and additionally the operators $\Omega^+\Omega^-$ composed of the determinant matter are gauge invariant. 

In the magnetic dual theory with $U(\widetilde{N}_c=N_f-N_c)$ gauge group, with field content and charges given in Table~\ref{tab:dual1},
we have the gauge singlets $\widetilde{M}_i$, as well as the magnetic baryons $\widetilde{b} = \widetilde{q}^{N_f-N_c}$, and $b^{(\widetilde{\alpha}_1,\widetilde{\alpha}_2,\widetilde{\alpha}_3)} = (q_1)^{\widetilde{\alpha}_1}  (q_2)^{\widetilde{\alpha}_2}  (q_3)^{\widetilde{\alpha}_3}$, where $\widetilde{\alpha}_1 + \widetilde{\alpha}_2 + \widetilde{\alpha}_3 = N_f-N_c$, and $\widetilde{\alpha}_i \leq N_i$. 
These operators carry the following charges under the magnetic $(\mathbb{Z}_{2N_f})_{\widetilde{A}}\times U(1)_{{R_1}} \times U(1)_{\widetilde{V}} \times U(1)_{\widetilde{x}}\times U(1)_{\widetilde{y}}$ symmetry,
	\begin{align}
	\begin{split}
	\widetilde{M}_1\sim \widetilde{a}^{-2}\,  \widetilde{x}^{-2n}\,,\qquad \widetilde{M}_2\sim  \widetilde{a}^{-2} \, r^1 \,\widetilde{x}^{N_1}\, \widetilde{y}^{-N_3}\,,\qquad \widetilde{M}_3\sim  \widetilde{a}^{-2} \, r^2\,  \widetilde{y}^{2n}\,, \\
	b^{(\widetilde{\alpha}_1,\widetilde{\alpha}_2,\widetilde{\alpha}_3)}\sim r^{2\widetilde{\alpha}_1 + \widetilde{\alpha}_2} \,\widetilde{a}^{N_f-N_c} \,\widetilde{v}^{N_f-N_c} \,\widetilde{x}^{2n\widetilde{\alpha}_1 - N_1 \widetilde{\alpha}_2} \, \widetilde{y}^{N_3\widetilde{\alpha}_2 - 2n \widetilde{\alpha}_3}\,,\qquad  \widetilde{b}\sim \left(\widetilde{a} / \widetilde{v}\right)^{N_f-N_c}\,.
	\end{split}
	\end{align}
The magnetic operators match onto those on the electric side of the duality as follows,
\ba{
M_i\,\leftrightarrow\, \widetilde{M}_i\,,\quad \widetilde{B}B^{ (\alpha_1,\alpha_2,\alpha_3)} \,\leftrightarrow\,\widetilde{b}b^{(\widetilde{\alpha}_1,\widetilde{\alpha}_2,\widetilde{\alpha}_3)} \quad \text{with}\quad \alpha_i + \widetilde{\alpha}_i = N_i\,,\quad i=1,2,3
}
as well as $\Omega^+\Omega^-\to \widetilde{\Omega}^+ \widetilde{\Omega}^-$, where again we have used the shorthand notation $N_2=2n$. 
This implies the following identification of symmetry generators,
\ba{
\label{map1}
\widetilde{x}=y\,,\qquad \widetilde{y}=x\,,\qquad \widetilde{a} = a^{-1}\,,\qquad \widetilde{v}= v^{\frac{N_c}{N_f-N_c}} a^{-\frac{N_f}{N_f-N_c}}\,,\qquad a^{2N_f}=1\,,
}
or equivalently, the following relation between the  field strengths,
\ba{
\label{map1b}
F_{\widetilde{A}} = - F_A\,,\qquad (N_f-N_c) f_{\widetilde{V}} = N_c f_V - N_f F_A\,\qquad F_{\widetilde{x}} = F_y\,,\qquad F_{\widetilde{y}} = F_x\,.
} 
As expected, this reproduces the same mixing between $U(1)_V$ and $U(1)_A$ that was obtained in  \eqref{curvemap} before breaking the $SU(N_f)_L$ flavor symmetry, and we have also learned that the $U(1)_x$ and $U(1)_y$ flavor symmetries are exchanged by the duality. The classical $U(1)_I$ symmetry that leads to the non-invertible symmetry is identified  on either side of the duality. 
  
  It is straightforward to generalize to the case of  $n=0$, with dual theories  enumerated in Tables~\ref{tab:s2red2}-\ref{tab:dual2}. In this case, the mapping  of the global symmetries between the two sides of the duality is as follows, 
	\begin{align}
	\label{map2}
\widetilde{z}=z\,,\qquad  \widetilde{a} = a^{-1}\,,\qquad \widetilde{v}\to v^{\frac{N_c}{N_f-N_c}} a^{-\frac{N_f}{N_f-N_c}}\,,\qquad a^{2N_f}=1\,,
	\end{align}
with the gauge-invariant operators mapped according to, 
	\begin{align} \label{map2ops}
	M_i\leftrightarrow \widetilde{M}_i\,,\quad \widetilde{B} B^{ (\alpha_1,\alpha_3)} \leftrightarrow\widetilde{b}b^{(\widetilde{\alpha}_1,\widetilde{\alpha}_3)} \quad \text{with}\quad \alpha_i + \widetilde{\alpha}_i = N_i\,,\quad i=1,3\,.
	\end{align}
The perturbative anomalies can be shown to match between the electric and magnetic dual theories, as we verify explicitly in Appendix \ref{sec:anommatch}.

We next consider the compactification of the dual 4d  theories on the sphere. The magnetic $U(\widetilde{N}_c)$ theory results in a 2d $(0,2)$ theory with matter content given in Table~\ref{tab:2dtheoriesdual}. One may verify that the 4d duality reduces to a duality between the two 2d theories, again with the symmetries matched on either side of the duality according to \eqref{map1} or \eqref{map2}. {For this it is crucial to also redefine the symmetries as in \eqref{eq:newU1A} for the original theory (and similarly in the dual theory up to the map \eqref{map2}), and to impose that the new $U(1)_{A'}$ symmetry is broken by the ABJ anomaly as in \eqref{eq:U1Abreaking}.
This is in line with our general expectation that as long as the compactification of the 4d theories to 2d is restricted to the zero-flux sector, they will yield a consistent duality amongst the resulting 2d theories. 

\subsubsection{Comments on the infrared behavior}

Finally, let us make some remarks about the IR behavior of the 2d theories obtained from the $S^2$ compactification of the $U(N_c)$ SQCD with determinant matter via  {\it twist 1}. As we discussed above, from  \eqref{abj2d} we have  that for $n>0$ the R-symmetry that was used to perform the twist is anomalous in 2d. However, this does not necessarily mean that the 2d theory breaks supersymmetry at low energies, since this R-symmetry  will in general mix with the other abelian symmetries to possibly yield a new non-anomalous R-symmetry in the IR. Instead, a good diagnostic tool for supersymmetry breaking is the vanishing of the elliptic genus. As we reviewed before, the elliptic genus takes the form of an integral of dimension equal to the rank of the gauge group, which will in general vanish if the integrand does not provide enough poles. The poles are in particular provided by chiral fields, while Fermi fields do not contribute any pole. Our analysis is thus similar to the one done in \cite{Gadde:2013lxa}, since our theory differs from the one of \cite{Gadde:2013lxa} only for the representation of the Fermi fields $\Gamma^\pm$ in the determinant representation.

The chiral fields $P$ and $\Phi$ only come in the fundamental and anti-fundamental representations of the $U(N_c)$ gauge group. When computing the integral, we can take the residues provided by either of the two. Thus, the integral will vanish whenever the rank of the gauge group $N_c$ is bigger than both the number of fundamental and anti-fundamental chirals. We can also exploit the duality to obtain similar constraints in the dual theory. Overall we arrive at the conditions,
\begin{align}
N_1<N_c\,,\quad N_f<N_c\,,\quad N_3<\widetilde{N}_c\,,\quad N_f<\widetilde{N}_c\,.
\end{align}
Using \eqref{constraintfirst} and that the dual rank is $\widetilde{N}_c=N_f-N_c$, we see that one condition is redundant, and that the remaining three are
\begin{align}
N_c<0\,,\quad N_f<N_c\,,\quad n>0\,.
\end{align}
The first two conditions coincide with the requirements for the original 4d theory and its Seiberg dual to exhibit supersymmetry breaking. The last condition suggests that in the theory for $n>0$ supersymmetry is broken, while for $n=0$ it is not.  For $n<0$ we also get a consistent 2d theory that does not exhibit supersymmetry breaking, however this case cannot be obtained by  compactification from 4d.

We can next investigate whether the $n=0$ theory flows to an SCFT or not. For this, we need to determine the superconformal R-symmetry using $c$-extremization \cite{Benini:2012cz,Benini:2013cda}. We define a trial $U(1)_{R_{\text{trial}}}$ R-symmetry by mixing the  $U(1)_{R_1}$ we used for the twist with all the other $U(1)$ symmetries listed in Table~\ref{tab:2dtheories},
\begin{align}
q_{R_{\text{trial}}}=q_R+q_tR_t+q_IR_I+q_zR_z\,,
\end{align}
where $q_i$ are the charges and $R_i$ the mixing coefficient for each symmetry. We then compute the trial right-moving central charge as,
\begin{align}
\begin{split}
c_{\text{trial}}=\tr\,U(1)_{R_{\text{trial}}}^2=&N_1N_c(-N_3R_z-1)^2-N_3N_c(1+N_1R_z)^2\\
&+N_fN_c(-1)^2-(1+R_I+R_t)^2-(1+R_I-R_t)^2-N_c^2\,.
\end{split}
\end{align}
Notice that due to the anomaly \eqref{abj2d} one combination of the mixing coefficients has to be set to zero, since this corresponds to the $U(1)_{A'}$ symmetry that is broken to a discrete subgroup and thus cannot mix,
\begin{align}
R_t+(N_f-N_c)(R_A+N_cR_z)=0\,.
\end{align}
Moreover, one must be careful with non-compact directions in the target space, which are associated with additional symmetries that cannot mix. To identify these, we have to impose that the gauge invariant chiral operators corresponding to such directions have zero R-charge. One such operator is $P\Phi$ (with the contraction of gauge indices understood), which leads to the constraint
\begin{equation}
2R_A-N_3R_z=0\,.
\end{equation}
Another operator is $\Gamma^+\Gamma^-$, which gives,
\begin{equation}
R_I=0\,.
\end{equation}
Overall, we have three conditions that the four mixing coefficients have to satisfy. Hence, $c_{\text{trial}}$ depends on a single mixing coefficient, with respect to which we  extremize  to determine the exact infrared R-symmetry at the putative fixed point. This gives the following non-trivial value for the right-moving central charge,
\begin{align}
c_{\text{IR}}=N_c^2\left(3-\frac{4N_c}{N_f+N_c}\right)-2\,,
\end{align}
which indicates that the theory flows to an interacting SCFT.

\subsection{Compactification with Negative R-charges} \label{sec:undetsneg}

 	\def\arraystretch{1.3} 
	\setlength\tabcolsep{4.5pt} 
	\begin{table}[t!]
	\centering
	\resizebox{\columnwidth}{!}{%
	\begin{tabular}{|c||c|c||c|c|c|c|c|c|c|c|c|c|}
	\hline
 & $SU(N_c)$ & $U(1)_V$ & $SU(N_1)$ & $SU(2n)$ & $SU(N_3)$ & $SU(N_f)_R$ & $U(1)_x$ & $U(1)_y$ & $U(1)_{R_2}$ & $U(1)_A$ & $U(1)_I$ & $U(1)_t$  \\ \hline \hline
 $Q_1$ & $\Box$ & $1$ & $\Box$ & $\bf 1$ & $\bf 1$ & $\bf 1$ & $0$ & $-2n$ & $1$ & 1 & 0 & 0 \\ \hline
  $Q_2$ & $\Box$ & $1$ & $\bf 1$ & $\Box$ & $\bf 1$ & $\bf 1$ & $-N_3$ & $N_1$ & $2$ & 1 & 0 & 0 \\ \hline
   $Q_3$ & $\Box$ & $1$ & $\bf 1$ & $\bf 1$ & $\Box$ & $\bf 1$ & $2n$ & $0$ & $3$ & 1 & 0 & 0 \\ \hline
  $\widetilde{Q}$ & $\overline{\Box}$ & $-1$ & $\bf 1$ & $\bf 1$ & $\bf 1$ & $\overline{\Box}$ & $0$ & $0$ & $-1$ & 1 & 0 & 0 \\ \hline
  $\Omega^\pm$ & $\bf 1$ & $\pm N_c$ & $\bf 1$ & $\bf 1$ & $\bf 1$ & $\bf 1$ & $0$ & $0$ & $2$ & $0$ & $1$ & $\pm1$ \\ \hline
	\end{tabular}
	}%
	\caption{The charges of the fields for 4d $U(N_c)$ SQCD with  determinant matter $\Omega^\pm$, with the new R-symmetry $U(1)_{R_2}$ ({\it twist 2}). Recall that $N_1+2n+N_3=N_f$, and $n+N_3 = N_f-N_c$.
		\label{tab:charges4dnew}
	}
	\def\arraystretch{1}
	\end{table}

We next consider the twisted compactification on $S^2$ of  $U(N_c)$ SQCD with $N_f$ flavors and determinant/anti-determinant matter, now with a choice of R-symmetry such that not all fields have a non-negative R-charge. As mentioned above, this implies that the summation over gauge fluxes $\vec{m}_G$ is not truncated to the zero flux sector, thus allowing us to study the effect of compactifying with a non-trivial gauge flux.

For this, we once again break the $SU(N_f)_L$ flavor symmetry as in \eqref{eq:SUNfLbreak} with $n>0$, but we now mix the various abelian flavor symmetries with the R-symmetry so as to obtain a new R-symmetry that we denote by $R_2$. Namely, we mix the R-symmetry with a subgroup $U(1)_x\times U(1)_y$ of the $SU(N_f)_L$ symmetry, so that the anti-fundamental chiral superfields have R-charge $-1$, and the rest of the chiral fields have R-charges listed in Table~\ref{tab:charges4dnew}. We shall refer to this twist of $U(N_c)$ SQCD as {\it twist 2}.

The perturbative anomalies of the resulting 4d theory are computed in Appendix \ref{sec:twist2}. We will not repeat the analysis of the gauge anomalies, which are the same as in \eqref{gaugeufirst}, but will highlight the 2-group terms involving the $U(1)_{R_2}$, which in this basis take the form,
	\begin{align} \label{2grouprprime}
	\begin{split}
	\CI_6\big|_{\text{2-group}}\, \supset  \,c_1(B) c_1(R_2) \Big( &  - c_1(R_2) (2N_c+n) + 2 c_1(A) (2 N_f-N_c) + 2 c_1(t) \\
	&  \quad + 2 n N_1 c_1(y) + 2 n N_3 c_1(x) \Big)\,.
	\end{split}
	\end{align}

We proceed to compactify this theory on a sphere with flavor flux for the R-symmetry  as in \eqref{twist}. Since with this twist there is no truncation to the zero gauge-flux sector, to determine the content of the 2d theory in a non-trivial flux sector we cannot  use the rule reviewed in Section~\ref{sec:gen2d}.
Instead, in order to study the compactification, we look explicitly at the $T^2\times S^2$ partition function of the theory with the assignment of symmetries of Table~\ref{tab:charges4dnew} (following the conventions of \cite{Gadde:2015wta}),
	\begin{align}\label{eq:T2S2UNneg}
	\mathcal{I}&=\sum_{\vec{m}_G\in\mathbb{Z}^{N_c}}g^{\sum_{k=1}^{N_c}m_k}\oint\prod_{k=1}^{N_c}\frac{\udl{z_k}}{2\pi iz_k}\frac{\qfac{q^{2N_c}}\prod_{k<l}^{N_c}\theta(z_kz_l^{-1})\theta(z_k^{-1}z_l)}{\prod_{k=1}^{N_c}\prod_{i=1}^{N_1}\theta(z_ku_i^{-1}y^{-2n}a)^{m_k}}\nn
	&\times\frac{1}{\prod_{k=1}^{N_c}\left[\prod_{i=1}^{2n}\theta(z_kv_i^{-1}x^{-N_3}y^{N_1}a)^{m_k-1}\prod_{i=1}^{N_3}\theta(z_kw_i^{-1}x^{2n}a)^{m_k-2}\right]}\nn
	&\times\frac{1}{\prod_{k=1}^{N_c}\prod_{i=1}^{N_f}\theta(z_k^{-1}f_ia)^{2-m_k}\theta\left(\prod_{k=1}^{N_c}\iota z_kt\right)^{\sum_k m_k-1}\theta\left(\prod_{k=1}^{N_c}\iota z_k^{-1}t^{-1}\right)^{-\sum_k m_k-1}}
	\end{align}
where $\theta(x)=\qfac{x}\qfac{qx^{-1}}$ and we have also turned on an FI parameter whose exponentiation we denoted by $g$, which will be useful in the next subsection to recover the triality. Moreover, we denoted the fugacities by $\vec{z}$ for the $U(N_c)$ gauge group, by $\vec{u}$ for $SU(N_1)_L$, by $\vec{v}$ for $SU(2n)$, by $\vec{w}$ for $SU(N_3)_L$, by $\vec{f}$ for $SU(N_f)_R$, and by $x,y,a,\iota,t$ for $U(1)_x\times U(1)_y\times U(1)_A\times U(1)_I\times U(1)_t$.

First of all, we notice that with this choice of R-symmetry the zero-flux sector $\vec{m}_G=\vec{0}$ is not the only one for which the integrand has poles that lead to a non-trivial contribution to the integral. This indicates that the sum over $\vec{m}_G$ does not truncate to the zero-flux sector, and that in the compactification we get a direct sum of distinct 2d theories. Moreover, we can easily read off the matter content of the 2d theory in each flux sector by remembering that the contribution to the elliptic genus of a 2d $\mathcal{N}=(0,2)$ chiral multiplet of R-charge $r$,  in a representation $\mathcal{R}_H$ with weights $\rho$ of the residual gauge group $H\subset U(N_c)$ preserved by $\vec{m}_G$, and in a representation $\mathcal{R}_F$ with weights $\tilde{\rho}$ of the flavor symmetry group $F$, is
	\begin{equation}\label{eq:T2chir}
	\mathcal{I}_{\text{chir}}^{(r)}(\vec{z};\vec{u};q)=\prod_{\rho\in\mathcal{R}_H}\prod_{\tilde{\rho}\in\mathcal{R}_F}\frac{1}{\thetafunc{q^{\frac{r}{2}}\vec{z}^\rho \vec{u}^{\tilde{\rho}}}}\,,
	\end{equation}
while that of a Fermi multiplet is
	\begin{equation}\label{eq:T2fermi}
\mathcal{I}_{\text{ferm}}^{(r)}(\vec{z};\vec{u};q)=\prod_{\rho\in\mathcal{R}_H}\prod_{\tilde{\rho}\in\mathcal{R}_F}\thetafunc{q^{\frac{r+1}{2}}\vec{z}^{\,\rho} \vec{u}^{\tilde{\rho}}}\,.
	\end{equation}
The multiplicity of these fields is given by the exponent of the corresponding theta function in \eqref{eq:T2S2UNneg}, and thus depends not only on their R-charge but also on their representation under the gauge group and the value of the gauge flux $\vec{m}_G$. Finally, the contribution to the elliptic genus of a 2d $\mathcal{N}=(0,2)$ vector multiplet is,
	\begin{equation}
	\mathcal{I}_{\text{vec}}(\vec{z};q)=\qfac{q}^{2\,\mathrm{rk}\,H}\prod_{\alpha\in\mathfrak{h}}\thetafunc{\vec{z}^{\,\alpha}}\,,
	\end{equation}
where $\alpha$ are the roots of the Lie algebra $\mathfrak{h}$ of the residual gauge group $H$. We can see that when the flux $\vec{m}_G$ breaks the gauge group $U(N_c)$ to a subgroup $H$, then the 4d vector multiplet in the adjoint of $U(N_c)$ whose contribution to the $T^2\times S^2$ partition function is given by the numerator in the first line of \eqref{eq:T2S2UNneg} gives rise not only to a 2d vector multiplet in the adjoint of $H$, but also to some Fermi multiplets according to the decomposition of the adjoint representation of $U(N_c)$ with respect to $H$.

In the following we will consider the 2d theories corresponding to a few representative choices of flux, and demonstrate how they can be interpreted as the twisted compactification on $S^2$ with magnetic gauge flux, by studying the corresponding anomalies. In particular, we will see that also the anomalies involving gauge symmetries are preserved under the compactification and determine the breaking of various symmetries in 2d, in accordance with our general discussions in Section~\ref{sec:intro2d} and Section~\ref{sec:symtft}.

  \def\arraystretch{1.3}
\setlength\tabcolsep{0.19em}
\begin{table}[t!]
\centering
	\resizebox{\columnwidth}{!}{
\begin{tabular}{|c||c|c||c|c|c|c|c|c|c|c|c|c|c|}
\hline
 & $SU(N_c)$ & $U(1)_V$ & $SU(2n)$   & $SU(N_3)$  & $SU(N_f)_R$ & $SU(2)_{\Psi_3}$ & $SU(2)_{\Phi}$ & $U(1)_{{A}}$  & $U(1)_{R_2}$   & $U(1)_{I}$  & $U(1)_{t}$  & $U(1)_x$  & $U(1)_y$   \\ \hline \hline
 $\Psi_2$ & $\Box$ & ${1}$ & $\Box$ & ${\bf 1}$ & ${\bf 1}$ & ${\bf 1}$ & ${\bf 1}$ & 1 & 1 & 0 & 0 & $-N_3$ & $N_1$ \\ \hline
  $\Psi_3$ & $\Box$ & ${1}$  & ${\bf 1}$ & $\Box$ & ${\bf 1}$ & ${\Box}$ & ${\bf 1}$ & 1 & 2 & 0 & 0 & $2n$ & $0$ \\ \hline
$\Phi$ & $\overline{\Box}$ & $-1$ & ${\bf 1}$  & ${\bf 1}$ & $\overline{\Box}$  & ${\bf 1}$  & ${\Box}$ & 1 &  $-1$ & $0$ & 0  & 0 & 0  \\ \hline
$\lambda$ & adj & $0$ & ${\bf 1}$ &  ${\bf 1}$ & ${\bf 1}$ &  ${\bf 1}$ & ${\bf 1}$ & 0  & 1 & 0 &  0 & $0$ & $0$   \\ \hline
$\Gamma^\pm$ & ${\bf 1}$ & $\pm N_c$ & ${\bf 1}$  & ${\bf 1}$ & ${\bf 1}$ & ${\bf 1}$ & ${\bf 1}$ & 0 & $1$   & $1$ & $\pm 1$ & $0$ & $0$  \\ \hline
\end{tabular}
}
\caption{The 2d matter content that results from compactifying 4d $U(N_c)$ SQCD with determinant matter on the sphere, using {\it twist 2} in Table~\ref{tab:charges4dnew}, in the $\vec{m}_G=(0,\dots,0)$ flux sector.  The parameters satisfy $N_1+2n+N_3=N_f$, and $2n+2N_3 = 2(N_f-N_c)$. 
\label{tab:charges2dm0}
} 
\vspace{0.35cm}
	\resizebox{\columnwidth}{!}{
\begin{tabular}{|c||c|c||c|c|c|c|c|c|c|c|c|c|c|}
\hline
 & $SU(N_c)$ & $U(1)_V$ & $SU(N_1)$   & $SU(N_3)$  & $SU(N_f)_R$ & $SU(N_c-1)_{P^+}$ & $SU(N_c+1)_{\Gamma^-}$ & $U(1)_{{A}}$  & $U(1)_{R_2}$   & $U(1)_{I}$  & $U(1)_{t}$  & $U(1)_x$  & $U(1)_y$   \\ \hline \hline
 $P_1$ & $\Box$ & ${1}$ & $\Box$ & ${\bf 1}$ & ${\bf 1}$ & ${\bf 1}$ & ${\bf 1}$ & 1 & 1 & 0 & 0 & $0$ & $-2n$ \\ \hline
  $\Psi_3$ & $\Box$ & ${1}$  & ${\bf 1}$ & $\Box$ & ${\bf 1}$ & ${\bf 1}$ & ${\bf 1}$ & 1 & 2 & 0 & 0 & $2n$ & $0$ \\ \hline
$\Phi$ & $\overline{\Box}$ & $-1$ & ${\bf 1}$  & ${\bf 1}$ & $\overline{\Box}$  & ${\bf 1}$  & ${\bf 1}$ & 1 &  $-1$ & $0$ & 0  & 0 & 0  \\ \hline
$\lambda$ & adj & $0$ & ${\bf 1}$ &  ${\bf 1}$ & ${\bf 1}$ &  ${\bf 1}$ & ${\bf 1}$ & 0  & 1 & 0 &  0 & $0$ & $0$   \\ \hline
$P^+$ & ${\bf 1}$ & $N_c$ & ${\bf 1}$  & ${\bf 1}$ & ${\bf 1}$ & ${\Box}$ & ${\bf 1}$ & 0 & $2$   & $1$ & $ 1$ & $0$ & $0$  \\ \hline
$\Gamma^-$ & ${\bf 1}$ & $- N_c$ & ${\bf 1}$  & ${\bf 1}$ & ${\bf 1}$ & ${\bf 1}$ & ${\Box}$ & 0 & $1$   & $1$ & $- 1$ & $0$ & $0$  \\ \hline
\end{tabular}
}
\caption{The 2d matter content that results from compactifying 4d $U(N_c)$ SQCD with determinant matter on $S^2$, using {\it twist 2} in Table~\ref{tab:charges4dnew}, in the $\vec{m}_G=(1,\dots,1)$ flux sector.   
\label{tab:charges2dm1}
} 
\vspace{0.35cm}
	\resizebox{\columnwidth}{!}{
\begin{tabular}{|c||c|c|c||c|c|c|c|c|c|c|c|c|c|c|c|c|}
\hline
 & $SU(N_c-1)$ & $U(1)_G$ & $U(1)_V$ & $SU(N_1)$  & $SU(2n)$  & $SU(N_3)$  & $SU(N_f)_R$ & $SU(2)_{\Psi_3}$ &  $SU(2)_{\Phi}$ &  $SU(2)_{\Gamma^-}$ & $U(1)_{{A}}$  & $U(1)_{R_2}$   & $U(1)_{I}$  & $U(1)_{t}$  & $U(1)_x$  & $U(1)_y$   \\ \hline \hline 
$P_1$ & ${\bf 1}$ & $1-N_c$ & 1 &  $\Box$ & ${\bf 1}$ & ${\bf 1}$ & ${\bf 1}$& ${\bf 1}$ & ${\bf 1}$ & ${\bf 1}$ & 1 &1 &0 &0 & 0 & $-2n$  \\ \hline
 $\Psi_2$ & $\Box$ & $1$ & 1 & ${\bf 1}$ &  $\Box$ & ${\bf 1}$ & ${\bf 1}$ & ${\bf 1}$ & ${\bf 1}$ & ${\bf 1}$ & 1& 1&0 &0 & $-N_3$ & $N_1$  \\ \hline
$\Psi_3^{(1)}$ & $\Box$ & $1$ & 1 &  ${\bf 1}$ &${\bf 1}$ & $\Box$ & ${\bf 1}$ & $\Box$ & ${\bf 1}$ & ${\bf 1}$ &1 & 2& 0& 0& $2n$ & 0  \\ \hline
$\Psi_3^{(2)}$ & ${\bf 1}$ & $1-N_c$ & 1 &  ${\bf 1}$ & ${\bf 1}$ & $\Box$ &${\bf 1}$ & ${\bf 1}$& ${\bf 1}$&${\bf 1}$ &1 & 2&0 &0 & $2n$  & 0  \\ \hline
 $\Phi^{(1)}$ & $\overline{\Box}$ & $1$ & $-1$ & ${\bf 1}$ &${\bf 1}$ &${\bf 1}$ & $\Box$  &  ${\bf 1}$&$\Box$ & ${\bf 1}$ &1 & $-1$& 0& 0& 0& 0 \\ \hline
$\Phi^{(2)}$ & ${\bf 1}$ & $N_c-1$ & $-1$ & ${\bf 1}$ &${\bf 1}$ & ${\bf 1}$&$\Box$ & ${\bf 1}$&${\bf 1}$ & ${\bf 1}$&1 &$-1$ &0 &0 &0 &0  \\ \hline
$\Sigma^+$ & ${\Box}$ & $N_c$ & $0$ & ${\bf 1}$ & ${\bf 1}$&${\bf 1}$ &${\bf 1}$ &${\bf 1}$ &${\bf 1}$ &${\bf 1}$ & 0& 1&0 & 0& 0 &0   \\ \hline
$\Sigma^-$ & $\overline{\Box}$ & $-N_c$ & $0$ &${\bf 1}$  &${\bf 1}$ &${\bf 1}$ & ${\bf 1}$&${\bf 1}$ & ${\bf 1}$& ${\bf 1}$&0 & $-1$& 0& 0 &0 &0  \\ \hline
$\Gamma^-$ & ${\bf 1}$ & $0$ & $-N_c$ & ${\bf 1}$ &${\bf 1}$ & ${\bf 1}$&${\bf 1}$ &${\bf 1}$ &${\bf 1}$ & $\Box$ &0 & 1& 1 & $-1$ & 0 & 0  \\ \hline
\end{tabular}
}
\caption{The 2d matter content that results from compactifying 4d $U(N_c)$ SQCD with determinant matter on $S^2$, using {\it twist 2} in Table~\ref{tab:charges4dnew}, in the $\vec{m}_G=(1,0,\dots,0)$ flux sector.   
\label{tab:charges2dm10bis}
} 
\end{table}

\subsubsection*{Flux $\vec{m}_G = (0,\dots,0)$}

We start by considering the sector with a vanishing magnetic flux for the gauge symmetry. The matter content of the theory in this flux sector is summarized in Table~\ref{tab:charges2dm0}, and consists of the $(0,2)$ vector multiplet; the Fermi fields $\Psi_2,\Psi_3$ descending from the 4d fundamental chirals $Q_2,Q_3$; the chiral multiplet $\Phi$ descending form the anti-fundamental chirals $\widetilde{Q}$; and the Fermi fields $\Gamma^{\pm}$ descending from the determinant matter. The fields $\Psi_3$ and $\Phi$ each appear in two copies after the compactification, transforming in the fundamental representation of an accidental $SU(2)_{\Psi_3}\times SU(2)_{\Phi}$  symmetry. 

We have verified that the anomaly polynomial of the 2d theory consisting of the matter content  in  Table~\ref{tab:charges2dm0} is reproduced by integrating the anomaly polynomial of the original 4d theory, with result given in \eqref{udiff2d}. Notably, due to the topological twist of the R-symmetry, the compactification of the 2-group terms in $\CI_6$ from \eqref{2grouprprime}  directly reproduces the gauge anomalies of the 2d theory, which take the form 
	\begin{align}
\CI_4\big|_{\text{gauge}} = 2c_1(B) \Big ( & \hat{c}_1(R_2) (2 N_c + n) + c_1(A) (N_c-2 N_f ) -  c_1(t) \nonumber \\
	&\qquad\qquad-  n \left( N_3 c_1(x)+N_1 c_1(y) \right)  \Big)\,.
	\end{align}
These imply that a combination of the various classical abelian symmetries is broken to a discrete subgroup, including the $U(1)_{R_2}$ symmetry which descends from the four-dimensional R-symmetry. However much as was the case with the {\it twist 1} example examined in Section~\ref{sec:undets2}, we expect that the anomaly-free R-symmetry in the 2d theory is given by mixing with the global abelian symmetries, and can be determined by a $c$-extremization analysis.

As we have noted, the 2d theory contains symmetries $SU(2)_{\Psi_3}\times SU(2)_\Phi$ that are accidental from the perspective of the compactification from 4d. These can be related to the $SU(2)$ isometry of the sphere, which we will denote by $SU(2)_{\text{ISO}}$, as follows. The Bott-Cattaneo formula implies that the integration of the Chern root satisfies,
	\begin{align}
	\int_{S^2} t^3 = - 8 c_2(SU(2)_{\text{ISO}})\,.
	\end{align}
This formula follows from \eqref{bott} in Appendix \ref{sec:bott}. Upon decomposing the Pontryagin class for the spacetime tangent bundle as $p_1(T_4) = p_1(T_2) +t^2$, and applying this formula in the integration of the six-form anomaly polynomial $\mathcal{I}_6$ over the sphere, we can identify the result with the anomalies   for the accidental $SU(2)_{\Psi_3}\times SU(2)_\Phi$ symmetry computed from Table~\ref{tab:charges2dm0}, yielding   the following identification,
	\begin{align}
	N_f c_2(SU(2)_\Phi) - N_3 c_2(SU(2)_{\Psi_3}) = (N_f-N_3)c_2(SU(2)_{\text{ISO}})\,.
	\end{align}
	This anomaly matching indicates that the isometry group of the sphere is identified with the diagonal subgroup of  the accidental $SU(2)_{\Psi_3}\times SU(2)_\Phi$ symmetry.

\subsubsection*{Flux $\vec{m}_G = (1,\dots,1)$}

The other sectors correspond to having non-trivial flux for the gauge symmetry. We will consider fluxes of increasing complexity, starting from one which is only in the $U(1)_B$ part of the gauge group and not the $SU(N_c)$ part. The minimal such flux is $\vec{m}_G=(1,\cdots,1)$, which corresponds to decomposing $c_1(B)$ as follows,
\begin{align}
c_1(B) = \frac{N_c}{2}  t  + \hat{c}_1(B)\qquad\Rightarrow\qquad\int_{S^2} c_1(B) = N_c\,.
\end{align}
The matter content of the 2d theory in this flux sector is summarized in Table~\ref{tab:charges2dm1}, and the result of integrating the 4d anomaly polynomial is given in \eqref{m111}. The 2d theory has an accidental $SU(N_c-1)_{P^+} \times SU(N_c+1)_{\Gamma^-}$ symmetry, as there are   $N_c-1$ identical copies of the chiral field $P^+$ and $N_c+1$ copies of the Fermi field $\Gamma^-$ in the compactification.

Again we find perfect matching between the anomaly polynomial of the 2d theory with matter content and charges given in Table~\ref{tab:charges2dm1}, and the one obtained by integrating the 4d anomaly polynomial over the $S^2$, with details presented in Appendix \ref{sec:twist2}. 
In particular, we again  observe that  the 2d gauge anomalies can be reproduced in this way, leading to
\begin{align}
\CI_4\big|_{\text{gauge}} = 2\hat{c}_1(B)\Big[  &(2 N_c + n )\hat{c}_1(R_2) +N_c c_1(I) +(N_c-N_f)c_1(A) \quad\  \nonumber \\
	&\qquad\quad- c_1(t) - n N_1 c_1(y) - n N_3 c_1(x)  \Big] \,.
\end{align}
Notice that these descend from the 4d anomalies that are both quadratic  in $U(1)_B$ from \eqref{gaugeufirst}, and linear in $U(1)_B$ from \eqref{2grouprprime}.
Also in this case there are accidental symmetries $SU(N_c-1)_{P^+}\times SU(N_c+1)_{\Gamma^-}$, with anomalies
\begin{align}\label{eq:accanm1}
\CI_4\supset c_2(SU(N_c-1)_{P^+})- c_2(SU(N_c+1)_{\Gamma^-})\,.
\end{align}
These should be compared with the anomaly for the sphere isometry (from \eqref{m111}), 
\begin{align}
N_c^2c_2(SU(2)_{\text{ISO}})\,.
\end{align}
Indeed, we can embed $SU(2)_{\text{ISO}}$ inside $SU(N_c-1)_{P^+}\times SU(N_c+1)_{\Gamma^-}$ such that the anomalies match as follows. We first consider the $SU(2)$ subgroup of $SU(N_c-1)_{P^+}$ such that the fundamental representation of the latter reduces to the $(N_c-1)$-dimensional representation of the former. Similarly, we consider the $SU(2)$ subgroup of $SU(N_c+1)_{\Gamma^-}$ such that the fundamental representation of the latter reduces to the $(N_c+1)$-dimensional representation of the former. Finally, $SU(2)_{\text{ISO}}$ is identified with the diagonal combination of these two $SU(2)$'s. At the level of the characteristic classes, this embedding implies
\begin{align}
&c_2(SU(N_c-1)_{P^+})=-\frac{(N_c-1)((N_c-1)^2-1)}{6}c_2(SU(2)_{\text{ISO}})\,,\nn
&c_2(SU(N_c+1)_{\Gamma^-})=-\frac{(N_c+1)((N_c+1)^2-1)}{6}c_2(SU(2)_{\text{ISO}})\,,
\end{align}
upon which \eqref{eq:accanm1}  reduces to the expected anomaly for the sphere isometry from the compactification.

\subsubsection*{Flux $\vec{m}_G = (1,0,\dots,0)$}

We finally consider a more complicated flux that breaks the gauge symmetry $U(N_c)$ to a subgroup $H$. The minimal such flux is $\vec{m}_G=(1,0,\cdots,0)$ and it breaks the $SU(N_c)$ part of the gauge symmetry as follows,
\begin{align}\label{eq:fluxbreak}
SU(N_c)\to SU(N_c-1)\times U(1)_G\,,
\end{align}
where the embedding is,
\begin{align}\label{eq:br}
{\bf N_c}\to({\bf N_c-1})^1\oplus ({\bf 1})^{1-N_c}\,.
\end{align}
This branching rule corresponds to the following decomposition of characteristic classes,
\begin{align} \label{branching}
c_2(SU(N_c)) = c_2(SU(N_c-1)) + \frac{N_c (N_c-1)}{2} c_1(G)^2\,.
\end{align}
The flux vector can be rewritten as
\begin{align}
\vec{m}_G=(1,0,\cdots,0)=\left(\frac{1}{N_c}+\frac{N_c-1}{N_c},\frac{1}{N_c}-\frac{1}{N_c},\cdots,\frac{1}{N_c}-\frac{1}{N_c}\right)\,,
\end{align}
and so we see that this corresponds to flux $\frac{1}{N_c}$ for the $U(1)_V$ part of the gauge group (or equivalently $1$ for $U(1)_B$), and $-\frac{1}{N_c}$ for the $U(1)_G$ in \eqref{eq:fluxbreak}, so that we may express
\begin{align} \label{bg}
\begin{split}
&c_1(B) = \frac{t}{2}  + \hat{c}_1(B)\,,\qquad  c_1(G) = - \frac{t}{2N_c}+ \hat{c}_1(G)\,,\\    \Rightarrow  & \quad   \int_{S^2} c_1(B) = 1\,,\qquad  \int_{S^2} c_1(G) = -\frac{1}{N_c}\,.
\end{split}
\end{align}
 The matter content of the theory in this flux sector is summarized in Table~\ref{tab:charges2dm10bis}. Notice that each 4d field can give rise to multiple fields in 2d due to the breaking of the gauge symmetry. In particular, the Fermi fields $\Sigma^\pm$ descend from the broken part of the 4d $U(N_c)$ vector multiplet.

Again we can match the anomaly polynomial of this 2d theory with the one obtained by integrating the anomaly polynomial of the 4d theory over the $S^2$, given in \eqref{m100}, although the latter computation now requires some additional steps due to the fact that we have a flux for the $SU(N_c)$ part of the gauge group which partially breaks it. Indeed, we need to first decompose the $SU(N_c)$ Chern class $c_2(SU(N_c))$ in accordance with the branching rule \eqref{branching}, and then substitute \eqref{bg} (see Appendix \ref{sec:twist2} for more details).
We stress again that in this way we are also able to reproduce the gauge anomalies of the theory obtained after the compactification. In particular, we find the following gauge anomalies involving $U(1)_B$, which descend from those that led to a 2-group and to a non-invertible symmetry in 4d,
\begin{align}
\CI_4\, \supset \, 2 \hat{c}_1(B) \Big[ &
	\hat{c}_1(R) (2 N_c+n) + c_1(A) \left( N_c-2N_f +\frac{N_f}{N_c} \right)\nonumber \\
	&\qquad\qquad+c_1(I) - c_1(t) - n\left(N_1 c_1(y) + N_3 c_1(x) \right) 	\Big]\,.
\end{align}
Moreover, we have the following gauge anomaly involving $U(1)_G$, which comes only from the anomaly for the 4d $SU(N_c)$ symmetry \eqref{gaugeufirst} that breaks $U(1)_A\to\mathbb{Z}_{2N_f}$ in 4d,
\begin{align}
\CI_4\, \supset\, -2N_f(N_c-1) \hat{c}_1(G)c_1(A)\,,
\end{align}
and indicates that only $\mathbb{Z}_{2N_f}$ is preserved as well in 2d. 

This case also features accidental symmetries,  $SU(2)_{\Psi_3}\times SU(2)_{\Phi}\times SU(2)_{\Gamma^-}\times SU(2)_{\Sigma}$,  with  anomalies,\footnote{~Recall that a Fermi multiplet with R-charge $r$ and in a representation $\mathcal{R}$ is equivalent to a Fermi multiplet with R-charge $-r$ and in the complex-conjugate representation $\mathcal{R}$. Hence, $\Sigma^{\pm}$ are identical fields and they are rotated by an accidental $SU(2)_{\Sigma}$ symmetry.}
\begin{align}
\CI_4\, \supset\, & -N_3(N_c-1)c_2(SU(2)_{\Psi_3})+N_f(N_c-1)c_2(SU(2)_{\Phi})\nn
&-c_2(SU(2)_{\Gamma^-})-(N_c-1)c_2(SU(2)_{{\Sigma}})\,.
\end{align}
These should be compared with the anomaly for $SU(2)_{\text{ISO}}$ from the compactification (from \eqref{m100}),
\begin{align}
\left[N_c-(N_c-1)(N_c+n)\right]c_2(SU(2)_{\text{ISO}})\,.
\end{align}
Again we can see that the two match if we identify $SU(2)_{\text{ISO}}$ with the diagonal combination of $SU(2)_{\Psi_3}\times SU(2)_{\Phi}\times SU(2)_{\Gamma^-}\times SU(2)_{\Sigma}$.

\subsection{Recovering 2d Triality}
\label{sec:triality}

In this subsection we revisit the derivation from \cite{Gadde:2015wta,Tachikawa:2018sae} of the 2d $\mathcal{N}=(0,2)$ triality \cite{Gadde:2013lxa} as starting from the 4d $\mathcal{N}=1$ Seiberg duality for $U(N_c)$ SQCD with determinant matter, placing particular emphasis on the structure of the 0-form symmetries of the 2d theories from the perspective of the compactification. In particular, we show that all the symmetries of the 2d theories are already manifest in 4d and that their mapping under the 2d triality can be understood from the one for the 4d duality. For this it will be crucial to focus on the compactification of the anomaly polynomial (whose details we relegate to Appendix~\ref{sec:untriality}), with particular attention paid to the fate of the gauge anomalies (similarly to the discussion of the previous subsections), since they lead to the correct parameterization of the 2d symmetries that is needed in order to obtain the triality. This also allows us to identify a discrete symmetry that is not affected by the gauge anomalies and to map it across the triality. Hence, the fact that also gauge anomalies are preserved under compactification, which we understood in Section~\ref{sec:discussion} in terms of the reduction of generalized symmetry structures, provides us a deeper understanding of the structure of 0-form symmetries in the 2d triality, and how they can be derived from 4d.

 	\def\arraystretch{1.3} 
	\setlength\tabcolsep{4.5pt} 
	\begin{table}[t!]
	\centering
	\resizebox{\columnwidth}{!}{%
	\begin{tabular}{|c||c|c||c|c|c|c|c|c|c|c|c|}
	\hline
 & $SU(N_c)$ & $U(1)_V$ & $SU(N_1-3)_L$ & $SU(N_1)_R$ & $U(1)_1$ & $U(1)_2$ & $U(1)_3$ & $U(1)_{R_3}$ & $U(1)_A$ & $U(1)_I$ & $U(1)_t$  \\ \hline \hline
 $Q_1$ & $\Box$ & $1$ & $\bf 1$ & $\bf 1$ & $0$ & $1$ & $0$ & $1-N_2$ & 1 & 0 & 0 \\ \hline
 $Q_2$ & $\Box$ & $1$ & $\bf 1$ & $\bf 1$ & $0$ & $0$ & $1$ & $1+N_3$ & 1 & 0 & 0 \\ \hline
 $Q_3$ & $\Box$ & $1$ & $\bf 1$ & $\bf 1$ & $3-N_1$ & $-1$ & $-1$ & 1 & 1 & 0 & 0 \\ \hline
 $Q_{i>4}$ & $\Box$ & $1$ & $\overline{\Box}$ & $\bf 1$ & $1$ & $0$ & $0$ & 1 & 1 & 0 & 0 \\ \hline
 $\widetilde{Q}$ & $\overline{\Box}$ & $-1$ & $\bf 1$ & $\Box$ & $0$ & $0$ & $0$ & 0 & 1 & 0 & 0 \\ \hline
 $\Omega^+$ & $\bf 1$ & $N_c$ & $\bf 1$ & $\bf 1$ & $0$ & $0$ & 0 & $3$ & $0$ & $1$ & $1$ \\ \hline
 $\Omega^-$ & $\bf 1$ & $-N_c$ & $\bf 1$ & $\bf 1$ & $0$ & $0$ & 0 & $1$ & $0$ & $1$ & $-1$ \\ \hline
 $M_1$ & $\bf 1$ & $0$ & $\bf 1$ & $\overline{\bf \Box}$ & $0$ & $-1$ & 0 & $1+N_2$ & $-2$ & $0$ & $0$ \\ \hline
	\end{tabular}
	}%
	\caption{A different parametrization of the symmetries for the 4d $U(N_c)$ SQCD with $N_1$ flavors (\emph{twist 3}). Here $N_1+N_2-N_3=2N_c$.
	\label{tab:4dchargetri}
	}
\vspace{0.2cm}
	\resizebox{\columnwidth}{!}{%
	\begin{tabular}{|c||c|c||c|c|c|c|c|c|c|c|c|}
	\hline
 & $SU(\widetilde{N}_c)$ & $U(1)_{\widetilde{V}}$ & $SU(N_1-3)_L$ & $SU(N_1)_R$ & $U(1)_1$ & $U(1)_2$ & $U(1)_3$ & $U(1)_{R_3}$ & $U(1)_A$ & $U(1)_I$ & $U(1)_t$  \\ \hline \hline
 $q_1$ & $\Box$ & $1$ & $\bf 1$ & $\bf 1$ & $0$ & $-1$ & $0$ & $1+N_2$ & $-1$ & 0 & 0 \\ \hline
 $q_2$ & $\Box$ & $1$ & $\bf 1$ & $\bf 1$ & $0$ & $0$ & $-1$ & $1-N_3$ & $-1$ & 0 & 0 \\ \hline
 $q_3$ & $\Box$ & $1$ & $\bf 1$ & $\bf 1$ & $N_1-3$ & $1$ & $1$ & 1 & $-1$ & 0 & 0 \\ \hline
 $q_{i>4}$ & $\Box$ & $1$ & $\Box$ & $\bf 1$ & $-1$ & $0$ & $0$ & 1 & $-1$ & 0 & 0 \\ \hline
 $\tilde{q}$ & $\overline{\Box}$ & $-1$ & $\bf 1$ & $\overline{\Box}$ & $0$ & $0$ & $0$ & 0 & $-1$ & 0 & 0 \\ \hline
 $\widetilde{\Omega}^+$ & $\bf 1$ & $\widetilde{N}_c$ & $\bf 1$ & $\bf 1$ & $0$ & $0$ & 0 & $3$ & $0$ & $1$ & $1$ \\ \hline
 $\widetilde{\Omega}^-$ & $\bf 1$ & $-\widetilde{N}_c$ & $\bf 1$ & $\bf 1$ & $0$ & $0$ & 0 & $1$ & $0$ & $1$ & $-1$ \\ \hline
 $\widetilde{M}_2$ & $\bf 1$ & $0$ & $\bf 1$ & $\Box$ & $0$ & $0$ & 1 & $1+N_3$ & $2$ & $0$ & $0$ \\ \hline
 $\widetilde{M}_3$ & $\bf 1$ & $0$ & $\bf 1$ & $\Box$ & $3-N_1$ & $-1$ & $-1$ & $1$ & $2$ & $0$ & $0$ \\ \hline
 $\widetilde{M}_{i>4}$ & $\bf 1$ & $0$ & $\overline{\Box}$ & $\Box$ & $1$ & $0$ & 0 & $1$ & $2$ & $0$ & $0$ \\ \hline
	\end{tabular}
	}%
	\caption{The $U(\widetilde{N}_c=N_1-N_c)$ Seiberg dual to Table~\ref{tab:4dchargetri}.
	\label{tab:4dchargetridual}
	}
	\def\arraystretch{1}
	\end{table}

For uniformity with the existing literature on the 2d $\mathcal{N}=(0,2)$ triality, in this section we denote the total number of flavors $Q$, $\widetilde{Q}$ of  4d SQCD by $N_1$ rather than $N_f$. We also need to perform yet a different twist, which we call \emph{twist 3}, as follows. We decompose the fundamental chirals $Q_i$ into four sets $Q_1$, $Q_2$, $Q_3$ and $Q_i$ for $i=4,\cdots,N_1$. Moreover, we introduce some singlets $M_1$ flipping the meson $Q_1\widetilde{Q}$. Accordingly, we parametrize the symmetries as in Table~\ref{tab:4dchargetri}, denoting the R-symmetry by $U(1)_{R_3}$. In particular, the symmetries $U(1)_{R_3}$, $U(1)_t$ and $U(1)_i$ for $i=1,2,3$ are chosen to be non-anomalous provided that the following constraint is satisfied,
\begin{align}\label{eq:constrtri}
N_1+N_2-N_3=2N_c\,,
\end{align}
while as usual $U(1)_A$ has a mixed anomaly with the non-abelian part of the gauge group that breaks it to $(\mathbb{Z}_{2N_1})_{A}$, and a combination of $U(1)_A$, and $U(1)_I$ is involved in the ABJ anomaly with the abelian part of the gauge group, see Eq.~\eqref{gaugetri}. Moreover, we have the 2-group anomalies \eqref{2grouptri}, of which we will only need the part involving the R-symmetry for the purposes of the twisted compactification,
	\begin{align}\label{2grouptriR}
	 \CI_6\big|_{\text{2group}}\,\supset \,& c_1(B)c_1({{R_3}}) \Big(
	  \frac{1}{2} (4-N_1+N_2^2+N_3^2)c_1({{R_3}}) + c_1(A) (N_1-N_2+N_3)  \nonumber\\
	  &\qquad\qquad\qquad+ 2 c_1(I) + 2 c_1(t)   - N_2 c_1({U(1)_2}) + N_3 c_1({U(1)_3})  \Big)\,.
  \end{align}
	
We accordingly perform a similar splitting of the chirals in the Seiberg dual theory, where now we only have the singlets $\widetilde{M}_2$, $\widetilde{M}_3$ and $\widetilde{M}_i$ for $i=4,\cdots,N_1$ because the singlets $M_1$ we have added give a mass to $\widetilde{M}_1$ in the dual. The charges of the dual fields are summarized in Table~\ref{tab:4dchargetridual}, where the symmetry map can as usual be determined by comparing the charges of the operators that map to each other under the duality. In particular one has
\begin{align}\label{eq:seibergmapU1V}
F_{\widetilde{V}}=\frac{N_c}{N_1-N_c}F_V-\frac{N_1}{N_1-N_c}F_A\,,
\end{align}
familiar from \eqref{curvemap}.
As usual, with this assignment of charges one can match the anomaly polynomials of the Seiberg dual theories, where again it is crucial to consider that $U(1)_A$ is broken to $(\mathbb{Z}_{2N_1})_{A}$ by the gauge anomaly.

As we have already seen, when we perform a twisted $S^2$ compactification by an R-symmetry under which some of the 4d chirals have a negative R-charge, we do not get a single 2d theory but rather a sum of theories, each of which corresponds to a sector with a different magnetic flux for the gauge symmetry through the $S^2$. However, it is possible to map the theories obtained sector by sector across Seiberg duality. Indeed, as mentioned around \eqref{eq:T2S2UNneg}, we can refine the $T^2\times S^2$ partition function with an FI parameter $g$ which appears differently in each flux sector, so that the matching of this refined partition function across Seiberg duality implies relations between the distinct flux sectors of the dual theories. 
In order to recover the triality, we consider the part of the $T^2\times S^2$ partition function with a trivial $g$ dependence, which receives contribution only from the zero gauge flux sector on each side of the duality. This implies a matching of the $T^2$ partition functions associated to these zero flux sectors that we can interpret as a 2d duality, pictured in Figure~\ref{fig:dualityframes}. From the R-charges in Table~\ref{tab:4dchargetri} we can read off the matter content of the 2d theory and the charge assignment under the same symmetries that we defined in 4d (see Table~\ref{tab:2dchargetri}).\footnote{~Notice that here we are treating the axial symmetry as a $U(1)_A$ symmetry. As we shall see momentarily, this is broken by gauge anomalies to the discrete group $\mathbb{Z}_{2(N_1-N_c)}$ which is a different group than the $\mathbb{Z}_{2N_1}$ that is preserved in 4d. However, the duality only requires the breaking  $U(1)_A\to \mathbb{Z}_{2(N_1-N_c),A}$.}

Similarly to what we did in the previous sections, the anomaly polynomial of this 2d theory can be reproduced by integrating that of the 4d theory over the $S^2$ and taking into account the non-trivial flux for the R-symmetry (see Appendix \ref{sec:untriality}). This includes the mixed global-gauge symmetry anomalies, which descend from the 2-group anomalies \eqref{2grouptriR} in 4d,
\begin{align}\label{eq:trigaugean}
\CI_4\big|_{\text{gauge}} &= c_1(B)\left(N_2c_1(U(1)_2)-N_3c_1(U(1)_3)+\left(N_1-N_2^2-N_3^2-4\right)c_1(R_3)\right.\nn
&\qquad\qquad\left.-(N_1-N_2+N_3)c_1(A)-2(c_1(I)+c_1(t))\right)\,.
\end{align}
Moreover, the $SU(2)_{\text{ISO}}$ isometry symmetry of the two-sphere is identified with a subgroup of the accidental $SU(N_2)_L\times SU(N_3)_L\times SU(2)_\Gamma$ flavor symmetry that appears in the 2d model. The embedding is similar to those of the previous examples; specifically we consider the $SU(2)$ subgroups of $SU(N_2)_L$ and $SU(N_3)_L$ such that their fundamental representations are mapped to the $N_2$-dimensional and $N_3$-dimensional representations of $SU(2)$ respectively, and we consider the diagonal combination of these two $SU(2)$'s and $SU(2)_\Gamma$ which gets identified with $SU(2)_{\text{ISO}}$. At the level of the Chern classes this implies,
\begin{align}
\label{ident}
\begin{split}
c_2(SU(N_i)_L)&=\frac{N_i(N_i^2-1)}{6}c_2(SU(2)_{\text{ISO}})\,,\quad i=2,3\\
c_2(SU(2)_\Gamma)&=c_2(SU(2)_{\text{ISO}})\,.
\end{split}
\end{align}

Another observation that we can make from Table~\ref{tab:2dchargetri} is that the symmetry $U(1)_1$ ends up acting trivially in 2d, while $U(1)_t$ is redundant since it coincides with $U(1)_I$. Moreover, one combination of the remaining abelian symmetries including the R-symmetry is anomalous due to \eqref{eq:trigaugean}, which leaves us with four independent non-anomalous $U(1)$ symmetries. In order to conform with some previous literature on the 2d triality, we decide to solve the anomaly cancellation constraint by redefining the symmetries as follows,\footnote{~In particular, the R-symmetry coincides with the superconformal R-symmetry as shown in \cite{Gadde:2013lxa}.}
\begin{align}
\label{solvecancel}
\begin{split}
F_2&=-F_a-F_b+\frac{N_1 (N_2-3)+N_2 (N_2+N_3+1)-N_3}{N_1+N_2+N_3}F_{R'}\,, \\
F_3&=-F_a+F_c-\frac{N_3 (N_1+N_2+1)+N_1-N_2+N_3^2}{N_1+N_2+N_3}F_{R'}\,, \\ 
F_A&=F_a+\frac{N_1-N_2+N_3}{N_1+N_2+N_3}F_{R'}\,, \\ 
F_I&=-\frac{1}{2}\left(N_1 F_a+N_2 F_b+N_3 F_c\right)-2F_{R'}\,,  
\end{split}
\end{align}
and $F_{R_3}=F_{R'}$, where the equalities are between the background field strengths. 
However, there is a discrete subgroup of the anomalous $U(1)$ symmetry which is not anomalous, which we can parameterize such that the charge assignment of the fields is the same as under the original $U(1)_A$ symmetry. From \eqref{eq:trigaugean} we then see that this would-be abelian symmetry has the gauge anomaly,
\begin{align}
\CI_4\big|_{\text{gauge}} &= -(N_1-N_2+N_3)c_1(B)c_1(A)=-2(N_1-N_c)c_1(B)c_1(A)\,.
\end{align}
This implies that $U(1)_A$ is broken by the gauge anomaly to the discrete subgroup,
\begin{equation}
U(1)_A\to \mathbb{Z}_{2(N_1-N_c),A}\,.
\end{equation}
The charges of the fields under these new symmetries are summarized in Table~\ref{tab:2dchargetrinew}. Note in particular that the 4d non-invertible symmetry, which on general grounds we argued in Section~\ref{sec:intro2d} should reduce to at least an invertible $\mathbb{Q}/\mathbb{Z}$ symmetry in 2d, evidently enhances back to a full $U(1)$, which mixes with the other abelian symmetries to yield the non-anomalous symmetries listed in the table.  

 	\def\arraystretch{1.3} 
	\setlength\tabcolsep{4.5pt} 
	\begin{table}[t!]
	\centering
	\resizebox{\columnwidth}{!}{%
	\begin{tabular}{|c||c|c||c|c|c|c|c|c|c|c|c|c|c|}
	\hline
 & $SU(N_c)$ & $U(1)_V$ & $SU(N_2)_L$ & $SU(N_3)_L$ & $SU(2)_\Gamma$ & $SU(N_1)_R$ & $U(1)_1$ & $U(1)_2$ & $U(1)_3$ & $U(1)_{R_3}$ & $U(1)_A$ & $U(1)_I$ & $U(1)_t$  \\ \hline \hline
 $P$ & $\Box$ & $1$ & $\overline{\Box}$ & $\bf 1$ & $\bf 1$ & $\bf 1$ & $0$ & $1$ & $0$ & $1-N_2$ & 1 & 0 & 0 \\ \hline
 $\Psi$ & $\Box$ & $1$ & $\bf 1$ & $\overline{\Box}$ & $\bf 1$ & $\bf 1$ & $0$ & $0$ & $1$ & $N_3$ & 1 & 0 & 0 \\ \hline
 $\Phi$ & $\overline{\Box}$ & $-1$ & $\bf 1$ & $\bf 1$ & $\bf 1$ & $\Box$ & $0$ & $0$ & $0$ & 0 & 1 & 0 & 0 \\ \hline
 $\Gamma$ & $\bf 1$ & $N_c$ & $\bf 1$ & $\bf 1$ & $\Box$ & $\bf 1$ & $0$ & $0$ & 0 & $2$ & $0$ & $1$ & $1$ \\ \hline
 $\mu$ & $\bf 1$ & $0$ & $\Box$ & $\bf 1$ & $\bf 1$ & $\overline{\Box}$ & $0$ & $-1$ & 0 & $N_2$ & $-2$ & $0$ & $0$ \\ \hline
	\end{tabular}
	}%
	\caption{Field content and charges of the 2d theory obtained from the $S^2$ compactification of 4d $U(N_c)$ SQCD with $N_1$ flavors, with \emph{twist 3}. Here $N_1+N_2-N_3=2N_c$.}
	\label{tab:2dchargetri}
\vspace{0.2cm}
	\resizebox{\columnwidth}{!}{%
	\begin{tabular}{|c||c|c||c|c|c|c|c|c|c|c|c|c|c|}
	\hline
 & $SU(\widetilde{N}_c)$ & $U(1)_{\widetilde{V}}$ & $SU(N_2)_L$ & $SU(N_3)_L$ & $SU(2)_\Gamma$ & $SU(N_1)_R$ & $U(1)_1$ & $U(1)_2$ & $U(1)_3$ & $U(1)_{R_3}$ & $U(1)_A$ & $U(1)_I$ & $U(1)_t$  \\ \hline \hline
 $\widetilde{\Psi}$ & $\Box$ & $1$ & $\Box$ & $\bf 1$ & $\bf 1$ & $\bf 1$ & $0$ & $-1$ & $0$ & $N_2$ & $-1$ & 0 & 0 \\ \hline
 $\widetilde{P}$ & $\Box$ & $1$ & $\bf 1$ & $\Box$ & $\bf 1$ & $\bf 1$ & $0$ & $0$ & $-1$ & $1-N_3$ & $-1$ & 0 & 0 \\ \hline
 $\widetilde{\Phi}$ & $\overline{\Box}$ & $-1$ & $\bf 1$ & $\bf 1$ & $\bf 1$ & $\overline{\Box}$ & $0$ & $0$ & $0$ & 0 & $-1$ & 0 & 0 \\ \hline
 $\widetilde{\Gamma}$ & $\bf 1$ & $\widetilde{N}_c$ & $\bf 1$ & $\bf 1$ & $\Box$ & $\bf 1$ & $0$ & $0$ & 0 & $2$ & $0$ & $1$ & $1$ \\ \hline
 $\tilde{\mu}$ & $\bf 1$ & $0$ & $\bf 1$ & $\overline{\Box}$ & $\bf 1$ & $\Box$ & $0$ & $0$ & 1 & $N_3$ & $2$ & $0$ & $0$ \\ \hline
	\end{tabular}
	}%
	\caption{Field content and charges of the 2d theory obtained from the $S^2$ compactification of the Seiberg dual 4d $U(\widetilde{N}_c=N_1-N_c)$ SQCD with $N_1$ flavors, with \emph{twist 3}.
	\label{tab:2dchargetridual}
	}
\vspace{0.2cm}
	\resizebox{\columnwidth}{!}{%
	\begin{tabular}{|c||c|c||c|c|c|c|c|c|c|c|c|}
	\hline
 & $SU(N_c)$ & $U(1)_V$ & $SU(N_2)_L$ & $SU(N_3)_L$ & $SU(2)_\Gamma$ & $SU(N_1)_R$ & $U(1)_a$ & $U(1)_b$ & $U(1)_c$ & $U(1)_{R'}$ & $\mathbb{Z}_{2(N_1-N_c),A}$ \\ \hline \hline
 $P$ & $\Box$ & $1$ & $\overline{\Box}$ & $\bf 1$ & $\bf 1$ & $\bf 1$ & $0$ & $-1$ & $0$ & $\frac{N_2-N_1+N_3}{N_1+N_2+N_3}$ & 1 \\ \hline
 $\Psi$ & $\Box$ & $1$ & $\bf 1$ & $\overline{\bf \Box}$ & $\bf 1$ & $\bf 1$ & $0$ & $0$ & $1$ & 0 & 1 \\ \hline
 $\Phi$ & $\overline{\Box}$ & $-1$ & $\bf 1$ & $\bf 1$ & $\bf 1$ & $\Box$ & $1$ & $0$ & $0$ & $\frac{N_1+N_3-N_2}{N_1+N_2+N_3}$ & 1 \\ \hline
 $\Gamma$ & $\bf 1$ & $N_c$ & $\bf 1$ & $\bf 1$ & $\Box$ & $\bf 1$ & $-\frac{N_1}{2}$ & $-\frac{N_2}{2}$ & $-\frac{N_3}{2}$ & 0 & 0 \\ \hline
 $\mu$ & $\bf 1$ & $0$ & $\Box$ & $\bf 1$ & $\bf 1$ & $\overline{\Box}$ & $-1$ & $1$ & 0 & $\frac{N_1+N_2-N_3}{N_1+N_2+N_3}$ & $-2$ \\ \hline
	\end{tabular}
	}%
	\caption{The independent non-anomalous symmetries of the 2d theory in Table~\ref{tab:2dchargetri}.
	\label{tab:2dchargetrinew}
	}
\vspace{0.2cm}
	\resizebox{\columnwidth}{!}{%
	\begin{tabular}{|c||c|c||c|c|c|c|c|c|c|c|c|}
	\hline
 & $SU(\widetilde{N}_c)$ & $U(1)_{\widetilde{V}}$ & $SU(N_2)_L$ & $SU(N_3)_L$ & $SU(2)_\Gamma$ & $SU(N_1)_R$ & $U(1)_a$ & $U(1)_b$ & $U(1)_c$ & $U(1)_{R'}$ & $\mathbb{Z}_{2(N_1-N_c),A}$ \\ \hline \hline
 $\widetilde{\Psi}$ & $\Box$ & $1$ & $\Box$ & $\bf 1$ & $\bf 1$ & $\bf 1$ & $0$ & $0$ & $1$ & $0$ & $-2-\frac{N_c}{N_1-N_c}$ \\ \hline
 $\widetilde{P}$ & $\Box$ & $1$ & $\bf 1$ & $\Box$ & $\bf 1$ & $\bf 1$ & $0$ & $-1$ & $0$ & $\frac{N_2-N_1+N_3}{N_1+N_2+N_3}$ & $-2-\frac{N_c}{N_1-N_c}$ \\ \hline
 $\widetilde{\Phi}$ & $\overline{\Box}$ & $-1$ & $\bf 1$ & $\bf 1$ & $\bf 1$ & $\overline{\Box}$ & $-1$ & $1$ & $-1$ & $\frac{N_1+N_2-N_3}{N_1+N_2+N_3}$ & $\frac{N_c}{N_1-N_c}$ \\ \hline
 $\widetilde{\Gamma}$ & $\bf 1$ & $\tilde{N}_c$ & $\bf 1$ & $\bf 1$ & $\Box$ & $\bf 1$ & $\frac{N_1}{2}$ & $-\frac{N_1+N_3}{2}$ & $\frac{N_1-N_2}{2}$ & $0$ & 0 \\ \hline
 $\tilde{\mu}$ & $\bf 1$ & $0$ & $\bf 1$ & $\overline{\Box}$ & $\bf 1$ & $\Box$ & $1$ & $0$ & 1 & $\frac{N_1+N_3-N_2}{N_1+N_2+N_3}$ & 2 \\ \hline
	\end{tabular}
	}%
	\caption{The independent non-anomalous symmetries of the 2d theory in Table~\ref{tab:2dchargetridual}.
	\label{tab:2dchargetridualnew}
	}
	\end{table}
	
We can proceed similarly with the compactification of the Seiberg dual theory. The field content and the charges under the symmetries descending from 4d are summarized in Table~\ref{tab:2dchargetridual}. As before, the anomalies of this 2d theory can be obtained from the 4d anomalies by integration over the $S^2$, including the mixed global-gauge symmetry anomalies descending from the 4d 2-group anomalies for the R-symmetry, leading to,
\begin{align}
\widetilde{\CI}_4\big|_{\text{gauge}} &=c_1(\widetilde{B})\left(N_2c_1(U(1)_2)-N_3c_1(U(1)_3)+(N_1-N_2^2N_3^2-4)c_1(R_3)\right.\nn
&\qquad\qquad\left.+(N_1+N_2-N_3)c_1(A)-2(c_1(I)+c_1(t))\right)\,.
\end{align}
Again we can notice that the symmetry $U(1)_1$ now acts trivially, and that $U(1)_t$ is not independent since it acts as $U(1)_I$. We decide to parameterize the non-anomalous symmetries as in the original theory so to facilitate the matching across the duality, which is achieved by specifying,
\begin{align}
\begin{split}
F_2&=-F_a+U(1)_b-2F_c+\frac{(N_2+1) (N_1+N_2)+(N_2-1) N_3}{N_1+N_2+N_3}F_{R'}\,,\\
F_3&=-F_a+2F_b-F_c-\frac{N_1 (N_3-3)+N_3 (N_2+N_3)-N_2+N_3}{N_1+N_2+N_3}F_{R'}\,,\\
F_A&=F_a-F_b+F_c+\frac{N_1+N_2-N_3}{N_1+N_2+N_3}F_{R'}\,,\\
F_I&=\frac{1}{2}\left(N_1F_a-(N_1+N_3)F_b+(N_1-N_2)F_c\right)-2F_{R'}\,.
\end{split}
\end{align}
We would like to parameterize the non-anomalous discrete subgroup of the anomalous abelian symmetry in such a way that it maps to the one we previously defined in the original theory, see Table~\ref{tab:2dchargetrinew}. For this, we exploit the mapping \eqref{eq:seibergmapU1V} of the $U(1)_A$ symmetry across Seiberg duality. The result is the charge assignment summarized in Table~\ref{tab:2dchargetridualnew}.\footnote{~While this charge assignment might look strange due to the fractional charges, one can verify that the charges of  gauge invariant operators are all integers so that the actual symmetry is indeed $\mathbb{Z}_{2(N_1-N_c),A}$.} Computing the gauge anomaly for this symmetry, we find,
\begin{align}
\begin{split}
\widetilde{\CI}_4\big|_{\text{gauge}} &=\frac{2N_1(N_2-N_3)-N_c(N_1+N_2-N_3)}{N_1-N_c}c_1(\widetilde{B})c_1(A) \\
&=-2(N_1-N_c)c_1(\widetilde{B})c_1(A)\,,
\end{split}
\end{align}
which indeed tells us that such an anomaly is trivialized if the symmetry is restricted to be $\mathbb{Z}_{2(N_1-N_c),A}$.

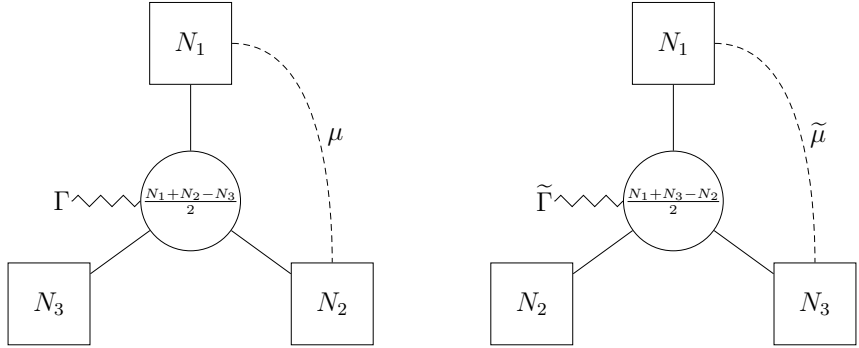
\begin{figure}
\centering
\resizebox{0.74\textwidth}{!}{
 \begin{tikzpicture}[square/.style={regular polygon,regular polygon sides=4}]
 
 \begin{scope}
         \node[minimum size=1.8cm] at (0,2.45) [draw,square]  (N1) {\large $N_1$};
          \node[minimum size=1.8cm] at (2.2,-1.6) [draw,square]  (N2) {\large $N_2$};
         \node[minimum size=1.8cm] at (-2.2,-1.6) [draw,square]  (N3) {\large $N_3$};
           \node[minimum size=1.55cm] at (0,0) [draw,circle] (Nc) {} ;
           
                   \node at (0,0) {\footnotesize ${{\frac{N_1+N_2-N_3}{2}}}$};  
                     \node at (-2,0.0) { \large ${\Gamma}$};
      		 \node at (2.25,1) { \large ${\mu}$};  
          
          \draw[decorate, decoration={zigzag}] (Nc) -- (-1.85,0);

   	 \draw[black](N1) -- (Nc) ;
	\draw[black](N2) -- (Nc) ;
	\draw[black](N3) -- (Nc) ;
	
           \draw[densely dashed] (N1) to [out=0,in=90,looseness=0.9] (N2);
\end{scope}
 \begin{scope}[xshift=7.5cm]
          \node[minimum size=1.8cm] at (0,2.45) [draw,square]  (N1) {\large $N_1$};
          \node[minimum size=1.8cm] at (2.2,-1.6) [draw,square]  (N3) {\large $N_3$};
         \node[minimum size=1.8cm] at (-2.2,-1.6) [draw,square]  (N2) {\large $N_2$};
           \node[minimum size=1.55cm] at (0,0) [draw,circle] (Nc) {} ;
           
                   \node at (0,0) {\footnotesize ${\frac{N_1+N_3-N_2}{2}}$};  
              \node at (-2,0.05) { \large $\widetilde{\Gamma}$};  
                          \node at (2.25,1.03) { \large $\widetilde{\mu}$};  
          
          \draw[decorate, decoration={zigzag}] (Nc) -- (-1.85,0);

   	 \draw[black](N1) -- (Nc) ;
	\draw[black](N2) -- (Nc) ;
	\draw[black](N3) -- (Nc) ;
	
           \draw[densely dashed] (N1) to [out=0,in=90,looseness=0.9] (N3);
 \end{scope}
	
  \end{tikzpicture}
  }
  \caption{The two triality frames reached by compactifying the 4d Seiberg-dual pairs with $N_f =N_1$ with {\it twist 3} in the $\vec{m}_G=0$ sector, with symmetries given in Tables~\ref{tab:2dchargetrinew} and \ref{tab:2dchargetridualnew}. 
  \label{fig:dualityframes}
  }
  \end{figure}
	
In this way, the anomaly polynomials of the theories in Tables~\ref{tab:2dchargetrinew} and \ref{tab:2dchargetridualnew} perfectly match, which is a highly non-trivial test of the 2d duality. 
These two duality frames are pictured in Figure~\ref{fig:dualityframes}.
Moreover, one can observe that applying this duality twice one does not go back to the original theory, but rather we obtain a third dual frame. This is because the duality acts as a cyclic permutation of the three elements $(N_1,N_2,N_3)$. Hence, in 2d we actually obtain a triality rather than a duality. 

Let us conclude by commenting about the compactification of the 4d dual theories in a non-vanishing gauge flux sector. First of all, one can still derive the content of the 2d theories and reproduce their anomalies (including the gauge anomalies) by compactifying  the anomaly polynomials of the 4d theories on $S^2$, similarly to what we have already discussed at length. Moreover, one can match between the two 2d theories the anomalies of the non-anomalous symmetries that are already manifest from 4d plus the $SU(2)_{\text{ISO}}$ isometry of the sphere, in accordance with the fact that the $T^2\times S^2$ partition function of the 4d dual theories matches in each magnetic flux sector. However, there are several more accidental symmetries (inside of which $SU(2)_{\text{ISO}}$ is embedded) as compared to the zero flux case, due to the fact that some of the 4d fields produce multiple fields in 2d because of the higher magnetic flux. When comparing the theories in the same flux sector on each side of the 4d duality, these accidental symmetries turn out to be different, thus preventing the 2d theories from being dual. In other words, the emergence of accidental non-abelian continuous symmetries in the compactification seems to spoil the duality in 2d. Notice that this is not in conflict with the matching of the $T^2\times S^2$ partition function mentioned above, since these symmetries are not manifest in 4d and thus we cannot refine the partition function by fugacities for them.

\subsection*{Acknowledgements}

We thank Yichul Choi, Christian Copetti, Thomas Dumitrescu, Po-Shen Hsin, Ken Intriligator, Zohar Komargodski, Pierluigi Niro, Shlomo Razamat, Shu-Heng Shao, and Siwei Zhong for valuable discussions. EN and YZ are partially supported by WPI Initiative, MEXT, Japan at IPMU, the University of Tokyo. MS is partially supported by the ERC Consolidator Grant \#864828 “Algebraic Foundations of Supersymmetric Quantum Field Theory (SCFTAlg)” and by the Simons Collaboration for the Nonperturbative Bootstrap under grant \#494786 from the Simons Foundation. GZ is partially supported by the Israel Science Foundation under grant no. 759/23. OS is supported by the Mani L. Bhaumik Institute for Theoretical Physics at UCLA.

\begin{appendix}

\section{Conventions}
\label{sec:appconventions}

\subsection{Group Theory and Lagrangians}

Let $T^a$ be the Hermitian generators of a Lie algebra satisfying $[T^a,T^b]=if^{abc}T^c$, where $f^{abc}$ are the real structure constants.  The quadratic Casimir $C(\CR)$ and index $T(\CR)$ of a given representation $\CR$ are defined by,
\begin{align}
	T^a T^a = C(\CR)\mathbbm{1}\,,\qquad \tr_{\CR} \,T^a T^b = T(\CR) \delta^{ab}\,,
	\end{align}
which are related as $\text{dim}(\CR) C(\CR) = \text{dim}(\text{adj}) T(\CR)$. In our chosen normalization, the generators of $SU(N)$ in the fundamental and adjoint representations satisfy $T(\Box) = 1/2$ and $T(\text{adj})=N$, respectively.

The non-abelian field strength $F_{\mu\nu}^a$ is given by,
	\begin{align}
	F_{\mu\nu}^a = \partial_\mu A_\nu^a - \partial_\nu A_\mu^a + f^{abc} A_\mu^b A_\nu^c \quad\ \leftrightarrow\quad\  F_{\mu\nu} = \partial_\mu A_\nu - \partial_\nu A_\mu - i [A_\mu,A_\nu]\,.
	\end{align}
We are using a notation in which an adjoint field $\phi$ is given by $\phi = \phi^a T^a$, and note that in the language of differential forms, $A = A_\mu dx^\mu = A_\mu^a T^a dx^\mu$, and $F = \frac{1}{2} F_{\mu\nu} dx^\mu dx^\nu = \frac{1}{2} F_{\mu\nu}^a T^a dx^\mu dx^\nu$. Then,  $F = dA - i A \wedge A$.
The covariant derivative for a field $Q$ that transforms in a representation $\CR$ is,
	\begin{align}
	D_\mu Q^\ell = \partial_\mu Q^\ell - i A_\mu^a (T^a_{\CR})^{\ell}_{\ k} Q^k\qquad\leftrightarrow\qquad D_\mu Q = \partial_\mu Q - i A_\mu^\CR Q
	\end{align}
where $A_\mu^\CR = A_\mu^a (T_{\CR}^a)^i_{\ j}$, and where $T^a=(T_{\CR}^a)^{\ell}_{\ k}$ is the symmetry group generator transforming in the $(\text{dim}\CR)\times( \text{dim}\CR)$ representation. In the adjoint representation the generators are $(T^a_{\text{adj}})^{bc}  =- i f^{abc}$, so that the covariant derivative of an adjoint field $\phi = \phi^a T^a$ is,
	\begin{align}
	D_\mu \phi^a = \partial_\mu \phi^a + f^{abc} A_\mu^b \phi^c\qquad \leftrightarrow\qquad D_\mu \phi = \partial_\mu \phi - i [ A_\mu,\phi]\,.
	\end{align}
In these conventions, the kinetic term for the non-abelian field strength of an $SU(N)$ gauge theory is,
$\CL \supset  - \frac{1}{2g^2}  \tr F^{\mu\nu} F_{\mu\nu}  = - \frac{1}{4g^2} F^{\mu\nu  a} F_{\mu\nu}^a$, 
although sometimes we absorb the coupling $g$ into the definition of $A$.

\subsection{Characteristic Classes and Anomaly Polynomials}

Let $F$ denote the Hermitian field strength of a $G=SU(N)$ or $U(N)$ bundle. The $2k$-form Chern classes $c_k(G)$ are polynomials in $F$ of degree $k$, begining with,
	\begin{align}
	c_1(G) = \frac{\tr F}{2\pi} \,,\qquad c_2(G) = \frac{1}{8\pi^2} \left[    \tr F^2-(\tr F)^2  \right]\,.
	\end{align}
Here and throughout the paper we suppress the wedge products. 
For an $SU(N)$ bundle $\tr F_{SU(N)}=0$, so $c_1({SU(N)})=0$ and $c_2(SU(N)) = \frac{1}{8\pi^2} \tr F_{SU(N)}^2$. 

One can embed $SU(N)$ in $U(N)$ as $U(N) = [SU(N) \times U(1)_V] /\mathbb{Z}_N$, with $U(1)_V/\mathbb{Z}_{N}=U(1)_B$, so that the $U(1)$ curvatures are related as $F_B = N F_V$. This embedding amounts to the following identification of field strengths,
\ba{
F_{SU(N)} = F_{U(N)} - \frac{1}{N} \mathbbm{1}_N  F_{B}\,,\qquad \tr F_{U(N)} = F_{B}\,,
}
so that the Chern classes are related as,
	\begin{align}
	\label{c2s}
	c_2({SU(N)})  = c_2({U(N)}) + \frac{1}{2} \left( 1 - \frac{1}{N} \right) c_1(U(1)_{B})^2\,,
	\end{align}
with $c_1(U(1)_B)=F_B/(2\pi)$. 
	
Perturbative chiral anomalies for 0-form symmetries in even $d$ spacetime dimensions are encapsulated by the $(d+2)$-form anomaly polynomial $\CI_{d+2}$, defined for a Weyl fermion in a representation of a symmetry $G$ by \cite{Alvarez-Gaume:1983ihn},
	\begin{align}
	\CI_{d+2} = \left[  \hat{A}(R)  \text{ch}(F)\right]_{d+2}\,.
	\end{align}
Here the $(d+2)$ subscript instructs us to extract the $(d+2)$-form in the expansion of the curvatures. 
 $\hat{A}(R)$ is the $A$-roof genus, whose expansion in terms of the curvature $R$ of the spacetime tangent bundle is $\hat{A}(R) = 1 - \frac{1}{24} p_1(R) + ...$, and $\text{ch}(F)$ is the Chern character for the $G$-bundle with curvature $F$.
The Pontryagin classes $p_k$ are $4k$-forms, with the first $p_1(R)$ for a real vector bundle given by $p_1(R) = - \frac{1}{8\pi^2} \tr R^2$.

The anomaly polynomial $\CI_{d+2}$ satisfies the descent equations,
	\begin{align} \label{descent}
	\CI_{d+2} = d\CI_{d+1}\,,\qquad \delta \CI_{d+1} = d\CI_d\,.
	\end{align}
The anomaly inflow action $\CA_{d+1}^{\text{inf}}$ and variation of the effective action $\delta S$ are given in terms of these quantities by,
	\begin{align} \label{inflow}
	\CA_{d+1}^{\text{inf}} = 2\pi \int_{M_{d+1}} \CI_{d+1}\,,\qquad \delta S = 2\pi \int_{M_d} \CI_d\,.
	\end{align}
	
\subsection{2-Groups and ABJ Anomalies}
\label{sec:abjconventions}

In this work we have focused on 4d theories and their compactifications. The anomaly polynomial of a 4d theory with an abelian flavor symmetry $F=U(1)_F$, non-abelian flavor symmetry $H=SU(N)$, and abelian gauge symmetry $G=U(1)_G$ might include the following terms,
	\begin{align}
	\CI_6 = \frac{k}{2} c_1(F) c_1(G)^2 + \frac{\kappa_F}{2}  c_1(F)^2 c_1(G)  +{\kappa_{H} }  c_2(H)  c_1(G)
	\end{align}
where the coefficients are computed by taking the traces,
	\begin{align}
k= \tr\,  U(1)_F U(1)_G^2\,,\qquad  \kappa_F = \tr \, U(1)_F^2 U(1)_G \,,\qquad \kappa_{H} = \tr \,  U(1)_GH\,.
 	\end{align}
The forms $\CI_5$ and $\CI_4$ can be obtained from \eqref{descent}, using the fact that an infinitesimal background gauge transformation acts on the fields as,
	\begin{align}
	\begin{split}
	A_F &\to A_F + d \lambda_F \,\\
	A_H &\to A_H + d\lambda_H - i [A_H, \lambda_H]\,,\qquad F_H \to F_H - i [F_H , \lambda_H]\,.
	\end{split}
	\end{align} 
Here the curvatures are given by $F_F = dA_F$, and $F_H = dA_H  - i A_H \wedge A_H$. It is also useful to note that $\tr (F_H^2) = - d \, \tr (A_H dA_H - \frac{2i}{3} A_H^3)$, where again we are suppressing wedge products.
	
	The corresponding current non-conservation equation is,
	\begin{align} \label{nonc}
	d\star j_F =  \frac{k}{8\pi^2}   F_G \wedge F_G+\frac{\kappa_{F}}{8\pi^2}  dA_F \wedge F_G  - \frac{\kappa_{H}}{16\pi^2} dA_H \wedge F_G\,,
	\end{align}
	where in our conventions a current $j$ couples to a background field $A$ in the action with a term $S \supset - \int d^4x A^\mu j_\mu = - \int A \wedge \star j$ (or for a non-abelian symmetry, $S\supset - \int d^4x A^{\mu a} j_\mu^a = - 2 \int \tr (A \wedge \star j)$). 
	
	The first term in \eqref{nonc} corresponds to the ABJ anomaly. 
The 2-group structure derives from the second two terms. The variation of the effective action is canceled by coupling to a 2-form background field $B$ which transforms as,\footnote{~For reference, our conventions differ from those in \cite{Cordova:2018cvg}, which we dub CDI, as follows. We work in Minkowski signature, so $S_{\text{CDI}} = i S$. Our currents are defined with an additional factor of $i$, and for the non-abelian symmetry, $\{A,\lambda,F\}_{\text{CDI}} = - i \{ A,\lambda,F\}$. Furthermore, they use $\tr\, T^a T^b = -1/2 \delta^{ab}$ with a minus sign different from our conventions.}
\ba{
 B\to B + d\Lambda - \frac{\kappa_F}{4\pi} \lambda_F dA_F + \frac{\kappa_{H}}{4\pi} \tr (\lambda_H dA_H)\,.
}

\section{Anomaly Matching for \texorpdfstring{$U(N_c)$}{U(Nc)} SQCD}
\label{sec:appumatch}

In this appendix we explicitly verify that the 't Hooft anomalies of the 4d $U(N_c)$ SQCD match across Seiberg duality, including the discrete $\mathbb{Z}_{2N_f}$ axial symmetry. For this purpose it is useful to use the ``primed'' symmetry basis \eqref{unprimeA} under which the fields have integer $U(1)_{R'}$ symmetry charges. We will denote the classical abelian symmetries of the electric theory by $U(1)_{V'}\times U(1)_{A'}\times U(1)_{R'}$, and those of the magnetic dual by $U(1)_{\widetilde{V}'}\times  U(1)_{\widetilde{A}'} \times U(1)_{R'}$, where in the theories with unitary gauge group, $U(1)_B=U(1)_{V'}/\mathbb{Z}_{N_c}$ is gauged on the electric side, and $U(1)_{\widetilde{B}} = U(1)_{\widetilde{V}'} /\mathbb{Z}_{N_f-N_c}$ is gauged on the magnetic side.  
By application of \eqref{unprimeR}, the $Q$ and $\widetilde{Q}$ chiral fields have zero $U(1)_{R'}$ charge, while the gaugino retains unit $U(1)_{R'}$ charge, and the $U(1)_{A'}$ and $U(1)_{V'}$ charges are unchanged from their unprimed counterparts in Table~\ref{tab:UNdetcharges}. The magnetic chiral superfields $q$ and $\widetilde{q}$ similarly have $U(1)_{R'}$ charge zero, the magnetic singlet $\widetilde{M}$ has $U(1)_{R'}$ charge 2, and the $U(1)_{\widetilde{A}'}$ and $U(1)_{\widetilde{V}'}$ charges are the same as their unprimed counterparts. 
The Dirac quantization conditions of the various classical gauge fields in this basis are as follows. 

\begin{itemize}
\item The  $U(1)_{R'}$ curvature satisfies the canonical quantization conditions $\int_{S^2} c_1(R')\in \mathbb{Z}$. 

\item Similarly,  the classical $U(1)_{A'}$ curvature satisfies $\int_{S^2} c_1(A')\in \mathbb{Z}$, as does its magnetic dual counterpart  $\int_{S^2} c_1(\widetilde{A}')\in \mathbb{Z}$. 

\item In the electric theory with $SU(N_c)$ gauge group, the background curvature for the $U(1)_{V'}$ global symmetry is canonically quantized as $\int_{S^2} c_1(V')\in \mathbb{Z}$, and similarly for the magnetic theory, $\int_{S^2} c_1(\widetilde{V}')\in \mathbb{Z}$.
 However in the electric $U(N_c)$ theory with $U(1)_B$ gauged, it is the gauge curvature $f_B = N_c f_{V'}$ which satisfies the canonical quantization condition $\int_{S^2} c_1(B)\in \mathbb{Z}$. In the magnetic $U(\widetilde{N}_c = N_f-N_c)$ theory, $U(1)_{\widetilde{B}} = U(1)_{\widetilde{V}'}/\mathbb{Z}_{N_f-N_c}$ is gauged, so that the gauge curvature $f_{\widetilde{B}} = (N_f-N_c) f_{\widetilde{V}'}$ satisfies  $\int_{S^2} c_1(\widetilde{B})\in \mathbb{Z}$. We are using lowercase letters to denote dynamical rather than background gauge fields.
 
 \item Combining \eqref{unprimeA} and \eqref{curvemap}, the claim to be verified below is that field strengths on the electric and magnetic sides of the duality are related as,
	\begin{align}\label{mapapp}
	F_{\widetilde{A}'} = - F_{A'}   + F_{R'} \,,\qquad  f_{\widetilde{B}} = f_B - N_f F_{A'}   + ({N_f-N_c}) F_{R'}    \,.
	\end{align}

\end{itemize}

The anomaly polynomial for $U(N_c)$ SQCD is evaluated in this basis as,
	\begin{align}\label{ael0}
	\begin{split}
\CI_6&= \left[  2(N_c-N_f) c_1({R'})+ 2N_fc_1(A') \right] c_2(U(N_c))\\
&+ \left[(N_c-N_f-1) c_1({R'})+    N_f c_1({A'})  \right]c_1(B)^2 \\
&  +\left[c_2(SU(N_f)_L) -c_2(SU(N_f)_R) \right]c_1(B) \\
& + \left[c_2(SU(N_f)_L) +c_2(SU(N_f)_R) \right]\left[N_c \, c_1({A'})  - N_c\,  c_1({R'}) \right] \\
&+ \tfrac{1}{3} {N_c N_f} c_1({A'})^3+ \tfrac{1}{6}{(N_c^2- 2 N_cN_f -1)} c_1({R'})^3 \\
&+ {N_c N_f} \left[ c_1({R'})^2 c_1({A'})- c_1({R'}) c_1({A'})^2 \right]\\
&+\tfrac{1}{3\cdot 16\pi^3} {N_c}  \tr\left(F_{SU(N_f)_L}^3 - F_{SU(N_f)_R}^3 \right)\,,   \\
	\end{split}
	\end{align}
and for the dual theory with $U(\widetilde{N}_c=N_f-N_c)$ gauge group,
	\begin{align} \label{amag0}
	\begin{split}
\widetilde{\CI}_6 &= 
  \left[   -2N_c c_1({{R}'})+ 2N_fc_1(\widetilde{A}')  \right] c_2(U(\widetilde{N}_c))\\
  &+ \left[  (-N_c-1)c_1({{R}'})  + N_f c_1({\widetilde{A}'})   \right]c_1({\widetilde{B}})^2\\
  &+\left[c_2(SU(N_f)_L) -c_2(SU(N_f)_R) \right] c_1(\widetilde{B})  \\
&+\left[c_2({SU(N_f)_L}) +c_2({SU(N_f)_R}) \right] \left[{(-N_f-N_c)}c_1({\widetilde{A}'}) + N_c \, c_1({{R}'}) \right]\\
&- N_f(N_f + \tfrac{1}{3} N_c)c_1(\widetilde{A}')^3 +  \tfrac{1}{6}(N_c^2- 1)c_1({R}')^3 -N_f N_c  c_1({R}')^2 c_1(\widetilde{A}') \\
&+  N_f( N_f +  N_c) c_1({R}')c_1(\widetilde{A}')^2
  +\tfrac{1}{3\cdot 16\pi^3}{N_c}\tr\left(F_{SU(N_f)_L}^3 - F_{SU(N_f)_R}^3 \right) \,.
	\end{split}
	\end{align}
The ABJ anomalies are trivialized by the subgroup $(\mathbb{Z}_2)_{R'}\times (\mathbb{Z}_{2N_f})_{A'}$ of the R symmetry and axial symmetry, and similarly by  $(\mathbb{Z}_2)_{R'}\times (\mathbb{Z}_{2N_f})_{\widetilde{A}'}$ for the magnetic theory.

One can verify that the anomalies \eqref{ael0} and \eqref{amag0} match as follows.
Substituting \eqref{mapapp} into $\widetilde{\mathcal{I}}_6$, subtracting $\widetilde{\mathcal{I}}_6$ from $\mathcal{I}_6$, and dropping the  gauge anomalies that have already been trivialized, yields an apparent mismatch  which contributes to the 5d inflow action as,
	\begin{align} \label{mismatch}
	\begin{split}
\Delta\CA_5^{\text{inf}}=\pi \int_{M_5} \Big[&- 2 \left(  (N_c-N_f)  \hat{A}_{R'}+  N_f  \hat{A}_{A'}\right) \left( (N_c-N_f+1) \beta_{R'}+N_f  \beta_{A'}  \right)  c_1(B)\\
&+  (N_c-N_f)^2 (N_c-N_f + 1)\hat{A}_{R'}  \beta_{R'}^2 +  N_f (N_f-1) \hat{A}_{A'} \beta_{A'}^2\\
&-  \left(3 N_f^2 (N_f-1) - N_c N_f (1-3N_f)\right)  \hat{A}_{R'} \beta_{A'}^2\\
& + 3   (N_c-N_f) (N_c-N_f+1)\hat{A}_{A'} \beta_{R'}^2\Big]\,.
	\end{split}
	\end{align}
We have replaced the background gauge fields for the discrete symmetries with the $\mathbb{Z}_{2N_f}$- and $\mathbb{Z}_2$-valued cocycles, 
	\begin{align}
A_{A'} \to \frac{2\pi}{2N_f} \hat{A}_{A'} \,\qquad F_{A'} \to 2\pi \text{Bock}(\hat{A}_{A'}) =2\pi  \beta_{A'}\,, \\
A_{R'} \to \pi \hat{A}_{R'} \,\qquad F_{R'} \to 2\pi \text{Bock}(\hat{A}_{R'}) = 2\pi \beta_{R'}\,.
	\end{align}
Each of the coefficients in \eqref{mismatch} can be shown to be an even integer times $\pi$ ({\it e.g.}~by noting that terms of the form $m(m-1)$  for $m\in \mathbb{Z}$ are even), so that $\Delta\CA_5^{\text{inf}} = 2\pi \mathbb{Z}$, and the apparent mismatch is trivialized.

\section{Charges of Monopole Operators in 3d}
\label{sec:appzero}

In this appendix we briefly review the formula to compute the charges  of monopole operators in a 3d $\mathcal{N}=2$ theory under the abelian global symmetries (see \emph{e.g.}~\cite{Borokhov:2002ib,Borokhov:2002cg,Borokhov:2003yu,Gaiotto:2008ak,Benna:2009xd,Bashkirov:2010kz,Cremonesi:2013lqa}), and apply it to the particular type of monopoles that show up in the compactification of 4d SQCD as  discussed in Section~\ref{sec:3dsqcd} of the main text. 

The set of allowed monopole operators in a 3d $\mathcal{N}=2$ gauge theory with gauge group $G$ is labelled by a magnetic flux $\vec{m}_G$ which lives in the co-weight lattice $\Lambda^\vee_w(G)$ of $G$ modulo Weyl transformations $W_G$, namely,
\begin{equation}
\vec{m}_G\,\in\,\Lambda^\vee_w(G)/W_G\,.
\end{equation}
Such a flux is an $r$-dimensional vector $\vec{m}_G=(m_1,\cdots,m_r)$, where $r=\mathrm{rk}(G)$ is the rank of $G$, \emph{i.e.}~the dimension of its Cartan subgroup.  The charge of a given such monopole operator under an abelian symmetry receives contributions from all the fermions in the theory. Denoting by $q_i$ the charge of the $i$-th fermion which transforms in a representation $\mathcal{R}_i$ with weight vector $\vec{\rho}_i$ of the gauge group $G$, we have,\footnote{~Here we assume that the symmetry for which we are computing the charge of the monopole is not a gauge symmetry. Otherwise, we should also consider additional contributions from Chern--Simons interactions, and moreover this gauge charge can also be non-trivial (so that the monopole is not gauge invariant) if the theory is chiral, \emph{i.e.}~the number of fermions in a representation $\mathcal{R}_i$ and those in the complex conjugate representation $\overline{\mathcal{R}}_i$ are not equal. Furthermore, we assume that there is no monopole superpotential, which would lead to a mixed Chern--Simons interaction between the symmetry and some abelian subgroup of $G$ that is not just the ordinary BF coupling with the FI parameter \cite{Pasquetti:2019uop}.}
\begin{align}
q(\vec{m}_G)=-\frac{1}{2}\sum_{i:\text{ ferm.}}q_i\sum_{\vec{\rho}_i\in\mathcal{R}_i}|\vec{\rho}_i(\vec{m}_G)|\,.
\end{align}
If the symmetry we are considering is not the R-symmetry, then the fermions inside the 3d $\mathcal{N}=2$ vector multiplets are uncharged so that the sum runs over the 3d $\mathcal{N}=2$ chiral multiplets in the theory, for which the charges of the fermions $q_i$ coincide with those of all the other components. In the case of the R-symmetry instead, the fermions in the vector multiplet have R-charge $1$, while the fermions in the $i$-th chiral multiplet have R-charge $(q_{R,i}-1)$, where $q_{R,i}$ is the R-charge of the chiral. Hence, we have that,
 \begin{align}
q_R(\vec{m}_G)=-\frac{1}{2}\sum_{i:\text{ chir.}}(q_{R,i}-1)\sum_{\vec{\rho}_i\in\mathcal{R}_i}|\vec{\rho}_i(\vec{m}_G)|-\sum_{\vec{\alpha}\in\Delta_+}|\vec{\alpha}(\vec{m}_G)|\,,
\end{align}
where $\vec{\alpha}\in\Delta_+$ are the positive roots of $G$.

As an example, let us consider  3d $U(N_c)$ SQCD with $N_f$ flavors $Q$, $\widetilde{Q}$ and determinant matter $\Omega^\pm$, whose field content and charge assignments are summarized in Table~\ref{tab:UN3ddet}. For convenience, we define a trial R-symmetry $U(1)_{R_\text{trial}}$ obtained by mixing $U(1)_R$ with all the other classical abelian symmetries,
\begin{align}
q_{R_{\text{trial}}}=q_R+q_AR_A+q_tR_t+q_IR_I\,,
\end{align}
where $q_i$ are the charges and $R_i$ the mixing coefficients for each symmetry. In this way we can keep track of the charges of the monopoles under all the symmetries simultaneously: the charge under one of the symmetries is given by the prefactor of the corresponding mixing coefficient. For a $U(N_c)$ gauge group the allowed magnetic fluxes $\vec{m}_G$ live in $\Lambda^\vee_w(U(N_c))=\mathbb{Z}^{N_c}$ modulo Weyl transformations $W_{U(N_c)}=S_{N_c}$, and the charge of a monopole under $U(1)_{R_\text{trial}}$ is computed as,
\begin{align}\label{eq:monochargeU}
q_{R_{\text{trial}}}(\vec{m}_G)&=-\frac{1}{2}\left(2+R_I+R_t-1\right)\left|\sum_{k=1}^{N_c}m_k\right|-\frac{1}{2}\left(2+R_I-R_t-1\right)\left|\sum_{k=1}^{N_c}m_k\right|\nonumber\\
&-N_f\left(\frac{N_f-N_c}{N_f}+R_A-1\right)\sum_{k=1}^{N_c}|m_k|-\sum_{k<l}^{N_c}|m_k-m_l|\,,
\end{align}
where in the first line we have the contributions of the determinant fields, and in the second line those of the flavors and of the vector. This theory also has a $U(1)_J$ topological symmetry, under which the monopoles have charge
\begin{align}
q_J(\vec{m}_G)&=\sum_{k=1}^{N_c}m_k\,.
\end{align}

The minimal monopoles that we can have for $U(N_c)$ are (up to Weyl transformations) of the form $\vec{m}_G=(\pm1,0,\cdots,0)$. These are usually denoted by $V_{\pm}$, and from the formula \eqref{eq:monochargeU} we find
\begin{align}
q_{R_{\text{trial}}}(\vec{m}_G=(\pm1,0,\cdots,0))=-N_fR_A-R_I\,,
\end{align}
meaning that they have charge $-N_f$ under $U(1)_A$, $-1$ under $U(1)_I$, and zero under all the other symmetries including $U(1)_R$. Moreover, they have charge $\pm1$ under the $U(1)_J$ topological symmetry.

The monopole operator that appeared in our discussion of the 4d to 3d compactification in Section~\ref{sec:3dsqcd}, which we called $V$, is instead the one with magnetic flux $\vec{m}_G=(1,0,\cdots,0,-1)$,
\begin{align}
q_{R_{\text{trial}}}(\vec{m}_G=(1,0,\cdots,0,-1))=2-2N_fR_A\,.
\end{align}
This reproduces the charges summarized in Table~\ref{tab:UN3ddet}, namely charge $2$ under $U(1)_R$ and $-2N_f$ under $U(1)_A$. Moreover, this monopole is uncharged under $U(1)_J$. Turning on this monopole in the superpotential forces it to have R-charge 2 and to be uncharged under all the other symmetries. This reproduces the breaking of $U(1)_A$ to $(\mathbb{Z}_{2N_f})_{A}$ that in 4d was due to the anomaly $\tr\,U(1)_ASU(N_c)^2$. In accordance with this, notice that this monopole corresponds to a flux for the $SU(N_c)$ part of the gauge group and not for its $U(1)_B$ part.

\section{Anomaly Polynomials of 4d Theories on \texorpdfstring{$S^2$}{S2}}
\label{sec:anom2d}

In this appendix we compute the  six-form anomaly polynomials of the various 4d theories whose compactification on $S^2$ are considered in the main text, as well as the four-form anomaly polynomials that result from their integration over $S^2$.    

\subsection{Flux Sectors and Bott--Cattaneo Formula}
\label{sec:bott}

Let us consider a 4d theory compactified on $S^2$. 
If a $U(1)$ symmetry $F$ of the 4d theory has $q$ units of flux through the internal space, then its first Chern class is related to the first Chern class of the 2d theory (denoted by a hat) by, 
	\begin{align}\label{c1flux}
	c_1^{4d}(F) = q e_2({S^2})+ \hat{c}_1(F)\,,
	\end{align}
where $e_2({S^2})$ is the global angular form of the $SO(3)$ isometry of the sphere, related to the Chern root $t$ of the tangent bundle to the sphere by $e_2({S^2})= \frac{t}{2}$. For the theories under consideration in Section~\ref{sec:2d}, we have introduced a non-trivial R-symmetry flux $q_R=-1$, so that this prescription amounts to replacing $c_1(R) = - e_2({S^2}) + \hat{c}_1(R)$.  By the Bott--Cattaneo formula, $e_2({S^2})$ satisfies \cite{bott1999integral},
	\begin{align}\label{bott}
	\int_{S^2} \left( e_2({S^2}) \right)^{2s+1} = 2^{-2s} \left[ p_1({SO(3)}) \right]^s\,,\qquad s = 0,1,2,\dots
	\end{align}
with integrals over even powers of the global angular form evaluating to zero. (We will also use the replacement $p_1({SO(3)} )= - 4c_2({SU(2)})$.) This prescription for taking into account fluxes through an internal $S^2$ was explained in \cite{Bah:2019rgq} (see also \cite{Hosseini:2020vgl}, and its application in \cite{Hwang:2021xyw}).

\subsection{\texorpdfstring{$SU(N_c)$}{SU(Nc)} with Non-negative R-charges ({\it twist 1})}
\label{sec:sunngtr0}

We begin with the compactification of $SU(N_c)$ SQCD on $S^2$, mixing the R-symmetry with a subgroup $U(1)_{x,y}$ of $SU(N_f)_L$ as,
	\begin{align} \label{eq:SUNfLbreak}
	SU(N_f)_L\to SU(N_1) \times SU(2n) \times SU(N_3) \times U(1)_x \times U(1)_y\,,
	\end{align}
where the $N_i$ are constrained as,
	\begin{align}
	\label{constraint}
	N_1+2n+N_3=N_f\,,\qquad n+  N_3 =  N_f-N_c\,.
	\end{align}
The twist is chosen so as to obtain non-negative integer R-charges for all the fields, with the charges of the fields under the classical 0-form symmetries given in Table~\ref{tab:s2red}, and the new R-symmetry denoted there by $U(1)_{R_1}$ (although, below to save on subscripts we will simply denote the symmetry by $U(1)_R$). We denote the first Chern class of the abelian $U(1)_B = U(1)_V/\mathbb{Z}_{N_c}$ gauge symmetry by $c_1(B)$, and similarly for the global symmetries  $U(1)_x$, $U(1)_y$, $U(1)_{R_1}$, $U(1)_A$, namely $c_1(i)$ for $U(1)_i$.  We will separately analyze the case $n=0$ in a subsequent section. 

\subsubsection{$n>0$}

The six-form anomaly polynomial contains the following contributions.  
There is a gauge anomaly involving the axial symmetry,
	\begin{align}
	\label{gaugesu}
	 \CI_6\big|_{\text{gauge}}\, =   \, 2 N_f  c_1(A)   c_2({SU(N_c)})   \,,
	\end{align}
which implies that $U(1)_A$ is broken to $\mathbb{Z}_{2N_f}$. The gravitational anomalies involving the Pontryagin class $p_1(T)$ for the spacetime tangent bundle are given by,
	\begin{align}
	\label{gravsu}
	\CI_6\big|_{\text{grav}}&\, = \, \frac{p_1(T)}{24} \left[ (N_c^2+1) c_1({{R}}) - 2 N_c N_f c_1(A)  \right]\,.
	\end{align}
Finally, there are the following purely global 't Hooft anomalies,
	\begingroup
	\allowdisplaybreaks
	\begin{align}\label{bsu}
	\CI_6\big|_{\text{global}}&\, =\,     c_1(B)^2  \left[ \frac{N_f }{N_c}c_1(A)  - c_1({{R}})  \right]  \nonumber \\
	&+c_1(B)
\Big[ 
 -nc_1({{R}})^2 +  2 c_1(R) \big(  (N_f-N_c)c_1({A})  +  nN_3 c_1(x) +nN_1 c_1(y)      \big)   \nonumber \\
	 &\qquad\quad +n  \left( N_3(N_3 + 2n)  c_1(x)^2  +N_1 (N_1+2n)c_1(y)^2 - 2 N_1 N_3 c_1(x) c_1(y)\right)   \nonumber \\ 
	 &\qquad\quad + c_2({SU(N_1)}) + c_2({SU(2n)}) +c_2({SU(N_3)}) -c_2({SU(N_f)_R}) 
  \Big]   \nonumber \\
&-\frac{1}{6} c_1({{R}})^3 \left(N_c^2+1 \right) +  N_c c_1({{R}})^2 \big[    (N_f-n) c_1(A)+ n \left(N_3 c_1(x) - N_1 c_1(y)\right) \big]   \nonumber \\
	&+c_1({{R}}) N_c  \Big[  
	-c_2(SU(N_1)) + c_2(SU(N_3)) -c_2({SU(N_f)_R}) -N_c c_1(A)^2 \nonumber \\
	&\qquad\qquad + 2n c_1(A) \left(N_3 c_1(x) + N_1 c_1(y)\right)   +  2n^2 (N_3 c_1(x)^2-N_1 c_1(y)^2 )  \Big]    \\
	 &+ \frac{N_c}{48\pi^3} \left[ \tr F_{SU(N_1)}^3 + \tr F_{SU(2n)}^3 + \tr F_{SU(N_3)}^3 - \tr F_{SU(N_f)}^3 \right]   \nonumber \\
	&+ c_2({SU(N_1)}) N_c \big(c_1(A) - 2n c_1(y)\big)     \nonumber  \\
	& +  c_2({SU(N_2)}) N_c \big(c_1(A) - N_3 c_1(x) + N_1 c_1(y)\big)   \nonumber\\
	&+ c_2({SU(N_3)})  N_c  \big(c_1(A) + 2n c_1(x)\big)    + c_2({SU(N_f)_R}) N_c  c_1(A)     \nonumber   \\
	& +  c_1(A) n N_c \left[ c_1(x)^2 N_3 N_f + c_1(y)^2 N_1N_f  - N_1 N_3 \left(c_1(x)+c_1(y)\right)^2  \right]   \nonumber \\
	&+\frac{1}{3} N_c N_f c_1(A)^3+ \frac{1}{3}n \big[ c_1(y)^3 N_1 (N_1^2-4n^2)+ c_1(x)^3 N_3 \left(4n^2-N_3^2\right) \big]   \nonumber \\
	&+ nN_1N_3 \big( -  N_1c_1(x) c_1(y)^2  +  N_3 c_1(x)^2 c_1(y)   \big) \,.   \nonumber
	\end{align}
	\endgroup

Integrating the total anomaly polynomial over $S^2$ with $U(1)_R$ flux $q_R=-1$, the pertinent integrals are as follows, 
	\begin{align}\label{s2int}
	\int_{S^2} c_1(R) = -1\,,\quad \int_{S^2} c_1(R)^2 = -2\hat{c}_1(R)\,,\quad \int_{S^2} c_1(R)^3 = - 3c_1(R)^2 - \tfrac{1}{4} p_1({SO(3)})\,.
	\end{align}
The Pontryagin class for the spacetime tangent bundle will also decompose in the reduction as $p_1(T) = p_1(T_2) + p_1(SO(3))$. Applying these formulae to \eqref{gaugesu}-\eqref{bsu} leads to the following 2d anomaly polynomial,
	\begin{align}
	\label{2dsun1}
	\begin{split}
	\int_{S^2} \CI_6&= 
	\frac{1}{2}  \left(N_c^2+1 \right) \hat{c}_1(R)^2\\
	&+2\hat{c}_1(R) \Big[ nc_1(B)   -      N_c (N_f-n)  c_1(A)- N_c  n \left(N_3 c_1(x) - N_1 c_1(y)\right) 
\Big]\\
	&+c_1(B)^2  -  c_1({B}) \left[  2 (N_f-N_c)   c_1({A}) + N_2  \left( N_3 c_1(x) +N_1 c_1(y)    \right)  \right] \\
	&+   N_c^2 c_1(A)^2  -  2n N_c c_1(A) \left(N_3 c_1(x) + N_1 c_1(y)\right) -  2n^2N_c  \left(N_3 c_1(x)^2-N_1 c_1(y)^2 \right)  \\
	&+N_c \big( c_2({SU(N_1)})    - c_2({SU(N_3)})  + c_2({SU(N_f)})         \big) - \frac{p_1(T_2)}{24}  (N_c^2+1) \,.
	\end{split}
	\end{align}

\subsubsection{$n=0$}

Taking $n$ equal to zero in the previous subsection leads to the symmetries and charges listed in Table~\ref{tab:s2red2}. In this case, the R-symmetry is twisted with the subgroup $U(1)_z\subset SU(N_f)_L$, in such a way that the R-charges of all fields are non-negative.

Both the gauge anomaly $\CI_6|_{\text{gauge}}$  and the gravitational anomaly $\CI_6|_{\text{grav}}$ are  unchanged from \eqref{gaugesu},\eqref{gravsu}, and so we will not repeat these below. The global anomalies are now computed as,
	\begingroup
	\allowdisplaybreaks
	\begin{align}
	\CI_6\big|_{\text{global}}&=   c_1(B)^2  \left[  \frac{N_f }{N_c}c_1(A)  - c_1({{R}})  \right]  \nonumber \\
	& + c_1(B) \Big[(N_f-N_c) \left(
2 c_1({{R}}) c_1(A) + 2 N_c c_1({{R}}) c_1(z) + \frac{1}{2} N_c N_f c_1(z)^2 
  \right)  \nonumber \\
  &\qquad \quad + c_2({SU(N_1)})+c_2({SU(N_3)}) -c_2({SU(N_f)_R})  \Big] \nonumber \\
&-\frac{1}{6} c_1({{R}})^3 \left(N_c^2+1 \right) + N_c N_f c_1({{R}})^2 c_1(A) \nonumber  \\
	&+ c_1({{R}}) N_c   \Big[ 
	 -c_2({SU(N_1)}) + c_2({SU(N_3)}) -c_2({SU(N_f)_R})
	 - N_c c_1(A)^2 \\
	&\qquad\qquad  + 2 N_c(N_f -N_c) c_1(A) c_1(z)  + \frac{1}{2} N_c(N_f-N_c)(2N_c-N_f)c_1(z)^2 \Big] \nonumber  \\
	  &+ \frac{N_c}{48\pi^3} \left( \tr F_{SU(N_1)}^3  + \tr F_{SU(N_3)}^3 - \tr F_{SU(N_f)}^3 \right) \nonumber \\
	 	&+   c_2({SU(N_1)}) N_c \big(c_1(A) + (N_c-N_f) c_1(z)\big) \nonumber \\
	&     + c_2({SU(N_3)}) N_c \big(  c_1(A) + N_c c_1(z)\big)   + c_2({SU(N_f)_R})N_c   c_1(A)       \nonumber     \\
	&+ \frac{1}{3} N_c N_f c_1(A)^3 + \tfrac{1}{2} N_c^2 N_f (N_f-N_c) c_1(A) c_1(z)^2 \nonumber  \\
	&+ \frac{1}{6} N_c^2 N_f ( N_f-N_c) (2N_c-N_f)c_1(z)^3 \,. \nonumber 
	\end{align}
	\endgroup
Integrating the total anomaly polynomial over $S^2$ and using \eqref{s2int} yields,
	\begin{align}\label{2dsun2}
	\begin{split}
	\int_{S^2} \CI_6 &= 
	\frac{1}{2} (N_c^2+1 ) \hat{c}_1(R)^2  -2N_c N_f \hat{c}_1(R)    c_1(A)  \\
	&+  c_1(B)^2- 2 (N_f-N_c) c_1(B) \left(c_1(A) + N_c c_1(z) \right)\\
&  +N_c^2 c_1(A)^2  - 2 N_c^2  (N_f -N_c) c_1(A) c_1(z) - \frac{1}{2} N_c^2 (N_f-N_c)(2N_c-N_f)c_1(z)^2 \\
&+ N_c \left( c_2({SU(N_1)}) - c_2({SU(N_3)}) + c_2({SU(N_f)_R}) \right) - \frac{p_1(T_2)}{24} (N_c^2+1)\,.
	\end{split}
	\end{align}

\subsection{\texorpdfstring{$U(N_c)$}{U(Nc)} with Non-negative R-charges ({\it twist 1})}
\label{sec:unngtr0}

We next discuss the anomalies of the theories with $U(1)_B$ gauged, and including the determinant matter $\Omega^{\pm}$ so as to cancel the R-symmetry gauge anomaly. These can be computed from the charges in Tables~\ref{tab:s2red}-\ref{tab:s2red2}, which were chosen so that all R-charges are non-negative. (Again, in the main text we denote the new R-symmetry by $U(1)_{R_1}$, although below we will drop the subscript.) Having already determined the anomalies of the $SU(N_c)$ theories in Section~\ref{sec:sunngtr0}, a straightforward way to proceed is as follows. First, we use  \eqref{c2s} to rewrite $c_2(SU(N_c))$ in terms of $c_2(U(N_c))$ and $c_1(B)$, where now $c_2(U(N_c))$ and $n_B=\frac{1}{2} c_1(B)^2$ are the integer-valued classes on spin manifolds. We then add to the $SU(N_c)$ anomalies the following contributions,
	\begin{align}\label{deltai}
	\begin{split}
	\Delta \CI_6&\, \supset \, 2( n_B +  c_1(B) c_1(t))\left(c_1({{R}}) + c_1(I)\right)+c_1(t)^2\left(c_1({{R}}) + c_1(I)\right) \\
	&+ \frac{1}{3}\left(c_1({{R}}) + c_1(I)\right)^3 - \frac{1}{12} p_1(T) \left[   c_1({{R}}) +  c_1(I)  \right] \\
	&+ \frac{1}{6} c_1(R)^3 - \frac{1}{24} p_1(T) c_1(R)\,.
	\end{split}
	\end{align}
The first two lines are due to the determinant matter $\Omega^\pm$, where $U(1)_t$ and $U(1)_I$ are global symmetries that act only on these fields, and the last line is due to the extra gaugino. (As the fields $\Omega^\pm$ are not charged under the symmetries that differ between the $n=0$ and $n>0$ cases, their contributions are the same for either case.)
Again, our conventions are to label the first Chern classes for abelian groups as $c_1(i)$ corresponding to $U(1)_i$. 
	
\subsubsection{$n>0$}
	
Let us first consider the case $n>0$. Summing \eqref{gaugesu}, \eqref{gravsu}, \eqref{bsu}, and \eqref{deltai}, the  total anomaly polynomial can be decomposed into the following terms. Firstly, the gauge-gauge-global anomalies are given by,
	\begin{align}
	\label{gaugeu}
	\CI_6\big|_{\text{gauge}} \,&= \,2 N_f  c_1(A)  \left( c_2({U(N_c)})   +   n_B \right)   + 2 n_B  c_1(I) \,,
	\end{align}
	implying that $U(1)_A\to \mathbb{Z}_{2N_f}$, and that the invertible part of $U(1)_I$ that is preserved is $\mathbb{Z}_2$. 
The gravitational anomalies are given by,
	\begin{align} \label{gravu}
	\CI_6\big|_{\text{grav}} &\,=\, \frac{p_1(T)}{24} \left[ (N_c^2-2) c_1({{R}}) - 2 N_c N_f c_1(A)   - 2 c_1(I)\right]\,.
	\end{align} 
The terms linear in the $U(1)_B$ curvature lead to a 2-group structure involving the magnetic 1-form symmetry, and are given by,
	\begin{align} \label{2groupan}
	\begin{split}
	 \CI_6\big|_{\text{2-group}}\,=\, & c_1(B) \Big[
 - n c_1({{R}})^2 +   2 (N_f-N_c) c_1({{R}}) c_1({A})  \\
 &\quad + c_1(R) \big( 2n \left( N_3 c_1(x) +N_1 c_1(y)    \right)  + 2 c_1(t)\big)  \\
 &\quad  +  nN_3(2n+N_3)  c_1(x)^2  +nN_1 (N_1+2n) c_1(y)^2\\
 &\quad  -  2 nN_1 N_3 c_1(x) c_1(y)+ 2 c_1(t) c_1(I) \\
 & \quad+ c_2({SU(N_1)}) + c_2({SU(2n)}) +c_2({SU(N_3)}) -c_2({SU(N_f)_R})
  \Big]\,.
  	\end{split}
 	 \end{align}
  Finally, the purely global anomalies are given by,
  	\begingroup
	\allowdisplaybreaks
	\begin{align}	\label{globalu}
	\CI_6\big|_{\text{global}}\,& =\,
	 -  \frac{1}{6} c_1({{R}})^3 \left(N_c^2-2 \right) \nonumber  \\
&+  c_1({{R}})^2   \big[   N_c (N_f-n) c_1(A)+ nN_c \left(N_3 c_1(x) - N_1 c_1(y)\right) + c_1(I) \big]  \nonumber  \\
	&+c_1({{R}}) \Big[
	  -N_c c_2({SU(N_1)}) + N_c c_2({SU(N_3)})- N_c c_2({SU(N_f)_R}) \nonumber
	  \\
	&\qquad\qquad - N_c^2 c_1(A)^2 + 2n N_c c_1(A) \left(N_3 c_1(x) + N_1 c_1(y) \right) \nonumber \\
	&\qquad\qquad + 2n^2 N_c(N_3 c_1(x)^2-N_1 c_1(y)^2 )  +  c_1(t)^2+c_1(I)^2  \Big] \nonumber \\
	&+	 \frac{N_c}{48\pi^3} \left[ \tr F_{SU(N_1)}^3 + \tr F_{SU(2n)}^3 + \tr F_{SU(N_3)}^3 - \tr F_{SU(N_f)}^3 \right]    \\
	&+ c_2({SU(N_1)}) N_c \big( c_1(A) - 2n c_1(y)\big)     \nonumber  \\
	& +  c_2({SU(N_2)}) N_c \big(c_1(A) - N_3 c_1(x) + N_1 c_1(y)\big) \nonumber   \\
	&+ c_2({SU(N_3)})  N_c  \big(c_1(A) + 2n c_1(x)\big)  + c_2({SU(N_f)_R}) N_c  c_1(A)  \nonumber \\
	&+ n N_c c_1(A) \left[ c_1(x)^2 N_3 N_f + c_1(y)^2 N_1N_f  - N_1 N_3 \left(c_1(x)+c_1(y)\right)^2  \right] \nonumber  \\
	&+\frac{1}{3} N_c N_f c_1(A)^3 + \frac{1}{3}n \left[ c_1(y)^3 N_1 (N_1^2-4n^2)+ c_1(x)^3 N_3 \left(4n^2-N_3^2\right) \right] \nonumber \\
	&+nN_1N_3 \left[-  N_1c_1(x) c_1(y)^2   + N_3  c_1(x)^2 c_1(y)  \right]+ c_1(t)^2 c_1(I) + \frac{1}{3} c_1(I)^3\,. \nonumber 
	\end{align}
Integration of the total anomaly polynomial obtained by summing  \eqref{gaugeu}, \eqref{gravu}, \eqref{2groupan}, and \eqref{globalu} over $S^2$ while utilizing \eqref{s2int} yields,
	\begin{align} \label{2dun1}
	\begin{split}
	\int_{S^2} \CI_6 &= 		
	2 c_1(B) \Big[n \hat{c}_1(R) - (N_f-N_c)  c_1({A})
  -  n \left( N_3 c_1(x) +N_1 c_1(y)    \right)  -   c_1(t) \Big]\\
  &+   \frac{1}{2}  \left(N_c^2-2 \right)  \hat{c}_1(R)^2  -2 N_c  (N_f-n)  \hat{c}_1(R) c_1(A)\\
  &  -2 \hat{c}_1(R)\Big[ n N_c  \left(N_3 c_1(x) - N_1 c_1(y)\right)   + c_1(I)
 \Big]\\
 	 &
    +  N_c^2 c_1(A)^2  - 2nN_c c_1(A) \left(N_3 c_1(x) + N_1 c_1(y)\right) \\
    &- 2n^2N_c  \left(N_3 c_1(x)^2-N_1 c_1(y)^2 \right)     - c_1(t)^2-c_1(I)^2 \\
	&N_c \left( c_2({SU(N_1)})    - c_2({SU(N_3)} ) + c_2({SU(N_f)})         \right)-\frac{p_1(T_2)}{24}  (N_c^2-2)\,.
	\end{split}
	\end{align}
	\endgroup

\subsubsection{$n=0$}

We can perform the same exercise for the case $n=0$, where the charges of the fields under the classical 0-form symmetries are given  in Table~\ref{tab:s2red2}, with the following results. The gauge anomalies $\CI_6|_{\text{gauge}}$ are the same as in \eqref{gaugeu}, and  the gravitational anomalies $\CI_6|_{\text{grav}}$ are the same as in \eqref{gravu}, which we will not repeat here. The 2-group terms linear in $c_1(B)$ are now given by,
	\begin{align} \label{2groupan0}
	\begin{split}
	 \CI_6\big|_{2\text{-group}}  &\, =\, c_1(B)\Big[
 (N_f-N_c) \left(
2 c_1({{R}}) c_1(A) + 2 N_c c_1({{R}}) c_1(z) + \frac{1}{2} N_c N_f c_1(z)^2 
  \right) \\
  &\quad+   2c_1(t) \left( c_1({{R}})   +   c_1(I) \right)  +  c_2({SU(N_1)})+c_2({SU(N_3)}) -c_2({SU(N_f)_R})  \Big]\,.
  	\end{split}
	\end{align}
The global anomalies consist of,
  	\begingroup 
	\allowdisplaybreaks
	\begin{align}\label{globalu0}
	\CI_6\big|_{\text{global}} \,&=\,
-\frac{1}{6} c_1({{R}})^3 \left(N_c^2-2 \right) +  c_1({{R}})^2 \left[N_c N_fc_1(A) + c_1(I)  \right] \nonumber \\
&+c_1(R) \Big[
- N_cc_2({SU(N_1)})  +N_c c_2({SU(N_3)}) -N_cc_2({SU(N_f)_R})
\nonumber \\
&\qquad\qquad +N_c^2 \left(  - c_1(A)^2  + 2 (N_f -N_c) c_1(A) c_1(z)\right) \nonumber \\
&\qquad\qquad + \frac{1}{2} N_c^2(N_f-N_c)(2N_c-N_f)c_1(z)^2  +  c_1(t)^2 + c_1(I)^2 \Big]  \\
	&+   \frac{N_c}{48\pi^3} \left( \tr F_{SU(N_1)}^3  + \tr F_{SU(N_3)}^3 - \tr F_{SU(N_f)}^3 \right) \nonumber \\
	&+   c_2({SU(N_1)}) N_c \big( c_1(A) + (N_c-N_f) c_1(z)\big) \nonumber \\
	&     + c_2({SU(N_3)}) N_c \big( c_1(A) + N_c c_1(z)\big)  + c_2({SU(N_f)_R})N_c  c_1(A)       \nonumber     \\
	& +c_1(t)^2 c_1(I) +  \frac{1}{3}c_1(I)^3 + \frac{1}{3} N_c N_f c_1(A)^3 + \frac{1}{2} N_c^2 N_f (N_f-N_c) c_1(A) c_1(z)^2 \nonumber \\
	& + \frac{1}{6} N_c^2 N_f ( N_f-N_c) (2N_c-N_f)c_1(z)^3\,.  \nonumber
	\end{align}
	\endgroup
Integrating the total anomaly polynomial over $S^2$ and using \eqref{s2int} yields,
	\begin{align}
	\begin{split} \label{2dun2}
	\int_{S^2} \CI_6 \,&= 	\,
	2 c_1(B)\Big[ - (N_f - N_c)  \left(c_1(A) + N_c c_1(z) \right) - c_1(t) \Big]\\
	&+\frac{1}{2} (N_c^2-2)\hat{c}_1(R)^2 - 2\hat{c}_1(R) \big(  N_c N_f c_1(A) + c_1(I)  \big) 
- c_1(t)^2 - c_1(I)^2 \\
	&  + N_c^2 c_1(A)^2  - 2N_c^2  (N_f -N_c) c_1(A) c_1(z) -  \frac{1}{2} N_c^2(N_f-N_c)(2N_c-N_f)c_1(z)^2 \\
	&+ N_c \big(   c_2({SU(N_1)}) - c_2({SU(N_3)}) + c_2({SU(N_f)_R}) \big) 
 - \frac{p_1(T_2)}{24} (N_c^2-2) \,.
	\end{split}
	\end{align}

\subsubsection{Matching anomalies across duality}
\label{sec:anommatch}

We will next verify that the perturbative anomalies of the $U(N_c)$ theories with determinant matter match for both the electric and magnetic dual theories. 	
The anomaly polynomials of the electric theories are given in \eqref{gaugeu}-\eqref{globalu} for $n>0$, and in  \eqref{2groupan0}-\eqref{globalu0}  for $n=0$. The anomaly polynomials of the dual theories can be computed similarly from the matter content in Tables~\ref{tab:dual1} and \ref{tab:dual2}, with result that their difference evaluates to (for any $n\geq 0$),
\begin{align}
\label{naivemis}
\begin{split}
\CI_6 - \widetilde{\CI}_6\, =\, &\frac{N_f^2}{N_c(N_f-N_c)} c_1(A) (c_1(B) - N_c c_1(A) )^2+ 2 N_f c_1(A) (c_1(R) + c_1(I) ) c_1(t)\,.
\end{split}
\end{align} 
Here we have utilized \eqref{map1b} to express the result in terms of the background fields in the electric basis, and we have also disregarded the gauge anomalies which lead to the same breaking pattern of the  global abelian symmetries in the two theories. 

This naive mismatch  can be understood as follows. 
The first term of \eqref{naivemis}  is cured by the observation that it contributes to the difference between the anomaly polynomials as,
\begin{align}
2 N_f N_c c_1(A) \frac{(c_1(A) - c_1(V) )^2}{2} - 2 N_f (N_f-N_c) c_1(\widetilde{A})  \frac{(c_1(\widetilde{A}) - c_1(\widetilde{V}) )^2}{2} \,.
\end{align}
The combination of curvatures $\frac{1}{2}(c_1(A) - c_1(V))^2$ is an integral class, and similarly for $\frac{1}{2}(c_1(\widetilde{A}) - c_1(\widetilde{V}))^2$, so that along with the fact that $2N_f c_1(A)$ integrates to an integer, these terms only contribute a trivial phase to the partition function. Moreover, the terms proportional to $c_1(t)$ in \eqref{naivemis} are trivialized by accounting for the breaking pattern $U(1)_I\to \mathbb{Z}_2$, and $U(1)_A\to \mathbb{Z}_{2N_f}$, so that the anomalies match as expected.

\subsection{\texorpdfstring{$U(N_c)$}{U(Nc)} with Negative R-charges ({\it twist 2})}
\label{sec:twist2}

We next consider the twisted compactification of  4d $U(N_c)$ SQCD with determinant matter on $S^2$, but this time with a choice of R-symmetry twist such that not all fields have non-negative R-charge.

Let us consider the same $SU(N_f)_L$ symmetry breaking pattern as in \eqref{eq:SUNfLbreak}, but we will now mix the various abelian symmetries with the R-symmetry as in Table~\ref{tab:charges4dnew} so as to obtain a new R-symmetry that we denote in the main text by $U(1)_{R_2}$ (for {\it twist 2}), while the charges of the fields under the other global symmetries remain the same. (Below, we will drop the $2$ subscript on $U(1)_{R_2}$ to simplify the notation.)
Doing so, the gauge anomalies and gravitational anomalies are the same as in \eqref{gaugeu} and \eqref{gravu}, which we will not repeat here. The 2-group anomalies are now given by,
	\begin{align}
	\begin{split}
	\CI_6\big|_{\text{2-group}} \,&=\, c_1(B) \Big[-c_1(R)^2 (2 N_c + n) \\
	&\qquad + 2 c_1(R) \big(   c_1(A) ( 2N_f-N_c) +c_1(t) +n N_1 c_1(y)  + n N_3 c_1(x)   \big)  \\
	&\qquad+c_2(SU(N_1))+c_2(SU(2n))+c_2(SU(N_3))-c_2(SU(N_f)_R)\\
	&\qquad + 2 c_1(I) c_1(t)+n N_1 (n+ N_c)  c_1(y)^2 -2nN_1 N_3 c_1(x)c_1(y)\\
	&\qquad+n c_1(x)^2 \left(  N_1 (n+N_c) + N_f (n N_f-2N_c) \right)  \Big]\,.
	\end{split}
	\end{align}
The global anomalies that involve the R-symmetry are as follows,
	\begin{align}
	\begin{split}
	\CI_6\big|_{\text{global}}\,&\supset \,- \frac{1}{6} c_1(R)^3 \left( 7 N_c^2 + 6 n N_c - 2\right) \\
	&+c_1(R)^2 \Big[  c_1(I)- N_c (n + 2N_c-4N_f) c_1(A)  + 3 n N_3 N_c c_1(x) + n N_1 N_c c_1(y) \Big]\\
	&+c_1(R) \Big[  N_c \left(  c_2(SU(2n)) + 2 c_2(SU(N_3)) - 2 c_2(SU(N_f)_R)  \right) \\
	&\qquad\qquad + c_1(I)^2 + c_1(t)^2 - N_c^2 c_1(A)^2 + 2 n N_c c_1(A) \left( N_1  c_1(y)   + N_3  c_1(x)  \right) \\
	&\qquad\qquad +  n N_c \left( (4 n + N_3) N_3 c_1(x)^2 + N_1^2 c_1(y)^2 - 2 N_1 N_3 c_1(y) c_1(x)  \right) 
	\Big]\,.
	\end{split}
	\end{align}
We will not need the global anomalies involving only non-R symmetries.

The 4d theory compactified on $S^2$ yields a sum over all possible gauge fluxes through the sphere. For illustrative purposes we will compute the anomaly polynomials of three different flux configurations. Firstly, when there is no gauge flux on the sphere we may simply use \eqref{s2int} along with the decomposition $p_1(T) = p_1(T_2) + p_1(SO(3))$ to obtain,
	  \begingroup
	\allowdisplaybreaks
	\begin{align}
	\int_{S^2} \CI_6\big|_{\vec{m}=0} \, &= \,c_1(B) \Big [ 2 \hat{c}_1(R) (2 N_c + n) +2 c_1(A) (N_c-2 N_f ) - 2 c_1(t) \nonumber \\
	&\qquad\quad- 2 n \left( N_3 c_1(x)+N_1 c_1(y) \right)  \Big]- \frac{p_1(T_2)}{24} (N_c^2 - 2)\nonumber \\
	&+ \hat{c}_1(R)^2 \frac{1}{2} (7 N_c^2 + 6 n N_c - 2) \nonumber \\
	&+2 \hat{c}_1(R) \Big[    N_c (n + 2 N_c - 4 N_f) c_1(A) - c_1(I)  \label{udiff2d}\\
	&\qquad\quad- n N_c \left( 3 N_3 c_1(x) + N_1 c_1(y) \right)  \Big]  +  N_c \frac{(n+N_c)}{4}  p_1(SO(3))  \nonumber \\
	&+ N_c \left[  -  c_2 (SU(2n)) - 2 c_2(SU(N_3)) + 2 c_2 (SU(N_f)_R) \right] \nonumber \\
	&-c_1(I)^2 - c_1(t)^2 + N_c^2 c_1(A)^2 - 2 n N_c c_1(A) \left( N_3 c_1(x) + N_1 c_1(y) \right)  \nonumber  \\
	& + n N_c \left( -N_1^2 c_1(y)^2 + 2 N_1 N_3 c_1(x) c_1(y) - (4n + N_3) N_3 c_1(x)^2    \right)\,. \nonumber 
	\end{align}
	\endgroup
The terms in the first two lines that are linear in $c_1(B)$ imply that the various classical symmetries are partially broken to discrete subgroups. 

We next consider the case of gauge flux $\vec{m}=(1,\dots,1)$, corresponding to $N_c$ units of flux in the $U(1)_B$ part of the gauge group,
\begin{align}
c_1(B) =  N_c e_2(S^2) + \hat{c}_1(B)\,.
\end{align} 
Now in addition to \eqref{s2int}, we require the following integrals,\footnote{~Note that since the $(\tr F)^2$ part of $c_2(U(N_c))$ contributes to the flux, we also must use \eqref{c2s}.}
\begin{align}
\begin{split}
&\int_{S^2} c_1(B) = N_c  \,, \quad \int_{S^2} c_1(B)^2 = 2 N_c \hat{c}_1(B)  \,,\quad \int_{S^2} c_1(B) c_1(R) = N_c \hat{c}_1(R) - \hat{c}_1(B)  \,,\\ 
& \int_{S^2} c_1(B) c_1(R)^2\, =\, N_c \hat{c}_1(R)^2 - 2 \hat{c}_1(B) \hat{c}_1(R) + \frac{N_c}{4} p_1(SO(3))\,.
\end{split}
\end{align}
The result of integrating the 4d anomaly polynomial using these formulae is,
	  \begingroup
	\allowdisplaybreaks
	\begin{align} \label{m111}
	\int_{S^2} \CI_6\big|_{\vec{m}=(1,\dots,1)} \, &= \,
	2\hat{c}_1(B)\Big[  (n+ 2 N_c )\hat{c}_1(R) +N_c c_1(I) +(N_c-N_f)c_1(A) \quad\  \nonumber \\
	&\qquad\quad- c_1(t) - n N_1 c_1(y) - n N_3 c_1(x)  \Big] \nonumber \\
	&+  \hat{c}_1(R)^2\left(\frac{3}{2} N_c^2 + 2nN_c - 1 \right)\nonumber\\
	&+2 \hat{c}_1(R) \Big[    N_c (n +  N_c - 2 N_f ) c_1(A) - c_1(I) +   N_c c_1(t) \\
	&\qquad\quad-   2n N_c N_3c_1(x) 
	 \Big] - \frac{p_1(T_2)}{24} (N_c^2-2)- \frac{N_c^2}{4} p_1(SO(3)) \nonumber\\
		&+ N_c \left[ c_2(SU(N_1)) -  c_2(SU(N_3)) +  c_2 (SU(N_f)_R) \right] \nonumber \\
	&-c_1(I)^2+ 2 N_c c_1(I) c_1(t) - c_1(t)^2\nonumber\\
	& + N_c^2 c_1(A)^2 - 2 n N_c c_1(A) \left(   N_1 c_1(y) +N_3 c_1(x)\right)  \nonumber  \\
	& + 2 n^2  N_c \left(  N_1c_1(y)^2 - N_3  c_1(x)^2 \right)\,.\nonumber 
	\end{align}
	\endgroup

Finally, we consider the case with gauge flux $\vec{m}=(1,0,\dots,0)$, breaking $SU(N_c)\to SU(N_c-1)\times U(1)_G$ such that the characteristic classes decompose as,
\begin{align}
c_2(SU(N_c)) = c_2(SU(N_c-1)) + \frac{N_c (N_c-1)}{2} c_1(G)^2\,,
\end{align}
 with the following gauge flux for $U(1)_B$ and $U(1)_G$,
\begin{align}
c_1(B) = e_2(S^2) + \hat{c}_1(B)\,,\qquad c_1(G) = - \frac{1}{N_c} e_2(S^2) +\hat{c}_1(G)\,.
\end{align}
The result is,
	  \begingroup
	\allowdisplaybreaks
	\begin{align}
	\label{m100}
	\int_{S^2} \CI_6\big|_{\vec{m}=(1,0\dots,0)} \,&=\, -2 N_f (N_c-1) \hat{c}_1(G) c_1(A) \nonumber \\
	&+2 \hat{c}_1(B) \Big[
	\hat{c}_1(R) (2 N_c+n) + c_1(A) \left( N_c-2N_f +\frac{N_f}{N_c} \right)\nonumber \\
	&\qquad\qquad+c_1(I) - c_1(t) - n(N_1 c_1(y) + N_3 c_1(x) 	\Big]\nonumber\\
	& - \frac{p_1(T_2)}{24} (N_c^2-2)+\frac{1}{4} \left(N_c^2-2N_c + n(N_c-1) \right)p_1(SO(3))\nonumber\\
	&+\frac{1}{2} \hat{c}_1(R)^2 \left( 7N_c^2 - 4 N_c + n(6N_c-2) -2\right) \nonumber \\
	&+2\hat{c}_1(R) \Big[
	c_1(A) \left( 2 N_f + N_c (2 N_c-4N_f + n-1) \right) - c_1(I) + c_1(t) \\
	&\qquad\qquad - n N_1 (N_c-1) c_1(y) -n N_3 (3 N_c-1) c_1(x)
	\Big]\nonumber\\
	&+ c_2(SU(N_1))-c_2(SU(2n)) (N_c-1) -c_2(SU(N_3)) (2N_c-1)\nonumber\\
	&+c_2(SU(N_f)_R) (2N_c-1) + N_c^2 c_1(A)^2 - c_1(I)^2 + 2 c_1(I) c_1(t)\nonumber\\
	& - c_1(t)^2 -2n N_c c_1(A) \left(  N_1 c_1(y) + N_3 c_1(x)  \right)\nonumber\\
	&+c_1(y)^2 n N_1 (n + N_c - N_1 N_c)+2 n N_1 N_3\left(N_c-1) \right)c_1(x) c_1(y)\nonumber \\
	&+ c_1(x)^2n N_3 \left( 2n(1-2N_c) + N_3 (1-N_c)  \right)\,.  \nonumber
	\end{align}
	\endgroup

\subsection{\texorpdfstring{$U(N_c)$}{U(Nc)} to Recover Triality ({\it twist 3})}
\label{sec:untriality}

Finally, we consider the 4d $U(N_c)$ theory discussed in Section~\ref{sec:triality}, which is parameterized in such a way as to result in the 2d theories that participate in a triality. In the main text we refer to the R-symmetry in this case by $U(1)_{R_3}$, although we will drop the $3$ subscript below. The matter content and symmetries of the 4d theory are listed in Table~\ref{tab:4dchargetri}, with the integers $N_i$ satisfying $N_1+N_2-N_3=2N_c$, leading to the following anomalies. Firstly, the gauge anomalies take the form,
	\begin{align} \label{gaugetri}
	\CI_6\big|_{\text{gauge}}\, &=\, 2 N_1  c_1(A)  \left( c_2({U(N_c)})   +   n_B \right)   + 2 n_B  c_1(I) \,.
	\end{align}
The gravitational anomalies are,
	\begin{align}\label{gravtri}
	\CI_6\big|_{\text{grav}}\, &=\, \frac{p_1(T)}{24} \left[ ( N_c^2-N_1N_2-2) c_1({{R}}) - 2 N_1 c_1(A) -2 c_1(I) + N_1 c_1({U(1)_2})\right]\,.
	\end{align} 
The anomalies linear in $U(1)_B$ that lead to the 2-group are,
	\begin{align}\label{2grouptri}
	\begin{split}
	 \CI_6\big|_{\text{2-group}}\,= \,& c_1(B) \Big[
	  -  c_2({SU(N_1)}) + \frac{1}{2} (4-N_1+N_2^2+N_3^2)c_1({{R}})^2
\\
 &\quad +
 c_1({{R}})  \Big( c_1(A) (N_1-N_2+N_3) + 2 c_1(I) + 2 c_1(t)  \\
 &\qquad\quad\quad - N_2 c_1({U(1)_2}) + N_3 c_1({U(1)_3})  \Big)  +2 c_1({I})c_1({t})\\
 &\quad +\frac{1}{2} (N_1-2)(N_1-3) c_1({U(1)_1})^2  + c_1({U(1)_2})^2+c_1({U(1)_3})^2\\
 &\quad  + (N_1-3)c_1({U(1)_1}) \big(c_1({U(1)_2}) + c_1({U(1)_3})\big) + c_1({U(1)_2})c_1({U(1)_3})
  \Big]\,.
  \end{split}
  \end{align}
  Finally, the global anomalies involving the R-symmetry are given by,
  	  \begingroup
	\allowdisplaybreaks
  	\begin{align}
	 \CI_6\big|_{\text{global}} &\supset 
\frac{1}{6} c_1({{R}})^3 \Big[ N_c^2 + N_c ( -N_1 - N_2^3 + N_3^3) + N_1 N_2^3 + 8\Big]  \nonumber\\
&\quad+   c_1({{R}})^2 \Big[  c_1(A) \left(  \frac{1}{2} N_c (N_1+N_2^2 + N_3^2) - N_1 N_2^2 \right) + 2 c_1({I}) + 2 c_1(t) \nonumber  \\
&\qquad\quad + \frac{1}{2} N_c N_3^2  c_1({U(1)_3})+ \frac{1}{2} N_2^2 ( N_c-N_1  )c_1({U(1)_2} )  \Big] \nonumber \\
&\quad+ c_1({{R}})\Big[ 
 \frac{1}{2} (-N_1+N_2+N_3)c_2({SU(N_1)} ) +
c_1(I)^2 + 2 c_1(I) c_1(t) + c_1(t)^2 \label{globaltri}\\
 &\qquad\quad + c_1(A) \left(  N_3 N_cc_1({U(1)_3} )  + N_2 (2 N_1-N_c)c_1({U(1)_2} )\right) \nonumber \\
     &\qquad \quad   + \frac{1}{2} \left( c_1({U(1)_2})^2 N_2 (N_1-N_c) +c_1({U(1)_3})^2N_3 N_c \right) \nonumber \\
     &\qquad\quad  +   c_1(A)^2(2 N_1 N_2 - N_c^2 ) \Big] \,. \nonumber
	\end{align}
	\endgroup
  There are a number of other global abelian anomalies that we have not written here. 
	
Integration of the total anomaly polynomial consisting of \eqref{gaugetri}, \eqref{gravtri}, \eqref{2grouptri}, and \eqref{globaltri} over $S^2$ in the zero flux sector yields the following,
	  \begingroup
	\allowdisplaybreaks
	\begin{align}
	\int_{S^2} \CI_6 \big|_{\vec{m}=0}&{=} c_1(B)\Big[
	 \hat{c}_1(R) (N_1-N_2^2-N_3^2-4) - c_1(A) (N_1-N_2+N_3)\nonumber \\
	 &\qquad\quad -2 ( c_1({I}) + c_1(t) )+ N_2c_1({U(1)_2} ) -N_3 c_1({U(1)_3} ) \Big]\nonumber \\
&+	\frac{1}{ 96} p_1(T_2)  \left[ 8-(N_1^2+N_2^2+N_3^2) + 2 (N_1N_2+N_2N_3+N_1N_3) \right]\nonumber \\
	&- \frac{1 }{48} p_1({SO(3)})\Big[ 12 + N_2 (N_2^2-1) (N_1-N_2+N_3) \nonumber\\
	&\qquad\quad+ N_3 (N_3^2-1) (N_1+N_2-N_3)   \Big] +\frac{1}{2} (N_1-N_2-N_3) c_2({SU(N_1)}) \nonumber \\
&-\frac{1}{4}\hat{c}_1(R)^2\left[  (N_2^3-N_c) (N_1-N_2+N_3) + N_3^3 (N_1+N_2-N_3) + 16   \right]  \\
	&+\hat{c}_1(R)\Big[ -4 (c_1(I) + c_1(t)) - \frac{1}{2}  N_3^2 (N_1+N_2-N_3)c_1({U(1)_3}) \nonumber \\
	&\qquad+ \frac{N_2^2}{2}  (N_1-N_2+N_3)c_1({U(1)_2})  + c_1(A) (2 N_1 N_2^2 - N_c (N_1+N_2^2 + N_3^2)\Big]\nonumber \\
& - (c_2(I) + c_1(t))^2 - \frac{1}{4}  N_3 (N_1+N_2-N_3) c_1({U(1)_3})^2\nonumber\\
&- \frac{1}{4}  N_2 (N_1-N_2+N_3)c_1({U(1)_2})^2 + \frac{1}{2} (-4 N_1 N_2 + N_c (N_1+N_2-N_3)) c_1(A)^2\nonumber \\
	&- c_1(A) (  c_1({U(1)_3}) N_3 N_c + c_1({U(1)_2}) N_2 (2 N_1-N_c)  ) \,.\nonumber
	\end{align}
	\endgroup
As discussed in the main text, $U(1)_t$ is a redundant symmetry since it coincides with $U(1)_I$.

\end{appendix}

\bibliographystyle{JHEP}
\bibliography{refs}

\end{document}